\documentclass[%
 aip,
 amsmath,amssymb,
 reprint,nofootinbib%
]{revtex4-2}
\usepackage{nomencl}
\makenomenclature

\usepackage{graphicx,array,booktabs,rotating}
\usepackage{dcolumn}
\usepackage{bm}
\usepackage{hyperref}
\usepackage{mathptmx}
\usepackage{multirow}
\usepackage{afterpage,float,color,xcolor}
\usepackage{makecell,mathtools,tabularx,tabulary}
\setcellgapes{3pt}
\usepackage[capitalize]{cleveref}
\usepackage{times}
\usepackage{etoolbox}
\hypersetup{
	colorlinks,
	linkcolor={blue!100!black},
	citecolor={blue!100!black},
	urlcolor={blue!100!black}
}
\newcommand{\PRLsep}{\noindent\makebox[\linewidth]{\resizebox{0.3333\linewidth}{1pt}{$\bullet$}}\bigskip}

\newcommand{\etal}{\textit{et al.}}
\usepackage[page]{totalcount}
\usepackage{etoolbox,fancyhdr,xcolor,soul}%
\usepackage[displaymath,pagewise]{lineno}

\begin{document}
\preprint{AIP/123-QED}

\title[\textbf{Prog. Energy Combust. Sci.} (2022) $\vert$ Manuscript Accepted \href{https://doi.org/10.1016/j.pecs.2022.101042}{(10.1016/j.pecs.2022.101042)}]{A review of diaphragmless shock tubes for interdisciplinary applications}
\author{S. Janardhanraj$^*$}
 \affiliation{Clean Combustion Research Center, King Abdullah University of Science and Technology, Thuwal 23955, Saudi Arabia}
 \email{janardhanraj.subburaj@kaust.edu.sa}
\author{S. K. Karthick}%
 \affiliation{Currently at the Faculty of Aerospace Engineering, Technion - Israel Institute of Technology, Haifa 3200003, Israel}%
\author{A. Farooq}
 \affiliation{Clean Combustion Research Center, King Abdullah University of Science and Technology, Thuwal 23955, Saudi Arabia}
\date{Received 2 December 2021; Received in revised form 18 September 2022; Accepted 25 September 2022}

\begin{abstract}
Shock tubes have emerged as an effective tool for applications in various fields of research and technology. The conventional mode of shock tube operation employs a frangible diaphragm to generate shock waves. The last half-century has witnessed significant efforts to replace this diaphragm-bursting method with fast-acting valves. These diaphragmless methods have good repeatability, quick turnaround time between experiments, and produce a clean flow, free of diaphragm fragments, in contrast to the conventional diaphragm-type operation. The constantly evolving valve designs target shorter opening times for improved performance and efficiency. The present review is a compilation of the different diaphragmless shock tubes that have been conceptualized, developed, and implemented for various research endeavors. The discussions focus on essential factors, including the actuation mechanism, driver-driven configurations, valve opening time, shock formation distance, and operating pressure range, that ultimately influence the shock wave parameters obtained in the shock tube. A generalized mathematical model to study the behavior of these valves is developed. The advantages, limitations, and challenges in improving the performance of the valves are described. Finally, the present-day applications of diaphragmless shock tubes have been discussed, and their potential scope in expanding the frontiers of shock wave research and technology is presented. 
\end{abstract}

\maketitle
\onecolumngrid
\tableofcontents
\vspace{5mm} 
\vspace{-5mm}
\twocolumngrid 
\mbox{}
\nomenclature[A, 01]{$P_0$, $p_0$}{Total pressure (Pa)}
\nomenclature[A, 02]{$P_{5-1}$}{Pressure in the corresponding regions in the shock tube (bar)}
\nomenclature[A, 03]{$P_{41}$}{Ratio of $P_4$ and $P_1$ (-)}
\nomenclature[A, 04]{$P_{51}$}{Ratio of $P_5$ and $P_1$ (-)}
\nomenclature[A, 05]{$T_{5-1}$}{Temperature in the corresponding regions in the shock tube (K)}
\nomenclature[A, 06]{$T_0$}{Total temperature (K)}
\nomenclature[A, 07]{$\gamma$}{Specific heat ratio (-)}
\nomenclature[A, 08]{$R$}{Specific gas constant (J/kg/K)}
\nomenclature[A, 09]{$M_\infty$}{Freestream Mach number (-)}
\nomenclature[A, 10]{$M_S$}{Shock Mach number (-)}   
\nomenclature[A, 11]{$x$, $y$}{Axial or radial distance (m)}
\nomenclature[A, 12]{$t$}{Time (s)}
\nomenclature[A, 13]{$t_r$}{Run time (s)}
\nomenclature[A, 14]{$t_{op}$}{Opening time (s)}
\nomenclature[A, 15]{$t_{ft}$}{Flow time (s)}
\nomenclature[A, 16]{}{}
\nomenclature[A, 17]{}{}
\nomenclature[A, 18]{}{}
\nomenclature[A, 19]{}{}
\nomenclature[A, 20]{}{}
\nomenclature[A, 21]{}{}
\nomenclature[A, 22]{}{}
\nomenclature[A, 23]{$u, v$}{$x$ and $y$ component of velocity (m/s)}
\nomenclature[A, 24]{$x_f$}{Axial shock formation distance (m)}
\nomenclature[A, 25]{$\Delta t_R$}{Steady time (s)}
\nomenclature[A, 26]{$F$}{Force (N)}
\nomenclature[A, 27]{$D$, $\phi$, $d$}{Diameter (m)}
\nomenclature[A, 28]{$r$}{Radius (m)}
\nomenclature[A, 29]{$a$}{Speed of sound (m/s)}
\nomenclature[A, 30]{$d_h$}{Hydraulic diameter (m)}
\nomenclature[A, 31]{$A$}{Area (m$^2$)}
\nomenclature[A, 32]{$PR$}{Perimeter (m)}
\nomenclature[A, 33]{$\mu$}{Mean of a quantity (-)}
\nomenclature[A, 34]{$\sigma$}{Standard deviation of a quantity (-)}
\nomenclature[A, 35]{$m$}{Mass of a system (kg)}
\nomenclature[A, 36]{$V$}{Volume of a system (m$^3$}
\nomenclature[A, 37]{$\rho$}{density of a medium/material (kg/m$^3$)}
\nomenclature[A, 38]{$a_p$}{Acceleration of the piston (m/s$^2$)}
\nomenclature[A, 39]{$L$}{Length (m)}
\printnomenclature[2cm]
\onecolumngrid
\vspace{5mm} 
\vspace{5mm}
\twocolumngrid
\section{\label{sec1:intro}Introduction}
Shock waves are a fascinating physical phenomenon resulting from rapid compression in matter. They propagate at supersonic velocities, and their effects are observed in all forms of matter. The formation and propagation of shock waves in a gaseous medium are of particular interest in this review. A typical shock wave in gas is characterized by a moving shock front that increases the fluid's pressure, temperature, and density in a very short interval of time (referred to as the rise time of the shock wave). These thermodynamic properties remain constant for a period called the steady time and gradually decrease to the equilibrium state over the decay time. The rise time is typically on the order of microseconds. In contrast, the steady and decay times can range from microseconds to milliseconds depending on the energy of the source, method of shock wave production, and propagation dynamics. Shock waves are multi-scale and are observed in length scales varying from micro- to macroscopic regimes. This property has led to a plethora of shock wave-related disciplines being established for over a century now \cite{Ben_2000}. The early applications of shock waves were in chemical kinetics and aerospace research to study processes in high-temperature gases and high-speed flows \cite{Bradley_1962, Gaydon_1963}. The emerging applications of shock waves in interdisciplinary fields of science and technology have opened up new avenues for collaborative research \cite{Takayama_2004, Takayama_2005a, Takayama_2005b}. Shock waves have also demonstrated the potential to address present-day industry challenges, such as preservative impregnation in bamboo, sandal oil extraction, removal of micron-size dust from silicon wafers, cell transformation, meat tenderization, and enhancing material properties \cite{Jagadeesh_2008, Jagadeesh_2009, Bolumar_2013, Murr_1988}. The prospect of developing novel disruptive technologies using shock waves is an added incentive for modern technologists and entrepreneurs.

The sudden release of energy in a confined space results in a supersonic displacement of gas and leads to the formation of shock waves. Chemical, mechanical, nuclear, or electrical energy can be a source of shock waves. Explosives\cite{Kinney_2013}, laser irradiation\cite{Director_1977}, electric discharge\cite{Honma_1991}, pressurized gases\cite{Ben_2000}, and detonable gaseous mixtures\cite{Ben_2000} are commonly used to produce shock waves in gases for research purposes. Among these methods, the use of pressurized gas in a device called the shock tube\cite{Bleakney_1949} is a simple, economical, and safe method to generate shock waves in a controlled manner. A simple shock tube comprises two sections separated by a diaphragm and is operated by pressurizing one section until the diaphragm ruptures to form a shock wave in the other section. Although shock tubes seem to have a simple operational procedure, the use of diaphragms has been found wanting for various reasons described in the following section. Therefore, a diaphragmless-mode of operation for shock tubes has been explored and implemented.    

The main idea behind a diaphragmless shock tube is to eliminate the diaphragm burst process and replace it with a quick-opening valve while retaining the performance capabilities of a diaphragm-type shock tube. Ideally, a diaphragmless shock tube has good repeatability, a high repetition rate, the ability to automate, and requires less manual intervention and physical effort. Also, in a diaphragmless shock tube, the valve opening process does not contaminate the flow downstream, unlike conventional diaphragm-type shock tubes. The opening time in the case of a diaphragm rupture generally varies from hundreds of microseconds to a few milliseconds depending on the diameter, pressure difference, material properties, and thickness of the diaphragm. Practically, it is an engineering and manufacturing challenge to design fast-acting valves with large diameters and opening times on the order of milliseconds. Moreover, producing a high-enthalpy shock wave using a diaphragmless valve as required in certain aerodynamic testing facilities\cite{Wagner2018,Capra2015} and materials research\cite{Nagaraja2011,Jayaram2011} is even more cumbersome. Numerous diaphragmless valve concepts have been proposed since the first concept of a `shock wave valve' was introduced by Condit in 1954\cite{Condit_1954}. With advances in manufacturing technology and high-performance actuation systems, fast-acting valves with improved performance and efficiency have been realized.

An assessment of diaphragmless shock tubes has been reported previously\cite{Kosing_1999,Alvarez_2015,McGivern_2019}, but the review was limited to only a few design concepts. The present work is a comprehensive compilation of diaphragmless valve concepts reported for over half a century, and it is the first such review to the best of our knowledge. The basic parameters and terminology used in diaphragmless shock tubes are initially defined and elucidated. Subsequently, the different design concepts reported in the literature are distinguished based on the mounting configuration, operating principle, and actuation mechanism. The advantages and shortcomings of specific diaphragmless valve designs have been identified. A generalized mathematical model and relations for opening time are presented based on the typical forces experienced by the moving element in a diaphragmless valve. A detailed procedure of the various design points that must be considered while developing such valve concepts has also been included, which helps analyze the valve performance analytically. The modern applications of shock waves that have unfolded with diaphragmless shock tubes in various fields of research and technology have been reviewed. Overall, the present communication attempts to identify the gaps, challenges, and opportunities in developing diaphragmless valves for interdisciplinary shock wave applications.

\section{\label{sec2:fastacting}Fast-acting valves in shock tubes}
The wave systems in a diaphragm-type and a diaphragmless shock tube are discussed considering a simple 1-D inviscid adiabatic flow. The use of diaphragms in a shock tube comes with numerous disadvantages. The major shortcomings of conventional shock tubes and how these are addressed using fast-acting valves are also described in this section.

\subsection{\label{sec2a:basicdialess}Wave system in a shock tube}
The flow field in a shock tube is complex, unsteady, and dependent on the initial conditions in the two sections of the shock tube. Generally, the pressurized gas filled in the high-pressure section of the shock tube is termed the ``driver gas," while the gas filled in the low-pressure section is called the ``test gas" or ``driven gas." When the barrier between the driver and driven gas is removed, a shock wave is formed that propagates in the low-pressure section. Simultaneously, expansion or rarefaction waves propagate in the opposite direction into the high-pressure chamber. The expansion fan region gradually reduces the pressure and temperature of the driver gas and is bound by the rarefaction head and tail. The rarefaction waves get reflected from the end-wall of the high-pressure section and then travel towards the driven gas region. Meanwhile, the shock wave travels towards the end-wall of the low-pressure section, where the incident shock wave compresses the gas behind it. The incident shock wave reflects off the end wall and travels into the onward flow, increasing the pressure and temperature a second time. For studies that utilize the incident shock wave, the steady time of the shock wave at a given location is the time interval between the arrival of the shock front and the contact surface (an imaginary surface that separates driven and driver gases). Most of the studies utilize the stationary shocked gas behind the reflected shock. In this case, the steady time window is the time interval between the reflection of the incident shock wave and the reflected waves from the contact surface. 

The entire flow in the shock tube is generally divided into five regions. Region 1 and 4 are the undisturbed gas in the shock tube's low- and high-pressure sections, respectively. Region 2 represents the gas between the incident shock wave and the contact surface, while region 3 is the flow behind the contact surface. The stagnant gas behind the reflected shock wave is in region 5. Thermodynamic parameters in these regions are indicated by using the corresponding number of the region as a subscript. For example, $P_4$ and $T_4$ are the pressure and temperature of the undisturbed driver gas, while $P_1$ and $T_1$ represent the pressure and temperature of the undisturbed driven gas, respectively. The ratio between the parameters in different regions is also generally represented using a subscript. For example, the ratio of $P_4$ and $P_1$ is indicated as $P_{41}$, ratio of $T_4$ and $T_1$ as $T_{41}$ and so on. The relationship between the initial conditions and the shock wave parameters can be solved exactly for a simple case assuming a one-dimensional, inviscid, and adiabatic flow\cite{Gaydon_1963}. The pressure ratio, $P_{41}$, is related to the shock Mach number, $M_S$, as,

\begin{equation}\label{eq1}
    P_{41} = \frac{2\gamma _1 M_S^2-(\gamma _1-1)}{\gamma _1 +1} \left(1-\frac{\gamma _4-1}{\gamma _1+1}\frac{a_1}{a_4}\left(M_S-\frac{1}{M_S}\right)\right)^{ - \frac{2\gamma _4}{\gamma _4 -1}}
\end{equation}

where $\gamma$ is the specific heat ratio and $a$ is the local speed of sound. A helpful method to represent the wave system in a shock tube is using an $x-t$ diagram or wave diagram. This diagram is constructed using a technique called the method of characteristics that considers flow perturbations to travel at the local speed of sound. For a stationary observer, the perturbations inside the shock tube travel with the sum of the local speed of sound and the gas velocity. It is also assumed that the shock wave instantly forms at the diaphragm location after the rupture. The wave diagram for a diaphragm-type shock tube is commonly used and can be found in multiple references\cite{Gaydon_1963,Bradley_1962}.

\begin{figure*}
\includegraphics[scale=0.55]{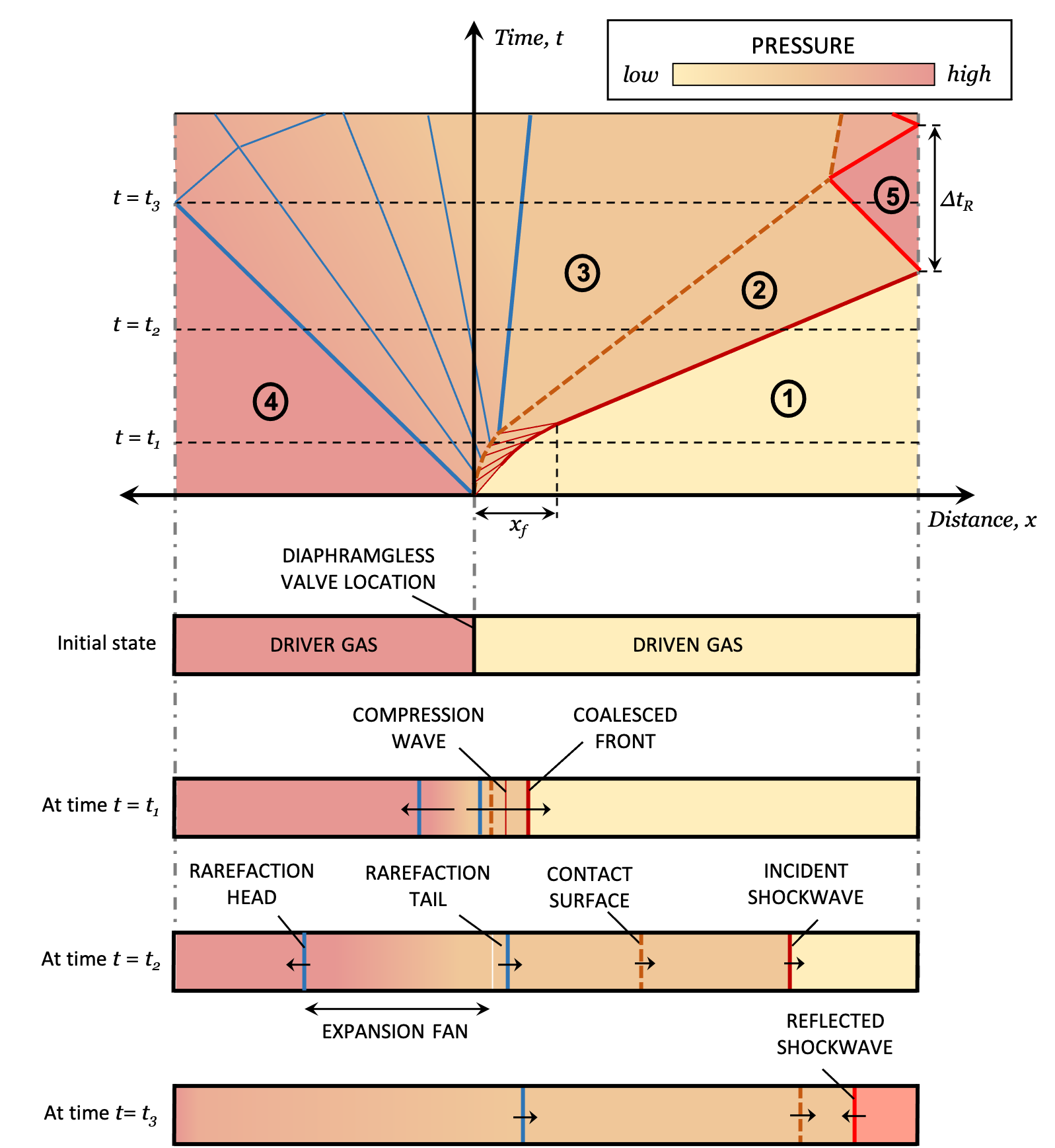}
\caption{\label{x-tdiag} A schematic diagram showing the distance-time ($x-t$) plot along with pressure contour for a diaphragmless shock tube and the corresponding wave system at time instants $t=0$, $t=t_1$, $t=t_2$ and $t=t_3$. $x_f$ indicates the shock formation distance and $\triangle t_R$ is the steady time of the reflected shock wave. {\color{black}(Adapted by permission from The Japan Society of Mechanical Engineers: Bulletin of JSME, Ikui et al.\cite{Ikui_1979}, copyright 1979)}}
\end{figure*}

The flow in a diaphragmless shock tube can also be represented using an $x-t$ diagram, assuming one-dimensional, inviscid, and adiabatic flow. Here, since a fast-acting valve replaces the diaphragm at the interface between the driver and the driven gas, the slower opening of the valve compared to the diaphragm rupture time results in a longer shock formation distance. Hence, the shock formation in a diaphragmless shock tube is an important process that has to be shown in the wave diagram. To understand the growth of the shock wave as a result of the finite opening time of the valve, the accelerating piston analogy described by Becker is useful\cite{Becker_1922}. Consider a tube inside which a piston accelerates from rest to a constant velocity, '$v$' ($v$ greater than the speed of sound). Let the piston reach velocity '$v$' through small increments over a finite time. The first increment in piston velocity causes a weak compression wave to propagate in the tube, compressing the gas uniformly and adiabatically behind it. The compression wave generated by the second velocity increment travels in a gas with a slightly higher sound speed due to the preceding compression wave's propagation. Subsequently, each compression wave produced by the piston motion moves in a gas, pressurized and heated by the previous compression wave, and eventually catch up with the preceding waves. The waves coalesce to form a shock wave that reaches a steady velocity when the piston's speed becomes steady. In a shock tube, the motion of the piston is analogous to the contact surface, which is a constant-pressure interface. 

Figure \ref{x-tdiag} shows the wave diagram for a diaphragmless shock tube with a pressure contour to highlight the typical pressure variations in the tube. The position of the corresponding waves in the shock tube at different time instants ($t=0$, $t=t_1$, $t=t_2$ and $t=t_3$) are shown below the wave diagram. The initial state of the shock tube is shown at time $t=0$. At the time $t=t_1$, the shock wave formation from compression waves is seen. The fully developed flow in the diaphragmless shock tube is seen at $t=t_2$. The reflections of the shock wave and expansion waves from their corresponding end-walls are seen at time $t=t_3$. The steady time duration in the reflected shock region is represented by $\triangle t_R$. Figure \ref{x-tdiag} also shows the development of the coalesced shock front from the merging of the characteristics developed due to the accelerating contact surface\cite{Ikui_1977}. The distance from the initial driver-driven gas interface to the location where the shock wave is formed is termed as the shock formation distance ($x_f$). The shock formation distance is also the distance from the driver-driven interface where the shock front reaches the maximum velocity. The shock formation process in the shock tube is directly proportional to the opening time of the shock tube. The slower the opening time of the valve, the longer the shock formation distance\cite{Janardhanraj_2021}. Therefore, diaphragmless shock tubes need longer driven sections, in general, compared to conventional diaphragm-type shock tubes.

\subsection{\label{sec2b:drawbacks}Drawbacks of diaphragm-type shock tubes}
The major reasons for exploring alternatives to conventional diaphragm-type shock tubes are listed below:
\begin{itemize}
    \item \textbf{Run-to-run variation} - The ability to replicate the burst process of the diaphragm determines the reproducibility of the shock wave conditions in the shock tube. The diaphragm opening can be very irregular, and, on many occasions, a part of the diaphragm can obstruct the flow of gas due to an incomplete opening\cite{Gaetani_2008,Nguyen_2014}. Therefore, every experiment performed in a diaphragm-type shock tube has a unique flow condition as the burst process differs for every run. The inability to obtain repeatable test conditions is a significant drawback of diaphragm-type shock tubes. 
    \item \textbf{Long turnaround times} - Diaphragm replacement is time-consuming in many shock tube facilities. In most cases, it might take a few minutes or even an hour in some large facilities. For investigations requiring shock waves to be produced at a high repetition rate (on the order of seconds or lower), the conventional diaphragm-type shock tube is unsuitable.
    \item \textbf{Manual effort} - Large-scale shock tubes typically have lengths of about 10–15 m and internal diameters in the range of 50-200 mm. There is a requirement for sufficient human resources or, in some cases, expensive hydraulic systems to disassemble bulky flanges at the diaphragm stations. In the case of miniature shock tubes, diaphragm changing becomes very cumbersome because of small-size fasteners. Also, manual effort is required to employ the best quality control methods in diaphragm manufacturing. Eliminating manual intervention is necessary to automate the shock tube facility. Automation can help shift the focus from spending time/energy operating the shock tube to the primary research/project goals.
    \item \textbf{Debris from diaphragm rupture} - The diaphragm rupture is a source of tiny fragments carried by the flow to the end of the shock tube. This debris produced by diaphragm rupture is particularly menacing for test samples, sensors, observation windows, and diagnostics in the shock tube. The fragments accumulate in the shock tube after several experiments and must be removed to avoid further damage to components in pneumatic lines. In some specific chemical kinetic studies, it has been observed that the debris can cause inhomogeneous ignition of fuels and hence lead to undesirable effects\cite{Tulgestke_2018}. The process of removal of debris and cleaning the shock tube is cumbersome and time-consuming. 
    \item \textbf{Unique opening time in every experiment} - Although instantaneous removal of the diaphragm is quintessential in an ideal shock tube, in reality, the diaphragm rupture process takes a finite time. The complex flow phenomena generated close to the diaphragm location depend on the diaphragm's opening time, which in turn is strongly related to the diaphragm's material properties of the diaphragm\cite{White_1958,Rothkopf_1974}. Different materials (polycarbonate, steel, aluminum, Mylar, Lexan, etc.) of varying thicknesses are used as diaphragms to vary the shock wave conditions. Therefore, the opening time of the diaphragm is unique for a particular combination of the diaphragm material, thickness, and shock tube dimensions\cite{Vianna_1999}.
    \item \textbf{Resistance behavior of thicker diaphragms} - In high-pressure shock tubes, thicker diaphragms have to be used to produce strong shock waves. It has been reported that thicker diaphragms exhibit significant resistance to opening due to the stresses at the hinge line of the diaphragm\cite{Hickman_1975}. Therefore, the flow produced in high-pressure shock tubes depends on thicker diaphragms' resistance behavior.
    \item \textbf{Extremely thin diaphragms for small shock tubes} - Engineering a small-scale shock tube poses many challenges as the diaphragm thickness would be on the order of micrometers or nanometers\cite{Brouillette_2003}. A very minute change in the thickness of the diaphragm is required to vary the burst pressure in small steps. In practice, it would be challenging and expensive to fabricate such diaphragms. Therefore, an alternative to using diaphragms is necessary for miniature shock tubes. For low-pressure applications, the flow behind the incident shock wave can be used instead of the conditions behind the reflected shock wave, depending on the nature of the experimental study. Nevertheless, the diaphragm choice for small increments in the burst pressure remains. 
    \item \textbf{Consumable and waste footprint} - Diaphragm-type shock tubes require the replacement of the frangible diaphragm after every single run. There is significant wastage of material during the fabrication of diaphragms and after the completion of experiments.
    \item \textbf{Impurities in ambient air} - In diaphragm-type shock tubes, an inert gas is filled in the tube when changing a diaphragm to avoid the release of harmful gases from previous experiments to be released into the ambient air. There are a few specific applications, such as in Gas Dynamic Lasers (GDL), for which it is essential to avoid the exposure of the inside of the shock tube to ambient air after every test\cite{Rego_2007_1}. Certain species in the ambient air can act as impurities and excite/de-excite the upper/lower laser levels. In such applications, a diaphragmless shock tube is a safer option.
\end{itemize}

Several methods have been reported to minimize shot-to-shot variation in diaphragm rupture. Generally, a V-groove notch is machined along two diameters at right angles to form a cross-shape ($\times$) on the exposed portion of the diaphragm facing the low-pressure section\cite{Gaydon_1963}. The high-pressure fractures the diaphragm along the preferential cross-shape and opens into the low-pressure section with the formation of four petals. Using electrical discharge to initiate diaphragm breaking gives short opening times and precise timing of the rupture\cite{Bradley_1965,Cole_1958}. A diaphragm-cutter can also obtain controlled diaphragm bursts, as demonstrated in a 432 mm diameter shock tube, to get an opening time of about one millisecond\cite{Roshko_1961}. The diaphragm bulges after pressurizing the high-pressure section, and the cutter is optimally placed so that the diaphragm is cut into four sections that open out like petals. A gas-operated clamp with a needle has also been reported to obtain repeatable diaphragm bursts and quick changing of the diaphragm ($\leq$ 1 min)\cite{Skews_1968}. Another novel method is using a double-diaphragm technique which is commonplace in high-pressure shock tubes, especially while avoiding the use of thicker diaphragms\cite{Stotz_2008}. By quickly evacuating a small buffer volume between the two diaphragms, the rupture of the two diaphragms is attained at the required pressure. These techniques work reasonably well in addressing run-to-run variations, but the other drawbacks of using diaphragms need to be tackled.

\section{\label{sec3:overview}Overview of fast-acting valve concepts}
Numerous diaphragmless valves with different configurations have been designed and implemented for research and technological applications. Each design has unique features in terms of operating principle, actuation techniques, mounting configuration, operating pressure range, overall size, and shock tube dimensions, making these suitable for the specific studies they facilitated. Diaphragmless valve designs are analyzed based on these features in the following sections.

\begin{figure*}
\includegraphics[scale=0.5]{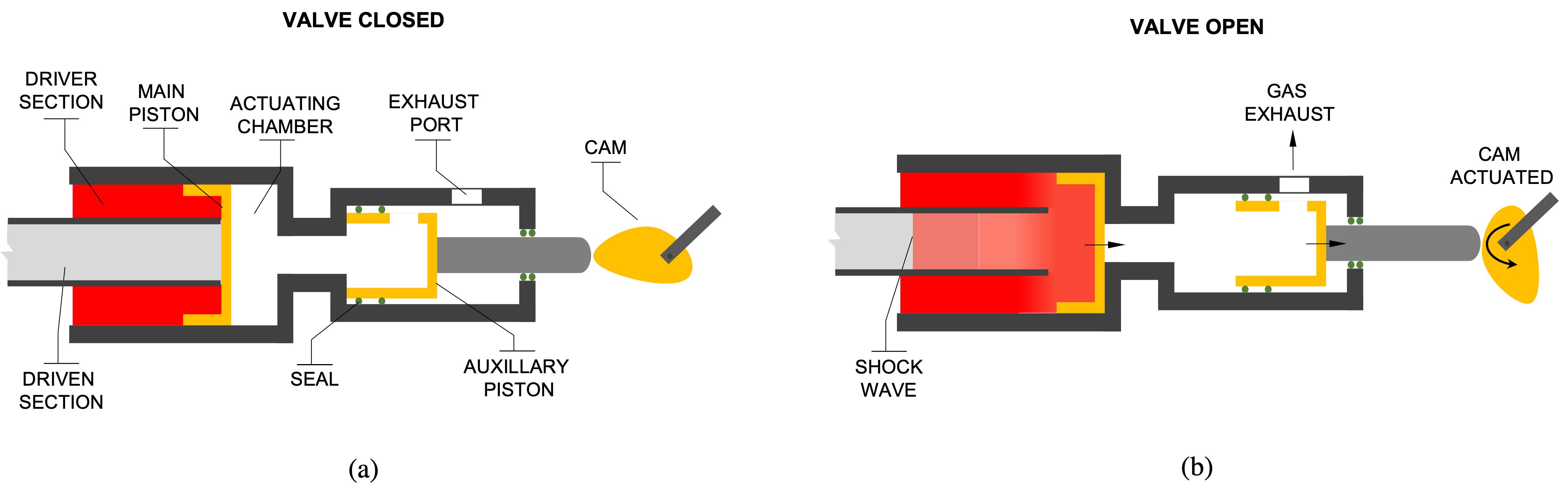}
\caption{\label{Muirhead1964}Schematic diagrams showing the valve design proposed by Muirhead et al.\cite{Muirhead_1964} {\color{black}(Adapted by permission from AIP Publishing: Review of Scientific Instruments, J. C. Muirhead and W. A. Jones\cite{Muirhead_1964}, copyright 1964)}. A double piston arrangement was used in a coaxial driver-driven configuration actuated using a cam. (a) Valve in the closed position with the high-pressure driver gas filled in the driver section. (b) The valve opens when the cam is actuated, retracting the main and auxiliary piston. Cam was replaced by a solenoid valve by Oguchi et al.\cite{Oguchi_1975}}.
\end{figure*}

\subsection{\label{sec3a:early}Early concepts (Prior to 1980)}
Condit\cite{Condit_1954} presented the idea of a shock wave valve that consisted of a piston held in position by pressurizing gas in an actuating chamber behind it (described in Muirhead et al.\cite{Muirhead_1964}). The design incorporated an annular driver section and a cam-actuated poppet, which quickly released the gas in the actuating chamber, allowing rapid retraction of the piston that initially sealed the driver gas. Condit's design was implemented in shock tubes with driven section diameters of 1 in. and 17 in. for maximum driver pressures of 600 and 200 psi, respectively. Muirhead et al. \cite{Muirhead_1964} modified Condit's design by replacing the poppet with an auxiliary piston that releases only a portion of the gas behind the main piston (see Fig \ref{Muirhead1964}). The remaining gas was used to bring the main piston back to the original position to control the positive duration of the shock wave. They also suggested a concept for higher driver pressures (up to 2000 psi) in which the piston had a smaller exposed area, and the driver section was placed in line with the driven section. Oguchi et al.\cite{Oguchi_1975} replaced the cam-release mechanism with a solenoid valve for quick action. Their design primarily consisted of a smaller auxiliary piston to control a larger main piston's motion. The solenoid valve releases a small volume of high-pressure gas behind the auxiliary piston to produce quicker retraction of the main piston. In this design, the size of the main piston was comparable to that of the driven section (smaller piston size compared to Condit's and Muirhead et al.'s designs). Shock Mach numbers of about 4.1 were obtained using this design, and many researchers widely used the solenoid-actuated double-sliding piston arrangement for future studies. Distefano et al.\cite{Distefano_1970} suggested an electromagnetically operated diaphragmless valve that utilizes two coils, one fixed and the other sliding, fixed to plates. When current flows through the coils, the seal plate slides into place and seals the driver section. When there is a sudden pressure increase due to combustion in the driver section or a short current interruption, the seal plate retracts and produces a shock wave. 

\begin{figure*}
\includegraphics[scale=0.5]{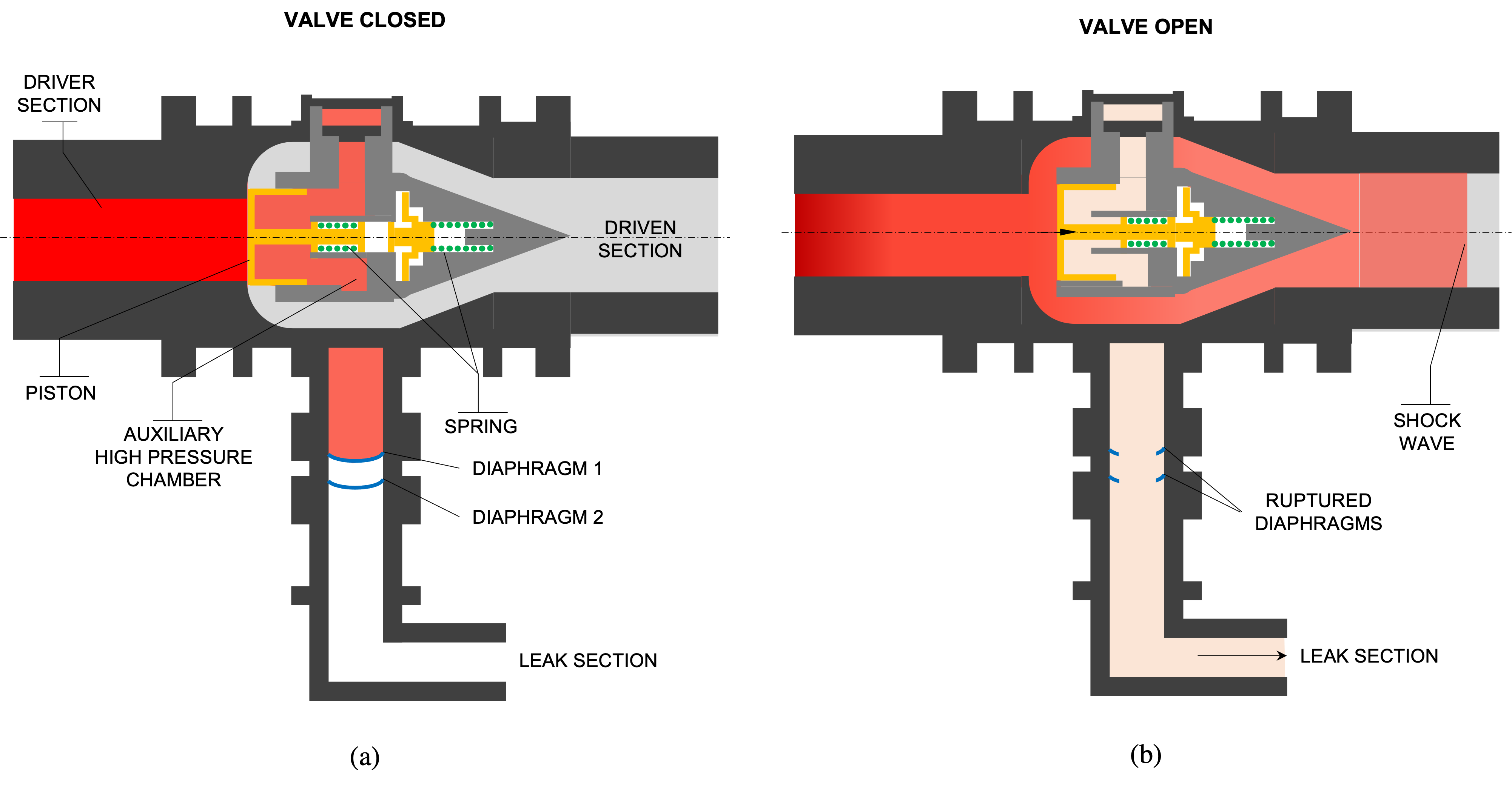}
\caption{\label{Oguchi1976} Schematic diagrams of the valve designed by Oguchi et al.\cite{Oguchi_1976} and Ikui et al.\cite{Ikui_1976}. Driver and driven sections are mounted inline and piston accelerates into a teardrop-shaped coaxial section. (a) When the valve is closed, the piston is held in place by high-pressure in an auxiliary chamber by means of diaphragms. (b) The valve opens when the diaphragms are ruptured and the high-pressure auxiliary chamber is evacuated, retracting the piston. {\color{black}(Adapted by permission from Springer Nature: Experimental Methods of Shock Wave Research by Igra, O., Seiler, F.\cite{Igra_2016}, copyright 2016)}}
\end{figure*}

Garen et al.\cite{Garen_1974} used a rubber membrane to seal the driven section from the driver section. The rubber membrane was inflated, by pressurizing a volume behind it, to block the entry into the low-pressure chamber. The bursting of a secondary diaphragm released the pressure behind the membrane. The retraction of the membrane led to the formation of shock waves in the driven section. This mechanism was used in a shock tube with an 18 mm square driven section and a 36 mm circular driven section. The replacement of the secondary diaphragm after every run and restriction to low driver pressures were significant drawbacks of this design. Matsuo et al.\cite{Matsuo_1975} suggested that the vertical movement of the piston was ideal for replacing the diaphragm in shock tubes. The low operating range of driver pressure ($\leq$ 1 bar) was a significant shortfall of these designs. Oguchi et al.\cite{Oguchi_1976}, and Ikui et al.\cite{Ikui_1976} presented a novel and sophisticated design for fast-acting pistons, which could be mounted inline with the driver and driven sections of a 100 mm by 180 mm diaphragmless shock tube (described in details in other reports \cite{Takayama_2013,Igra_2016,Timofeev_1997}). Therefore, there was no major redirection of gas flow from the driver to the driven section, as seen in Figure \ref{Oguchi1976}. The piston was accelerated into a teardrop-shaped coaxial section by venting the chamber behind it due to an auxiliary diaphragm rupture. A spring mechanism assisted this motion. A spring and gas damper was utilized to prevent piston damage due to impact and also helped bring the piston back to its original position. The valve produced shock Mach numbers in the range of 1.2-5, and the shock formation length was about 50 tube diameters. A clearance distance was incorporated to obtain an immediate opening of the valve in the piston-based valves in Ikui et al.'s designs\cite{Ikui_1977}. The piston accelerated and traversed the clearance distance before it breached the seal between the driver and the driven section. They designed a valve that opens along the shock tube axis, a double-piston sliding arrangement, and electromagnets for actuation. Since the piston moves horizontally, they called it a type-H valve. They also described a valve that opens perpendicular to the shock tube axis (called a type-V valve because of the vertical motion of the piston), which required the piston to move a more considerable distance than the type-H valve. In both cases, the valve's performance depended on the pressures in the actuating and driver chambers. The opening time and shock formation distance of the type-H valve were investigated at different pressures in the chambers\cite{Ikui_1979}.

\subsection{\label{sec3b:oguchi}Variants of Oguchi et al.'s double-piston design}
Oguchi et al.'s design principle was used in a shock tube for experiments in low-temperature gases\cite{Maeno_1990} (as low as 150 K) and gas dynamic laser applications\cite{Oguchi_1978,Maeno_1981}. The low temperatures were obtained by cooling the driven section with liquid nitrogen\cite{Maeno_1990}. The snap-action shock tube had a main and auxiliary piston made of nylon and actuated using electromagnetic valves. The maximum driver pressure was 5 bar, and the driven section had an inner diameter of 19.4 mm. Maeno and Oguchi\cite{Maeno_1980} performed studies using the synchronized operation of two diaphragmless shock tubes (with internal diameters of 20 mm and 50 mm) that implemented the double-piston arrangement actuated by electromagnetic valves. Solenoid valves helped time the actuation precisely to obtain the required delay in shock wave generation. A modified piston-driven shock tube actuated by solenoid valves and having a similar arrangement to Oguchi et al.'s design was also reported by Yamauchi and coworkers\cite{Yamauchi_1987}. This design implemented the main piston made of aluminum and a nylon auxiliary piston. Driver pressures of up to 20 bar were used in the shock tube with a 30 mm driven tube diameter. Hurst et al.\cite{Hurst_1993} adopted Yamauchi et al.'s design to develop an enlarged version of the double-piston arrangement. They significantly shortened the turn-around time between runs by automating the facility and showed good repeatability. The principle of Oguchi et al.'s design, with minor modifications in the supply of pressure, was used by Matsui et al.\cite{Matsui_1994} to obtain good reproducibility in shock wave conditions at low operating pressures with a temperature scatter behind the reflected shock as low as $\pm$20 K. Takano et al. \cite{Takano_1984} demonstrated a lightweight piston arrangement that was operated by magnetic valves and the system required little time for the initial setup to run the shock tube. The maximum driver pressure was 9 bar, and the design employed an annular driver-driven configuration. They also incorporated a lip in the piston to accelerate before breaching the seal between the driver and driven sections. 

\begin{figure*}
\includegraphics[scale=0.4]{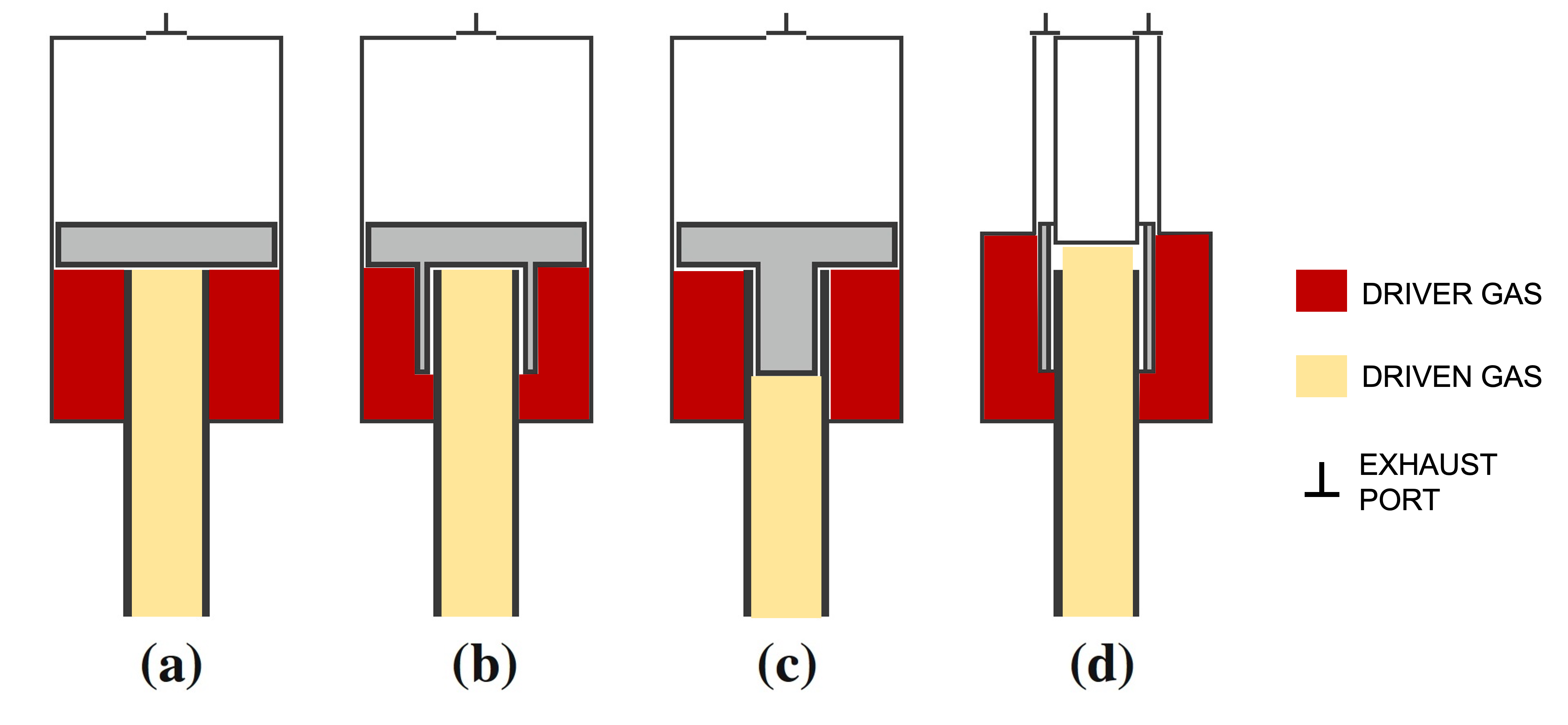}
\caption{\label{O-types} Variants of Oguchi et al.'s design as suggested by Mejia-Alvarez et al.\cite{Alvarez_2015} (a) Piston, (b) Piston with lip, (c) Piston with plug, and (d) Sleeve-type. {\color{black}(Adapted by permission from Springer Nature: Shock Waves, Mejia-Alvarez et al.\cite{Alvarez_2015}, copyright 2015)}}
\end{figure*}

A vertical shock tube system implementing the double-piston arrangement actuated by magnetic valves was reported by Teshima\cite{Teshima_1995}. In this design, an annular driver section was used, and the generated shock wave traveled vertically down in the driven section. shock waves with Mach numbers up to 2 were generated with a high-repetition-rate in a 16 mm diameter driven tube. The double-piston actuated design was further developed and implemented by Rego et al. in their large diameter diaphragmless shock tube\cite{Rego_2007_1, Rego_2007_2}. Improvements were suggested in the choice of piston material, seals, piston shape, and damping element to achieve good cycle life of operation. Onodera \cite{Onodera_1992} improved the double-sliding piston design of Oguchi et al. by combining the functions of the auxiliary and main piston into a more complex single composite piston. The composite piston had a large front end that sealed entry to the driven section and a smaller back end connected through a stem. The high-pressure gas in the small volume behind the piston, which initially keeps the piston in place, was rapidly exhausted by a solenoid valve. The gas surrounding the piston stem was evacuated to ensure minimum resistance to the piston movement. The complex geometry of the piston, the intricate sealing requirements, and the considerable weight of the piston were some disadvantages of this design. The small exhaust volume in the design helped achieve quick retraction to produce shock waves with a Mach number of 1.2. Mejia-Alvarez et al.\cite{Alvarez_2015} developed a sophisticated design for a double-sliding piston vertical shock tube based on a one-dimensional compressible flow model. This design had an annular driver section with many unique features that helped it outperform previous similar configurations. They analyzed all the variants of Oguchi et al.'s design before optimizing their structure. Figure \ref{O-types} shows the different possible moving elements used in Oguchi et al.'s concept. They also highlighted the crucial role of the discharge orifice in providing the quick retraction motion of the piston. A diaphragmless driver using a double-sliding piston arrangement was also reported for studies in a 19.4 mm diameter driven tube\cite{Zhang_2020}. 

\subsection{\label{sec3b:improvised}Improvised designs for specific applications}
Kosing et al.\cite{Kosing_1999} used an annular driver section with a single large piston while introducing a new actuation mechanism that did not utilize an auxiliary pressure chamber. The chamber behind the piston was evacuated to prevent the retraction movement, and the driver gas pressure aided the piston's acceleration. A steel brake pad mechanism on the piston side was installed to provide enough frictional force to hold the piston in place, which helps seal the driven section. The brake pad was operated using a hydraulic actuator with a small pressurized reservoir to provide sufficient force to hold the brake pad in place. Since the hydraulic fluid is incompressible, the pressure drop in the reservoir was very rapid, and the frictional force holding the piston dropped to zero. Different piston materials were experimented with, and the system produced a good performance for shock Mach numbers up to 2. A vertical co-axial shock tube with the double-piston arrangement was developed to generate toroidal shock waves traveling upwards in the driven section\cite{Watanabe_1995}. The pistons were ring-shaped and relatively heavy, resulting in slower opening times and a more considerable formation distance. A vertical diaphragmless shock tube with a 60 mm by 150 mm low-pressure channel was also developed. This facility was used to quantitatively visualize shock waves in a holographic interferometric system\cite{Ojima_2001, Takayama_2005a, Takayama_2005b}. Miyachi et al.\cite{Miyachi_2012} proposed two piston-driven diaphragmless shock tubes; the first design used five neodymium magnets for actuation, while the second valve was actuated electropneumatically to release the gas from a pressurized chamber behind the piston. Both designs incorporated a clearance distance in the piston movement for quicker opening. These designs were demonstrated in a 10 mm diameter shock tube for up to 8 bar driver pressures. 

A rapid opening valve assisted by magnetic force was used in a diaphragmless shock tube with a 10 mm internal diameter\cite{Abe_2015}. The axis of the driver and the driven sections were perpendicular to each other, and the maximum driver pressure used was 9 bar. Abe et al.\cite{Abe_1997} developed a high-speed valve to replace the use of diaphragms in a free-piston shock tube. Conventionally, the diaphragm in a free-piston shock tube ruptures when a piston moves towards it, accelerated by high-pressure gas from a reservoir and compresses the gas ahead. Abe et al. designed a valve with two-piston cylinders that moved perpendicular to the axis of the shock tube. An electromagnetic valve that released the high-pressure gas to drive the free piston actuated the entire facility. The fast-acting valve operates automatically when the free piston compresses the gas. Bredin and Skews\cite{Bredin_2007} used a three-piston configuration valve in a 50-meter long diaphragmless shock tube in which the opening times of the valve could be varied to obtain compression waves with rise times in the range of 5 to 20 milliseconds. The main piston was actuated using a secondary piston, and a tertiary piston was used to provide forward motion for the secondary piston. The use of multiple pistons made the operation of the shock tube very complicated. Another study compared a diaphragmless shock tube operated by a double-acting pneumatic cylinder and using a membrane-based fast-acting valve\cite{Hariharan_2010}. The valve opening speed of the pneumatic-cylinder-based valve varied from 0.325 to 1.15 m/s, while it was around 8.3 m/s for the membrane-based valve. Shock Mach numbers of up to 2.125 were obtained with good repeatability using the double-acting pneumatic cylinder-based valve. A valve similar to that reported by Oguchi et al.\cite{Oguchi_1976} and Ikui et al.\cite{Ikui_1976} consisting of a piston, spring dumper, and piston driver was operated using either a small diaphragm or a solenoid valve \cite{Taguchi_2018,Nishiyama_2018}. Itahashi et al. \cite{Itahashi_1996} optimized the opening of the driver gas to the driven section, thus resulting in a more efficient driver than most previous designs. 

\begin{figure*}
\includegraphics[width=0.8\textwidth]{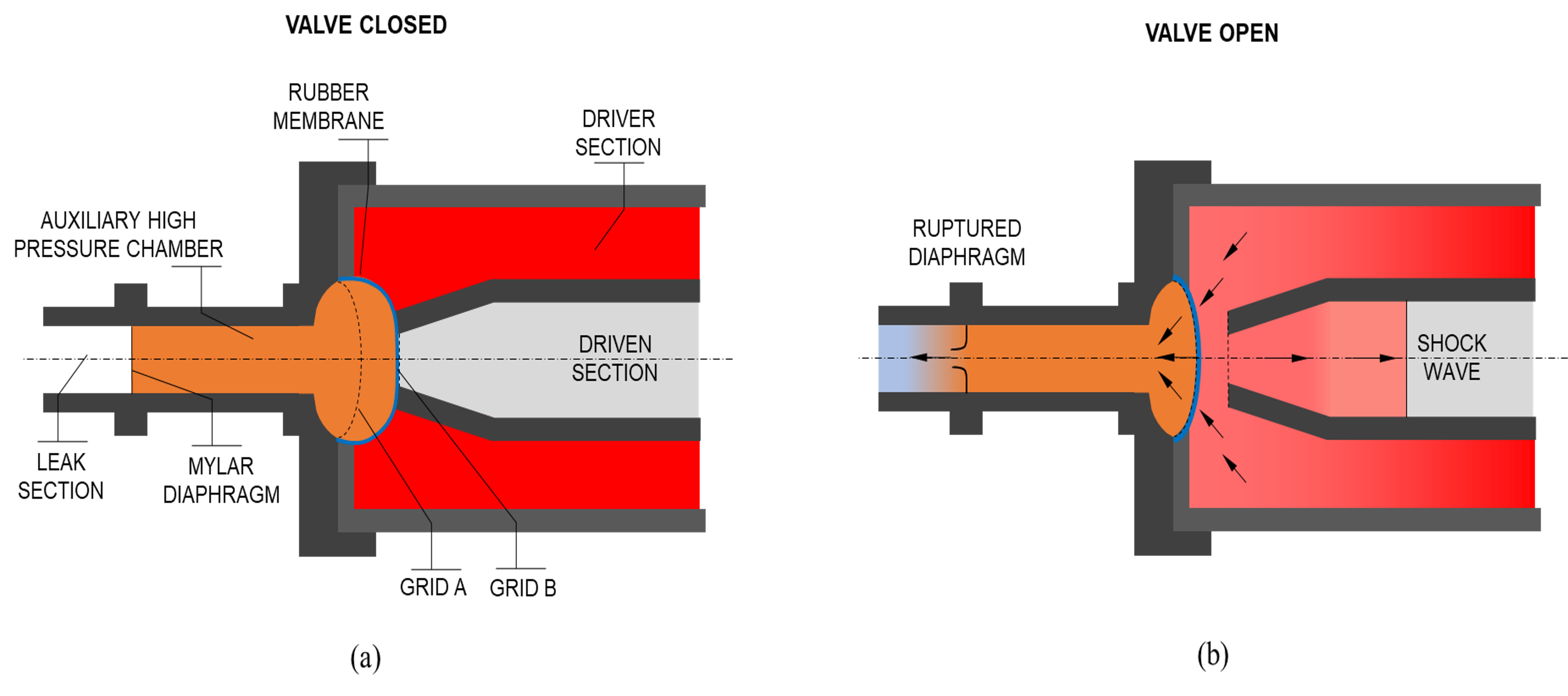}
\caption{\label{Yang1994} Fast-acting valve utilizing a rubber membrane proposed by Yang et al.\cite{Yang_1994} (a) Valve is closed: The rubber membrane bulges and blocks the path between the driver and driven section when an auxiliary chamber is pressurized. (b) Valve is open: When a diaphragm holding the pressurized gas in the auxiliary chamber is ruptured, the membrane retracts and opens the driver volume to the driven section. {\color{black}(Adapted with permission from The Japan Society of Mechanical Engineers: Transactions of the Japan Society of Mechanical Engineers Series B, Yang et al.\cite{Yang_1994}, copyright 1994)}}
\end{figure*}

\begin{figure*}
\includegraphics[width=0.85\textwidth]{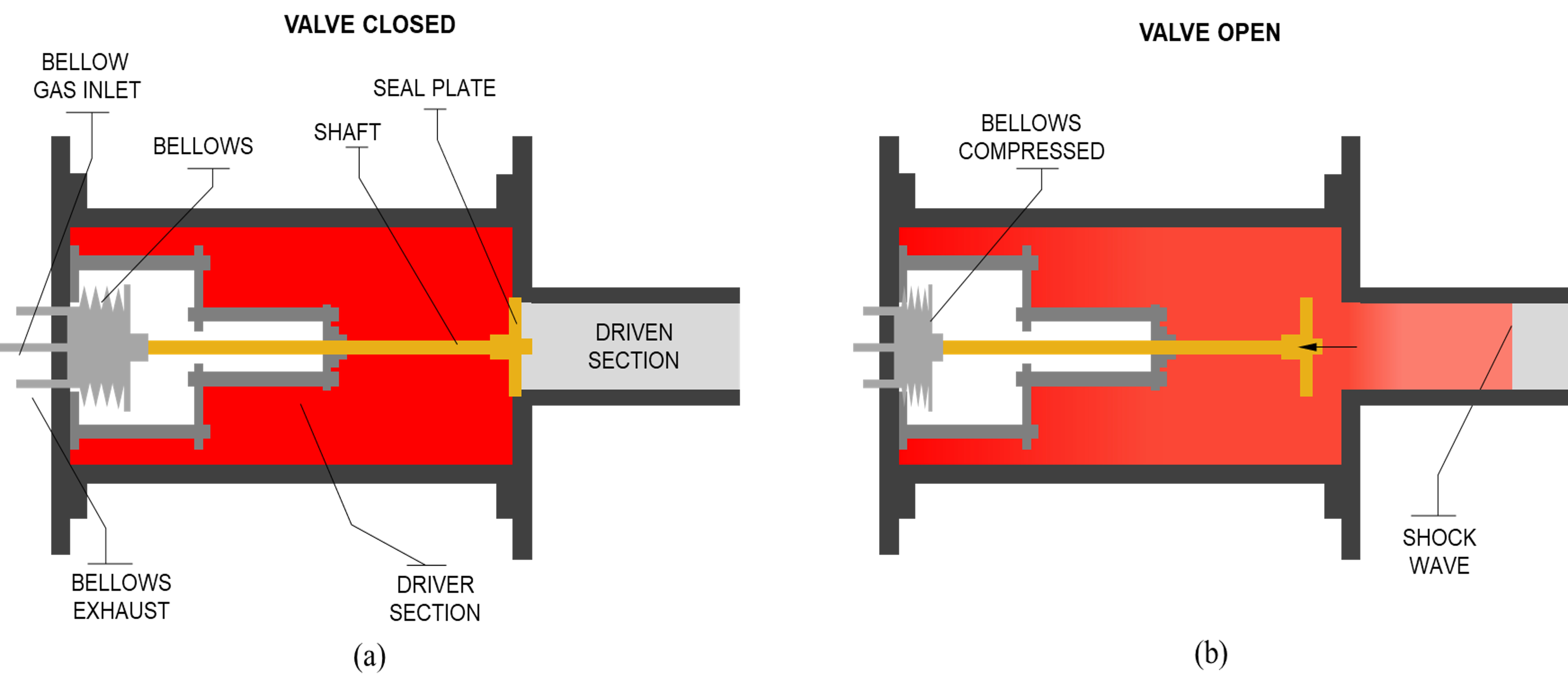}
\caption{\label{Tranter2008} Fast-acting valve design used by Tranter et al.\cite{Tranter_2008} that incorporated bellows for quick retraction of the piston. Different variations of this arrangement have been later designed by researchers. (a) A long shaft attached to the bellow moves forward to block the driver section from the driven section. (b) The piston retracts due to the bellow action to open the path between the driver and driven section. {\color{black}(Adapted with permission from AIP Publishing: Review of Scientific Instruments, Tranter et al.\cite{Tranter_2008}, copyright 2008)}}
\end{figure*}

Garen et al.'s design \cite{Garen_1974} was implemented in a 1 mm driven tube made of glass to study shock wave phenomena in miniature scales \cite{Udagawa_2007, Garen_2009}. The propagation velocities of the shock waves were measured with a specially designed laser interferometer, and experiments were performed up to driver pressures of 2 bar. The investigations were also extended using the same diaphragmless driver in a 3 mm driven tube, and the motion analysis of the rubber membrane was performed to evaluate the performance of the valve\cite{Udagawa_2008,Udagawa_2009}. Another membrane-based design with repeatability of 99 $\%$ for shock Mach numbers in the range of 1.02-1.55 was proposed\cite{Yang_1994}. Support blocks and grids were provided to limit the displacement of the rubber sheet and eliminate the expansion waves due to the rapid movement of the rubber sheet (see Fig \ref{Yang1994}). The piston-based design \cite{Watanabe_1995}to generate toroidal shock waves was improved by using a re-usable rubber membrane in a vertical shock tube\cite{Hosseini_1999}. Perforated ring-shaped steel plates were used to limit the rubber membrane's motion and to ensure the deformation was within the elastic limits. Consistent operation and a high degree of repeatability were ensured utilizing this technique, although the maximum shock Mach number was limited to 1.8\cite{Hosseini_2000}. A bellow-actuated diaphragmless valve was used to obtain reasonable control over the opening and closing of the valve \cite{Kim_1995}. Bellows were pressurized to provide a forward motion to seal the driven section from the annular driver section. The sudden release of the gas inside the bellows rapidly opened the pathway between the driver and the driven chamber. An alternative design based on Kim's design\cite{Kim_1995} was presented\cite{Tranter_2008}, as shown in Figure \ref{Tranter2008}. The problem of vibrations and alignment during the operation was minimized in the 71 mm internal diameter shock tube using linear bearings to support the bellow-piston arrangement. The design showed good shot-to-shot reproducibility over a range of operating conditions. An improved driver design was later suggested \cite{Randazzo_2015} where the bellow is placed in a manner such that it is compressed rather than extended when the driver gas is filled. With the growing interest in miniature shock wave applications, significant efforts were made to build small-scale high-repetition-rate automated shock tube systems. Shiozaki et al.\cite{Shiozaki_2005} built a 2 mm internal diameter diaphragmless shock tube that generated shock waves with Mach numbers of up to 2.8 and a temperature behind the reflected shock of about 1200 K. The solenoid-operated poppet valve produced high-repetition-rate shock waves at 5 Hz. 

\begin{figure*}
\includegraphics[scale=0.5]{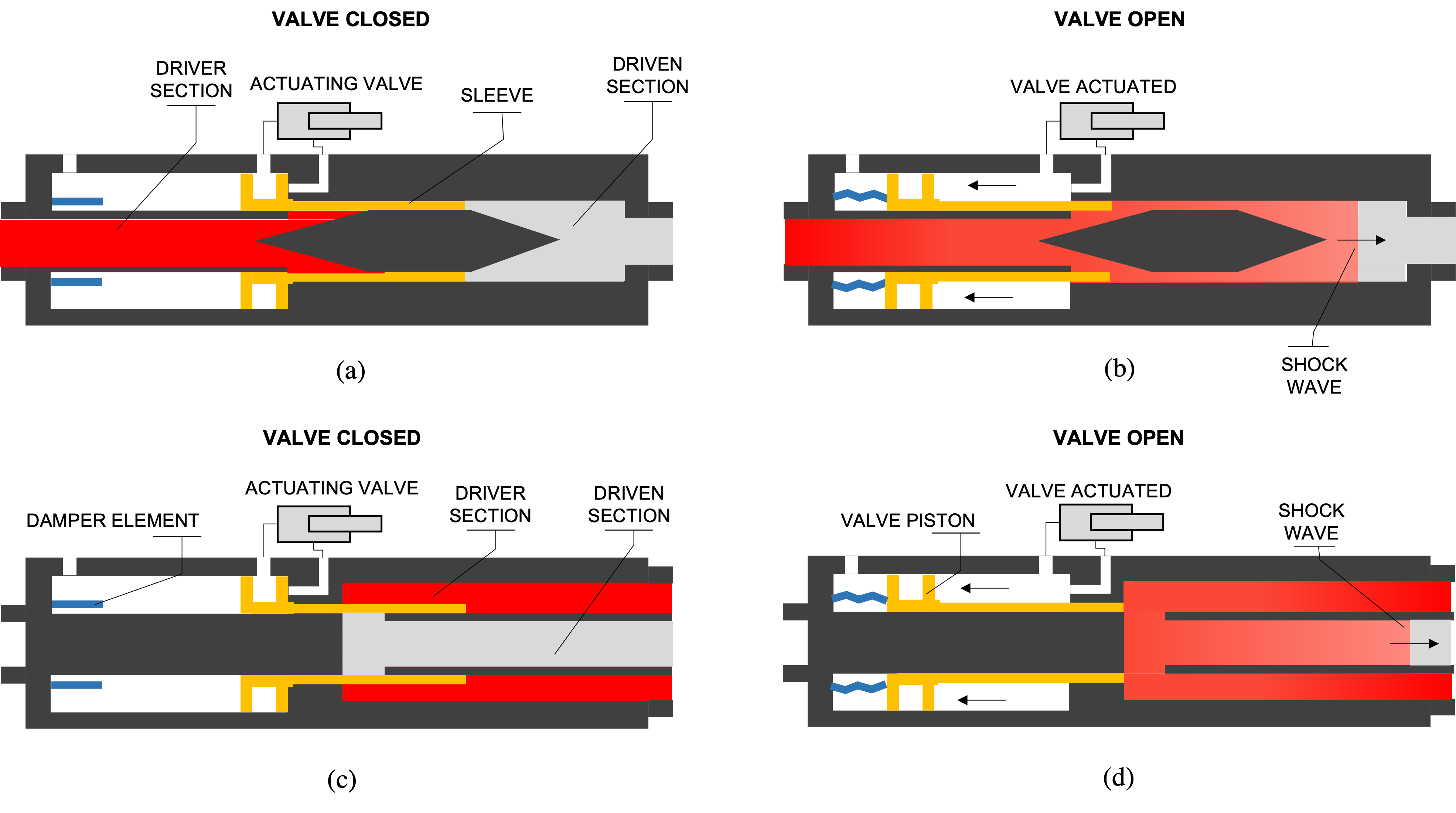}
\caption{\label{HeuferDowney2012} Sleeve-type fast-acting valve that is operated with actuating solenoid valves. The design proposed by Heufer et al.\cite{Heufer_2012} is shown in (a) when the valve is closed and (b) when the valve is open {\color{black}(Adapted with permission from Springer Nature: Heufer et al.\cite{Heufer_2012}, copyright 2012)}. Unlike the inline driver-driven arrangement, Downey et al.\cite{Downey_2011} used the sleeve in a coaxial driver-driven configuration. Schematic diagrams of Downey et al.'s valve design in the (c) closed state and in the (d) open state {\color{black}(Adapted with permission from Springer Nature: Shock Waves, Downey et al.\cite{Downey_2011}, copyright 2011)}.} 
\end{figure*}

\subsection{\label{sec3b:early}Recent designs of fast acting valves (Post 2010)}
In contrast to previous designs, Heufer et al.\cite{Heufer_2012} suggested a sleeve-type fast-acting valve in which the driver and driven sections have the same cross-section area and are mounted in line with each other (see Fig \ref{HeuferDowney2012}). Shock Mach numbers of up to 3.4 were obtained for driver pressures of 30 bar. Downey et al.\cite{Downey_2011} introduced a fast-opening valve that uses a sleeve to block the driver gas from entering the driven section initially. They incorporated multiple new features that improved the operating pressure range of the valve (up to 200 bar) while achieving short opening times. The sleeve also had a lip similar to Takano et al.'s design \cite{Takano_1984} for faster opening times. The sleeve was made of aluminum alloy, and the annular driver section had a streamlined flow path to minimize losses. The double-piston arrangement was adopted for studies in a miniature shock tube\cite{Udagawa_2012} as well with the development of a valve called the Maeno-Oguchi valve (described in Udagawa et al.\cite{Udagawa_2015}). This valve was used for studies in a miniature 2 mm and 3 mm internal diameter shock tube with driver pressures up to 9 bar. An improved version called the Udagawa-Maeno-Oguchi valve was also reported that used a clearance distance for piston movement that helped in faster retraction speeds and smaller opening times\cite{Udagawa_2015}. Based on the Tranter et al.'s bellow valve design \cite{Tranter_2008}, a new diaphragmless shock tube, called the Brown Shock Tube (BST), with software control and actuation of valving was presented\cite{Fuller_2019}. The operating pressures were increased up to 100 bar in the diaphragmless shock tube with a driven section diameter of 100 mm. McGivern et al. \cite{McGivern_2019} used a piston that was connected directly to the movable end of fixed stainless steel bellows for low-pressure application. This design incorporated a plug instead of a flat piston so that the piston accelerates before breaching the seal. 

\begin{figure*}
\includegraphics[scale=0.5]{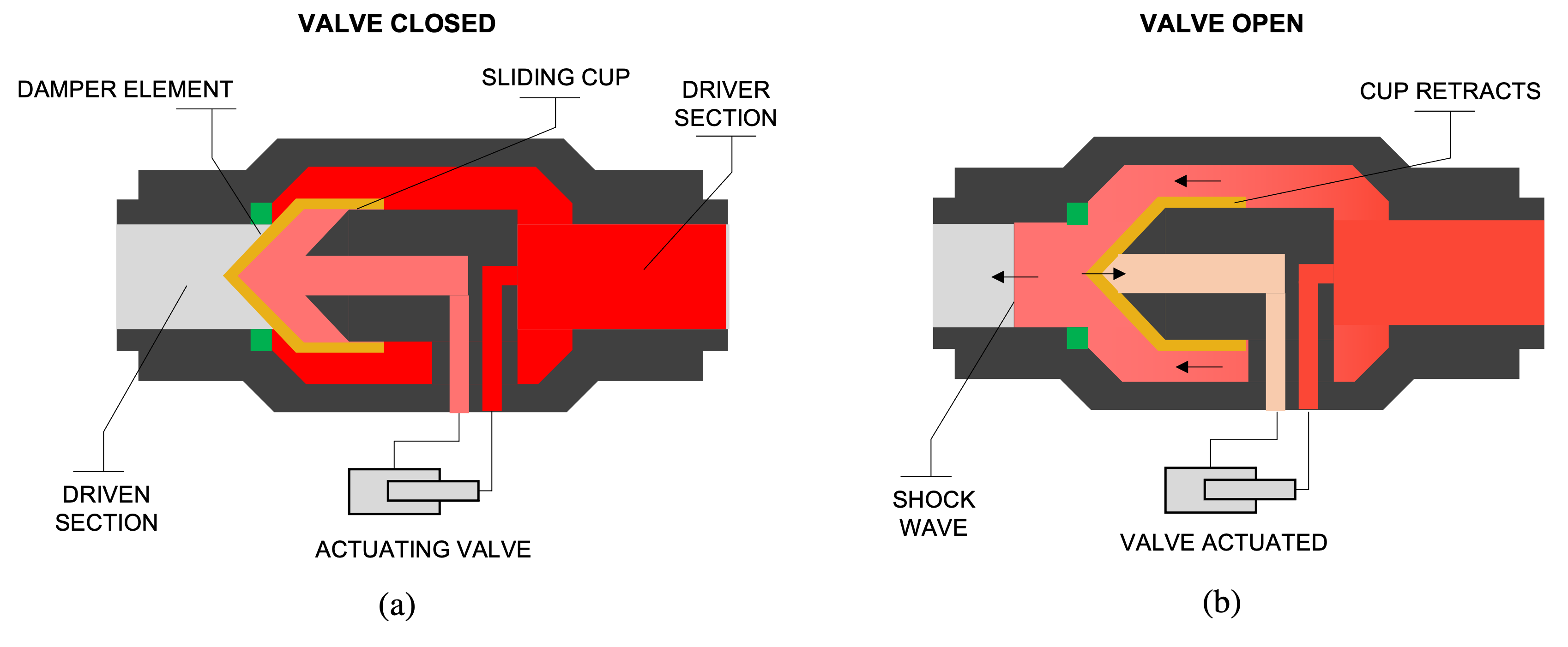}
\caption{\label{ISTA2020} A commercially available high-speed valve design that can be mounted in line with the driver and driven section\cite{Janardhanraj_2022}. (a) In the closed state, a sliding cup blocks the driver gas from entering the driven section. (b) When the volume of gas behind the sliding cup is exhausted, the cup retracts and shock wave is formed in the driven section.}
\end{figure*}

A 6.35 mm bore diaphragmless shock tube called the high repetition rate shock tube (HRRST) was designed for operational pressures up to 100 bar and observation times of about 100 microseconds\cite{Tranter_2013}. This diaphragmless driver was actuated using a solenoid valve, had a cycle rate of up to 4 Hz, and was designed for good reproducibility over thousands of shots. An improvement of the solenoid driver used in HRRST was suggested for a 12.7 mm bore shock tube\cite{Lynch_2016}. Improved performance and cycle life were obtained using a new face o-ring sealing design, increasing the internal volume and providing inserts near the solenoid core. A more recent development of the solenoid actuated valve improved the longevity and simplified the maintenance and manufacture of the solenoid valve\cite{Tranter_2020}. A pneumatic gas-driven diaphragmless shock tube with an operating mechanism similar to air gun technology was developed to produce weak shock waves\cite{Swietek_2019}. The table-top arrangement had a dumbbell-shaped piston whose movement was controlled using a trigger and reservoir chamber. The actuator could be quickly reset, thus decreasing the experimental turnaround time. Fast-acting valves procured commercially have been used in diaphragmless shock tubes\cite{Svete_2020_1,Svete_2020_2,Sembian_2020,Amer_2021,Janardhanraj_2022} (see Fig \ref{ISTA2020}). Amer and co-workers\cite{Amer_2021} compared the performance of fast-acting valves with conventional single and double diaphragm techniques. The shock-wave formation, repeatability, valve efficiency, and ease of operation were evaluated in their study. These valves were connected between the driver and driven section, replacing the diaphragm. The valves are available for different internal diameters and can operate at high driver pressures of up to 100 bar. Recently, a commercial valve was used in a diaphragmless shock tube for chemical kinetics studies which showed that the repeatability of the reflected shock temperature was similar to that obtained using a double-diaphragm technique\cite{Janardhanraj_2022}. A rapid opening shutter valve was developed and demonstrated in a 60 mm internal diameter diaphragmless shock tube\cite{Samimi_2020}. The innovative shutter valve was actuated electropneumatically and incorporated a feature that opened the valve from the center of the tube. 

\section{\label{sec3:concepts}Design features of fast-acting valves}
The main design features of a fast-acting valve include a closure element, an actuation mechanism, and the orientation of the driver and driven sections. These design features of various diaphragmless shock tube designs reported in the literature are consolidated in Table \ref{tab:table1}. The design aspects of the valves are discussed in the following sections.

\subsection{\label{sec3b:Element}Closure element}
Every valve design has an internal element that initially seals the pathway between the driver and the driven section. This element retracts on actuation and opens the pathway between the two sections of the shock tube. This movable obstruction is termed as a closure element\cite{Skousen_2011}. The actuator sets the closure element in position to separate the driver gas from the driven gas. When the actuator is toggled, the closure element quickly retracts to create an opening between the driver and the driven section. The closure element is either a piston (or seal plate), sleeve, cap, or membrane based on the various valve designs. The single and double piston arrangements have been widely used in diaphragmless shock tube designs. The piston is exposed to substantial forces, especially in the annular driver section, which requires stronger and heavy pistons. Kosing et al.\cite{Kosing_1999} performed experiments with pistons made of solid brass (4.4 kg), solid PVC (0.71 kg) and hollow aluminium (0.38 kg). The aluminum piston was used only for low driver pressures. The piston designed by Alvarez et al.\cite{Alvarez_2015} weighed 6.6 kgs and was used for driver pressures of up to 6 bar. Figure \ref{O-types} shows different closure elements that can be utilized in the Oguchi-type valve. Although the piston with lip and plug provides good sealing, the seals get damaged frequently as they leave the sliding face during the operation. A sleeve valve member has lesser weight as compared to the pistons used in diaphragmless shock tubes. Heufer et al.\cite{Heufer_2012} designed a sleeve made of aluminum which weighed about 3.124 kgs. Membranes are by far the lightest closure element for fast-acting valves. The elastic property of the membrane helps in achieving fast opening times. Rubber membranes are also self-sealing, unlike metallic pistons in which a groove has to be designed to accommodate a gasket or an o-ring. One of the main disadvantages of using a membrane is that the pressure difference across the membrane cannot be large, which restricts the maximum pressure used in the driver section. Most of the rubber membrane valves have been tested for driver pressures less than 25 bar\cite{Hosseini_2000,Hariharan_2010,Garen_1974,Udagawa_2007}. Membranes also lose their elastic property with repeated use, limiting the valve's cycle life. Caps are used in compact valve designs, especially in scenarios where the driver section is in line with the driven section. The mass of caps is a few hundred grams, and caps have been used in commercial valves up to pressures of 100 bar. The advances in material science have brought about numerous lightweight metal alloys and composites with high strength and durability. These materials present a wide range of options for high-performance closure elements.

\begin{table*}
\caption{\label{tab:table1}A consolidated table of various diaphragmless shock tube designs reported in literature.}
\begin{ruledtabular}
\begin{tabular}{ccccccc}
\multirow{2}{*}{Author/Year (Ref.)} & \multirow{2}{*}{Configuration} & \multirow{2}{*}{Closure element} & \multirow{2}{*}{Actuation elements} & \multirow{2}{*}{Control element} & Driven tube & Max. driver \\
&&&&& dimensions\footnote{$\phi$ indicates circular cross-section, $\square$ indicates square/rectangular cross-section, and Ann. indicates annulus between two diameters.} & pressure\footnote{The max. driver pressure listed is not necessarily the maximum operating pressure of the driver but the max. pressure listed in the cited source.} \\ \midrule
Condit, 1954 \cite{Condit_1954}         & Type-I(a)  & Piston       & Poppet                    & Cam              & $\phi$ 25.4 mm       & $\approx$ 41 bar   \\
Condit, 1954 \cite{Condit_1954}         & Type-I(a)  & Piston       & Poppet                    & Cam              & $\phi$ 431.8 mm      & $\approx$ 14 bar   \\
Muirhead, 1964 \cite{Muirhead_1964}     & Type-I(a)  & Piston       & Auxiliary piston-cylinder & Cam              & $\phi$ 50.8 mm       & $\approx$ 41 bar   \\
Takano, 1984 \cite{Takano_1984}         & Type-I(a)  & Piston       & Auxiliary piston-cylinder & Magnetic valve   & $\phi$ 40 mm         & 9 bar     \\
Kim, 1995 \cite{Kim_1995}               & Type-I(a)  & Seal Plate   & Opposed bellows           & Pneumatic valves & $\square$ 44.5 by 88.9 mm  & 3 bar     \\
Rego, 2007 \cite{Rego_2007_1,Rego_2007_2}& Type-I(a) & Piston       & Auxiliary piston-cylinder & Solenoid valve   & $\phi$ 100 mm        & 20 bar    \\
Miyachi, 2012 \cite{Miyachi_2012}       & Type-I(a)  & Piston       & Piston-cylinder/Magnets   & Solenoid valve   & $\phi$ 10 mm         & 8 bar     \\
Tranter, 2013 \cite{Tranter_2013}       & Type-I(b)  & Piston       & Vespel Poppet             & Solenoid         & $\phi$ 6.35 mm       & 102 bar   \\
Lynch, 2016 \cite{Lynch_2016}           & Type-I(b)  & Piston       & Vespel Poppet             & Solenoid         & $\phi$ 12.7 mm       & 102 bar   \\
Downey, 2011 \cite{Downey_2011}         & Type-I(c)  & Sleeve       & Piston-cylinder           & Trigger valve    & $\phi$ 50 mm         & 200 bar   \\
Yang, 1994 \cite{Yang_1994}             & Type-I(d)  & Membrane     & Diaphragm burst           & Vacuum pump      & $\square$ 60 by 150 mm     &   NA       \\
Hariharan, 2010 \cite{Hariharan_2010}   & Type-I(d)  & Membrane     & Diaphragm burst           & Vacuum pump      & $\phi$ 50 mm         & 25 bar    \\
Teshima, 1993 \cite{Teshima_1995}       & Type-I(e)  & Piston       & Auxiliary piston-cylinder & Magnetic valves  & $\phi$ 16 mm         & 16 bar    \\
Kosing, 1999 \cite{Kosing_1999}         & Type-I(e)  & Piston       & Brake pad                 & Hydraulic cylinder & $\phi$ 56 mm       & 50 bar    \\
Alvarez, 2015 \cite{Alvarez_2015}       & Type-I(e)  & Piston       & Auxiliary piston-cylinder & Solenoid valve   & $\phi$ 165 mm        & 6 bar     \\
Watanabe, 1995 \cite{Watanabe_1995}     & Type-I(f)  & Piston       & Diaphragm burst           & Release valve    & Ann. 210 \& 230 mm   & 5 bar      \\
Hosseini, 2000 \cite{Hosseini_2000}     & Type-I(g)  & Membrane     & Diaphragm burst           & Vacuum pump      & Ann. 80 \& 100 mm     & 5.4 bar   \\ \midrule
Svete, 2020 \cite{Svete_2020_1}         & Type-II(a) & Cap          & Piston-cylinder           & Solenoid valve   & $\phi$ 40 mm         & 70 bar    \\
Sembian, 2020 \cite{Sembian_2020}       & Type-II(a) & Cap          & Piston-cylinder           & Solenoid valve   & $\phi$ 80 mm         & 50 bar    \\
Amer, 2021 \cite{Amer_2021}             & Type-II(a) & Cap          & Piston-cylinder           & Solenoid valve   & $\phi$ 80 mm         & 20 bar    \\
Distefano, 1970 \cite{Distefano_1970}   & Type-II(b) & Seal plate   & Electromagnetic coils     & Current in coil  & $\phi$ 70 mm         & NA\footnote{Used in combustion drivers to produce $M_S$ in the range of 8-14.}    \\
Oguchi \& Ikui, 1976  \cite{Oguchi_1976,Ikui_1976} & Type-II(b) & Piston & Piston-cylinder      & Diaphragm burst   & $\square$ 100 by 180 mm         & NA\footnote{$M_S$ ranging from 1.2 to 5.0 in air}    \\
Taguchi, 2018  \cite{Taguchi_2018}      & Type-II(b) & Piston          & Piston-cylinder           & Solenoid valve   & $\square$ 60 by 150 mm  & 3 bar    \\
Heufer, 2012 \cite{Heufer_2012}         & Type-II(c) & Sleeve       & Piston-cylinder           & Solenoid valve   & $\phi$ 45 mm         & 20 bar    \\ 
Muirhead, 1964 \cite{Muirhead_1964}     & Type-II(d) & Piston       & Auxiliary piston-cylinder & Cam              & $\phi$ 50.8 mm       & $\approx$ 138 bar  \\
Ikui, 1977 \cite{Ikui_1977}             & Type-II(d) & Piston       & Auxiliary piston-cylinder & Electromagnet    & $\square$ 38 mm      & 1 bar  \\
Maeno, 1980 \cite{Maeno_1980}           & Type-II(d) & Piston       & Auxiliary piston-cylinder & Solenoid valve   & $\phi$ 50 mm         & 20 bar    \\
Onodera, 1992 \cite{Onodera_1992}       & Type-II(d) & Piston       & Piston-cylinder           & Release valve    & $\square$ 60 by 150 mm  & NA\footnote{Although the value of the driver pressure is not specified, the ratio $P_{41}$ used in the experiments was 2.5} \\
Shiozaki, 2005 \cite{Shiozaki_2005}     & Type-II(d) & Piston       & Poppet                    & Solenoid         & $\phi$ 2 mm          & 10 bar    \\
Tranter, 2008 \cite{Tranter_2008}       & Type-II(d) & Piston       & Bellow                    & Ball valve       & $\phi$ 71 mm         & $\approx$ 2 bar    \\
Hariharan, 2010 \cite{Hariharan_2010}   & Type-II(d) & Piston       & Piston-cylinder           & Solenoid valve   & $\phi$ 50 mm         & 25 bar    \\
Fuller, 2019 \cite{Fuller_2019}         & Type-II(d) & Piston       & Bellow                    & Solenoid         & $\phi$ 102 mm        & 100 bar   \\
Swietek, 2019 \cite{Swietek_2019}         & Type-II(d) & Piston       & Piston-cylinder         & Solenoid valve    & $\phi$ 76.2 mm        & 5.5 bar   \\
Zhang, 2020 \cite{Zhang_2020}           & Type-II(d) & Piston       & Auxiliary piston-cylinder & Ball valve       & $\phi$ 19.4 mm       & NA        \\ \midrule
Yamauchi, 1987 \cite{Yamauchi_1987}     & Type-III(a)& Piston       & Auxiliary piston-cylinder & Solenoid valve   & $\phi$ 30 mm         & 20 bar    \\
Hurst, 1993 \cite{Hurst_1993}           & Type-III(a)& Piston       & Auxiliary piston-cylinder & Solenoid valve   & $\square$ 62 by 44 mm  & $\approx$ 21 bar   \\
Matsui, 1994 \cite{Matsui_1994}         & Type-III(a)& Piston       & Auxiliary piston-cylinder & Magnetic valve   & $\phi$ 50 mm         & NA        \\
Udagawa, 2012 \cite{Udagawa_2012}       & Type-III(a)& Piston       & Auxiliary piston-cylinder & Solenoid valve   & $\phi$ 2 mm / $\phi$ 3 mm  & 9 bar     \\
Udagawa, 2015 \cite{Udagawa_2015}       & Type-III(a)& Piston       & Auxiliary piston-cylinder & Solenoid valve   & $\phi$ 2 mm / $\phi$ 3 mm  & 9 bar     \\
Garen, 1974 \cite{Garen_1974}           & Type-III(b)& Membrane     & Diaphragm burst           & Vacuum pump      & $\square$ 18 mm / $\phi$ 36 mm    & 1 bar  \\
Udagawa, 2007 \cite{Udagawa_2007}       & Type-III(b)& Membrane     & Diaphragm burst           & Diaphragm puncture & $\phi$ 1 mm        & 2 bar     \\ \midrule
Bredin, 2007 \cite{Bredin_2007}         & Type-IV(a) & Piston       & Auxiliary piston-cylinder & Ball valve       & $\phi$ 135 mm        & NA        \\
Miyachi, 2012 \cite{Miyachi_2012}       & Type-IV(a) & Piston       & Piston-cylinder           & Solenoid valve   & $\phi$ 10 mm         & 8 bar     \\
McGivern, 2019 \cite{McGivern_2019}     & Type-IV(b) & Piston       & Bellow                    & Ball valve       & $\phi$ 31.8 mm       & 8 bar     \\
Abe, 2015 \cite{Abe_2015}               & Type-IV(c) & Piston       & Piston-cylinder           & Ball valve       & $\phi$ 10 mm         & 9 bar     \\
Ojima, 2001 \cite{Ojima_2001}           & Type-IV(d) & Piston          & Piston-cylinder           & Diaphragm burst  & $\square$ 60 by 150 mm  & 9 bar     \\\midrule
Ikui, 1977 \cite{Ikui_1977}             & Type-V(a)  & Piston       & Auxiliary piston-cylinder & Electromagnet    & $\square$ 38 mm      & 1 bar  \\
Abe, 1997 \cite{Abe_1997}               & Type-V(b)  & Piston       & Piston-cylinder           & Solenoid valve   & $\phi$ 82 mm         & 176 bar   \\
Samimi, 2020 \cite{Samimi_2020}         & Type-V(b)  & Shutter      & Linear actuator           & Pneumatic drive  & $\phi$ 60 mm         & 20 bar
\end{tabular}
\end{ruledtabular}
\end{table*}

\begin{table*}
\caption{\label{tab:actuators} {\color{black}Comparison of different systems used for valve actuation.}}
\begin{ruledtabular}
\begin{tabular}{lcccc}
 & Hydraulic  & Pneumatic  & Electric & Electromagnetic\\
\midrule
\multirow{3}{*}{Advantages} & Powerful & Fast over long strokes & Quick and fast response & Reliable and robust\\
& Safe  & Economical & Precise control & Miniature and remote operation\\
& Self-contained & Simple design & Clean operation and no leaks & Cost-effective\\ \hline
\multirow{3}{*}{Disadvantages} & Fast over short strokes & Limited power & Low power & Electromagnetic interference\\
& High maintenance & Short cycle life & Complicated design & Sensitive to voltage\\
& Risk of leaks &  Gas requirement & Expensive & Limited force\\
\end{tabular}
\end{ruledtabular}
\end{table*}

\subsection{\label{sec3c:act}Actuation and control elements}
The actuation and control elements are responsible for the movement of the closure element. The valve's actuator translates the input energy to the motion of the control element. The actuator should be powerful enough to overcome the force to move the closure element at a very high speed to match the time scales of a diaphragm rupture. Skousen\cite{Skousen_2011} defined that the sizing of the actuator is dependent on the total force required to open the valve,  given by,
\begin{equation}
    F_{total} = F_{process} + F_{seat} + F_{friction} + F_{misc.}
\end{equation}
where $F_{process}$ is the force to overcome unbalanced process pressures, $F_{seat}$ is the force to provide correct seat load, $F_{friction}$ is the force to overcome frictional forces, and $F_{misc.}$ is the force to overcome certain design factors, such as the weight of the closure element, etc. {\color{black}The main types of systems that are used for valve actuation include hydraulic, pneumatic, electric, and electromagnetic\cite{Sotoodeh_2019}. Table \ref{tab:actuators} highlights the advantages and disadvantages of these actuators. Hydraulic systems are suitable for heavy-duty purposes since compressing a fluid, such as oil, produces much more motion power than compressing a gas. Pneumatic systems cannot produce the power that hydraulic systems generate, but they are stronger than purely electric actuators. Pneumatic systems also tend to work faster than hydraulic and electric systems over the stroke of movement. Pure electric actuators (motor-driven) cannot match the power of hydraulic and pneumatic systems, though they are cleaner and can provide precise control over movement. Electromagnetic actuators include solenoid-based devices that are very reliable, quick, and commonly used. Overall, pneumatic and electromagnetic systems are preferred for diaphragmless shock tubes because they are fast, and their retraction motion can produce short opening times.} The most common control element used for operating the fast-acting valves is solenoid valves (as seen in table \ref{tab:table1}) that use electromagnetic actuation for quick action (about 1 – 5 milliseconds). The use of bellows has yielded good results for repeated operation as the motion of the closure element is consistent and reliable\cite{Kim_1995,Tranter_2008, Fuller_2019, McGivern_2019}. 

\begin{figure*}
\includegraphics[scale=0.5]{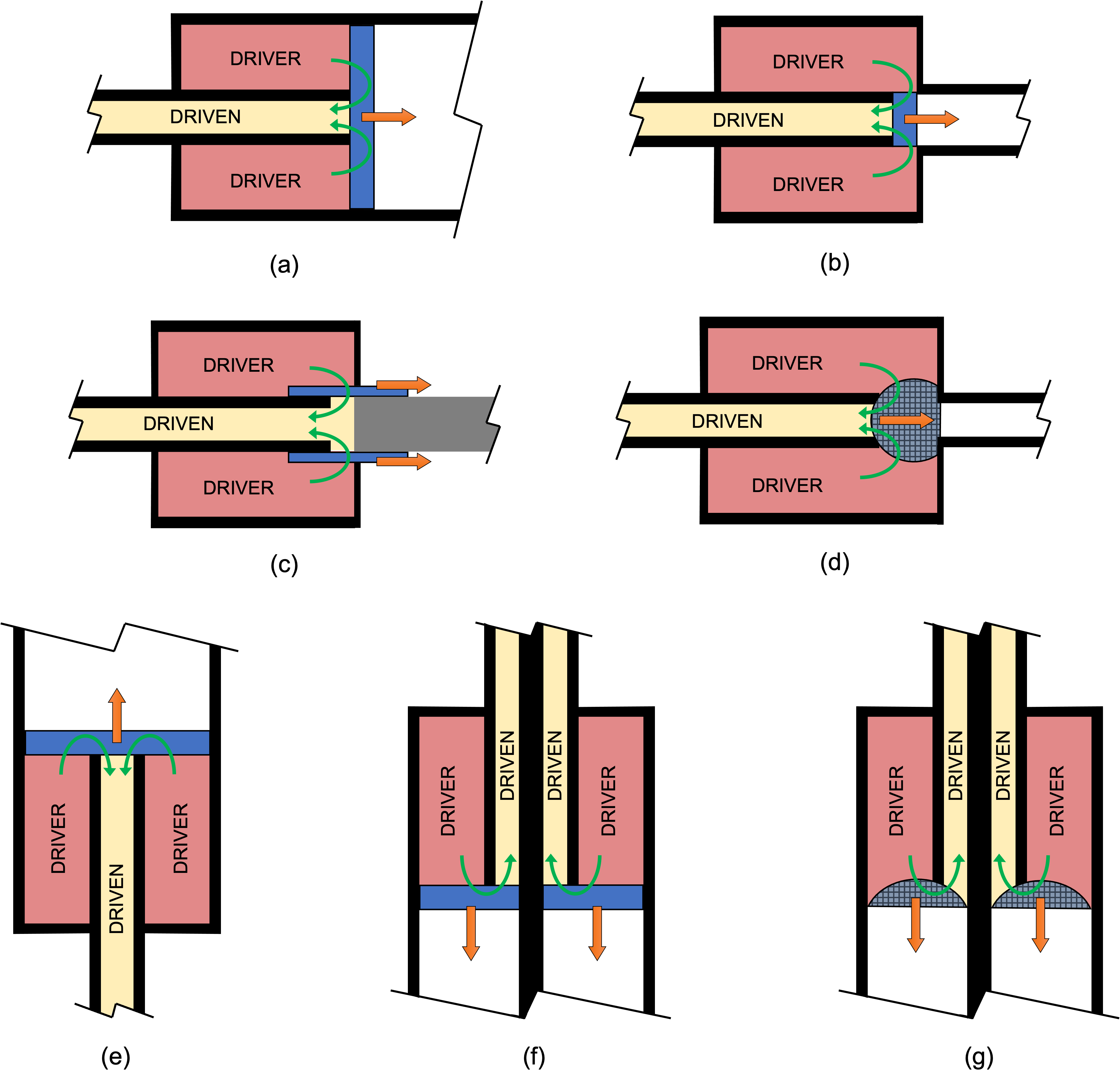}
\caption{\label{typeI}Schematic diagrams illustrating various Type-I configurations. The orange arrow indicates the movement of the element blocking the driver and driven section while the green arrows show the direction of flow after the movement of the element. (a) Piston-type configuration where piston is the size of driver section, (b) Piston-type configuration where the piston is the size of the driven section, (c) Sleeve-type configuration, (d) Membrane-type configuration,  (e) Vertical configuration where piston moves against gravity, (f) Vertical configuration where piston moves in the direction of gravity, and (g) Vertical configuration where membrane retracts in the direction of gravity.}
\end{figure*}

\begin{figure*}
\includegraphics[scale=0.5]{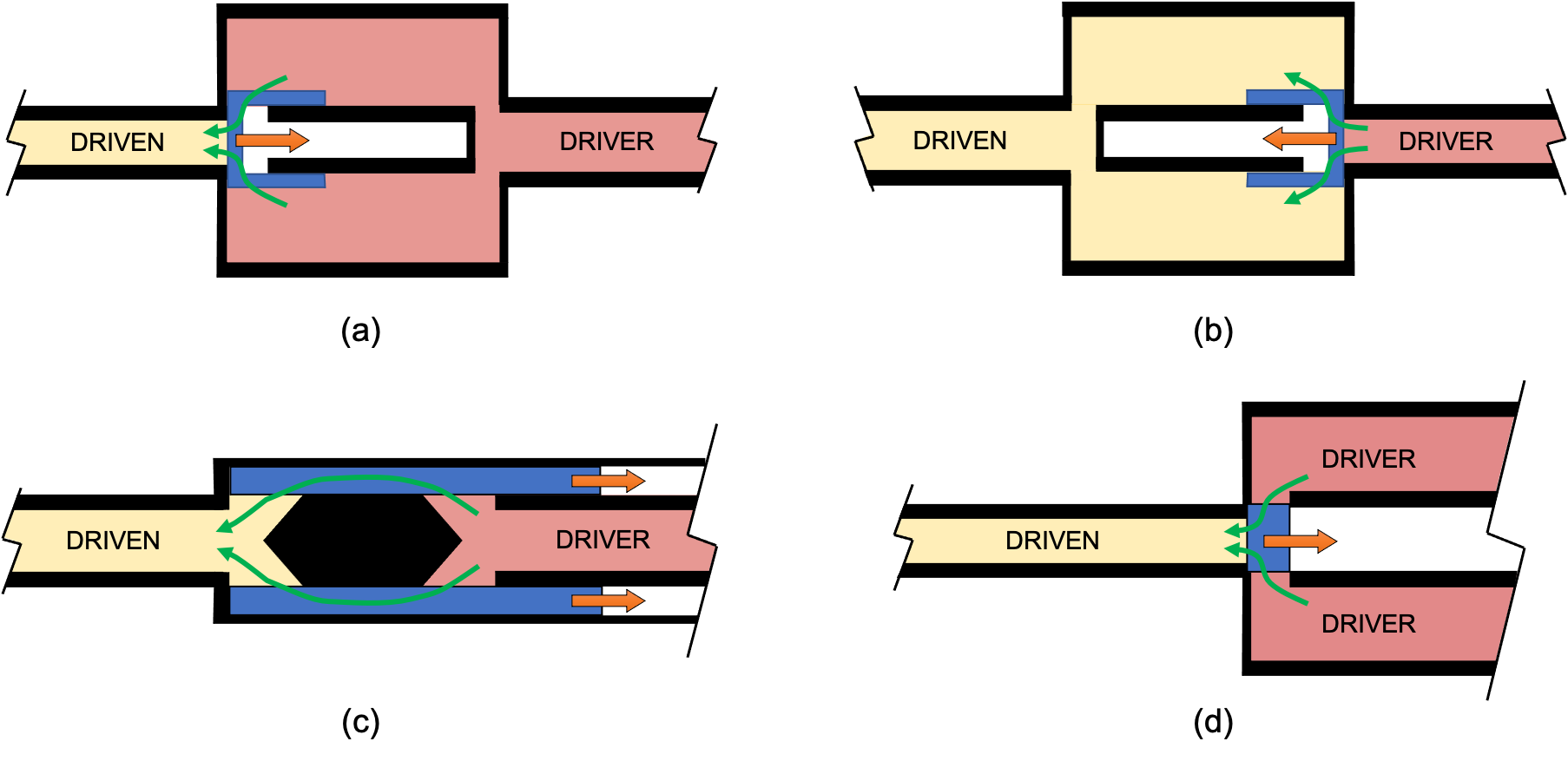}
\caption{\label{typeII}Schematic diagrams illustrating various Type-II configuration. The orange arrow indicates the movement of the element blocking the driver and driven section while the green arrows show the direction of flow after the movement of the element. (a) Cap-type configuration with variable cross-section driver section, (b) Cap-type configuration with variable cross-section driven section, (c) Sleeve-type configuration, and (d) Piston-type configuration.}
\end{figure*}

\begin{figure*}
\includegraphics[scale=0.5]{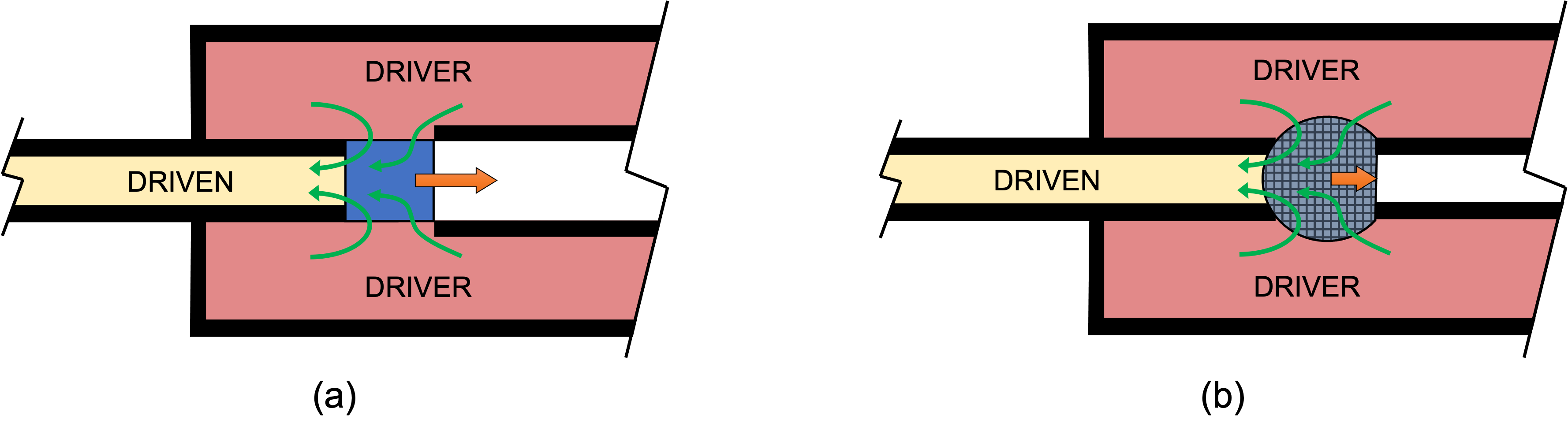}
\caption{\label{typeIII}Schematic diagrams illustrating various Type-III configuration. The orange arrow indicates the movement of the element blocking the driver and driven section while the green arrows show the direction of flow after the movement of the element. (a) Piston-type configuration and (b) Membrane-type configuration.}
\end{figure*}

\begin{figure*}
\includegraphics[scale=0.5]{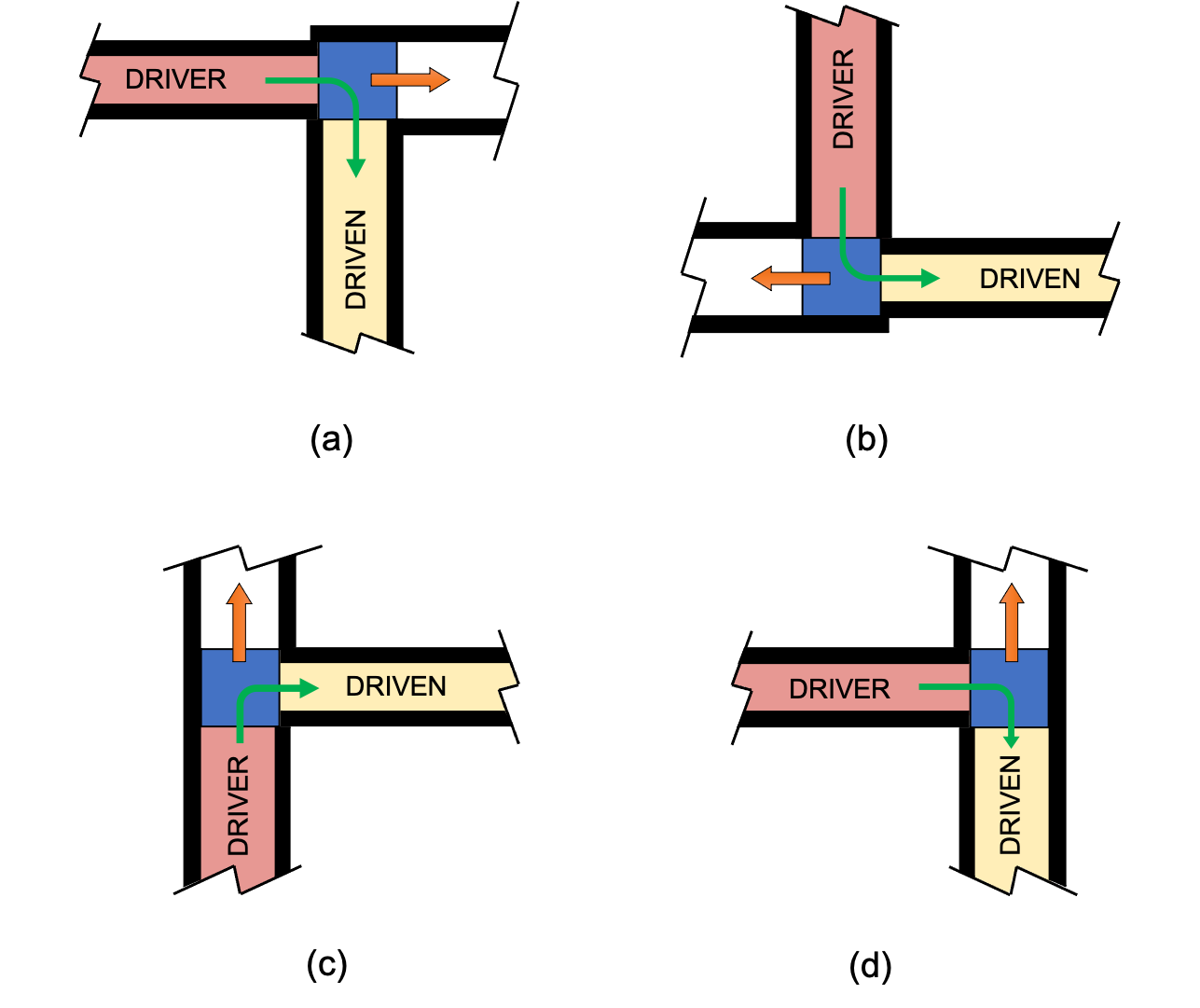}
\caption{\label{typeIV}Schematic diagrams illustrating various Type-IV configuration. The orange arrow indicates the movement of the element blocking the driver and driven section while the green arrow shows the direction of flow after the movement of the element. Horizontal piston movement with (a) vertical driven section and (b) horizontal driven section. Vertical piston movement with (c) horizontal driven section and (d) vertical driven section.}
\end{figure*}

\begin{figure*}
\includegraphics[scale=0.5]{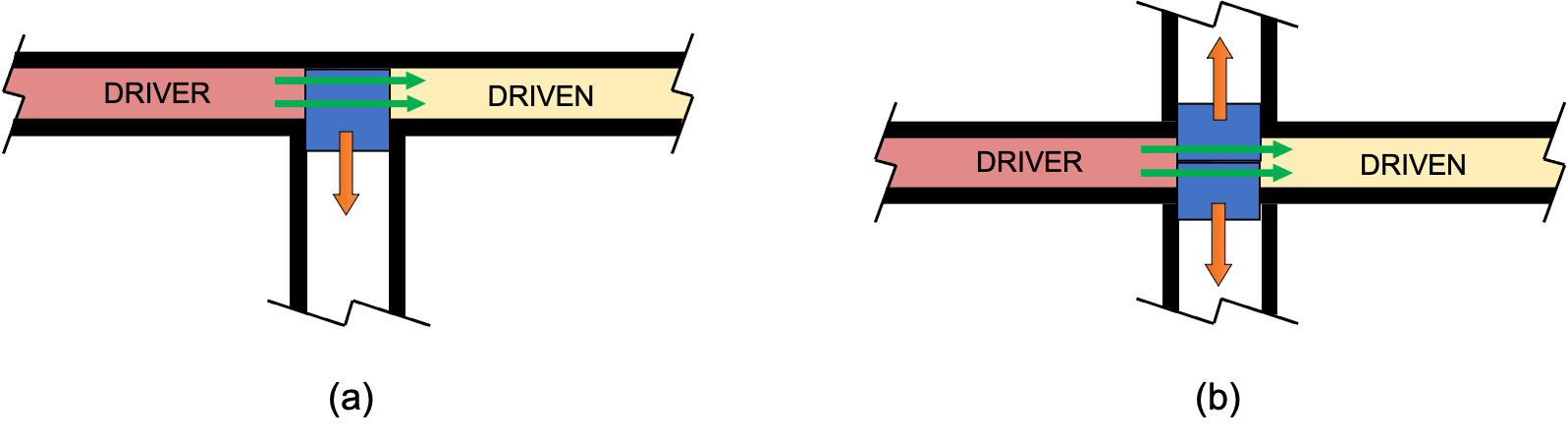}
\caption{\label{typeV}Schematic diagrams illustrating various Type-V configuration. The orange arrow indicates the movement of the element blocking the driver and driven section while the green arrows show the direction of flow after the movement of the element. (a) Movement of single element and (b) Movement of multiple elements.}
\end{figure*}

\subsection{\label{sec3c:config}Driver-Driven configurations}
The mounting configuration of the fast-acting valve and the driver's placement relative to the driven section determines the ease of gas flow from the high- to the low-pressure chamber. Although it would be ideal to have the two sections in line, the movement of the closure element may sometimes block the pathway. Based on the different designs reported in the literature, the driver-driven configurations can be arranged into five categories.

\subsubsection{\label{sec3c1:type1}Type-I driver-driven configurations}
In this configuration, the driver gas is forced to take a 180$^\circ$ turn at the valve location. The flow redirection could lead to losses that eventually affect the strength of the shock wave generated in the driven section. The design proposed by Downey et al.\cite{Downey_2011} and Mejia-Alvarez et al.\cite{Alvarez_2015} incorporated a curvature to streamline the flow to minimize the losses. Figure \ref{typeI} shows schematic diagrams of variants under the Type-I configuration. In Type-I(a) configuration (see figure \ref{typeI}a), a large piston with diameter same as the inner diameter of the driver section seals the gases in the respective sections\cite{Condit_1954, Muirhead_1964,Takano_1984, Kim_1995, Rego_2007_1, Miyachi_2012}. The large piston size implies a larger force is required to move the piston to compensate for the mass. Takano et al.\cite{Takano_1984} suggested using a lightweight hollow piston made of aluminum for faster retraction speeds. Also, additional seals must be used on the sliding face of the piston to prevent driver gas leaks. An important advantage of type-Ia is that the piston experiences a large force due to the driver gas in the direction of piston retraction. The piston size is smaller and comparable to the diameter of the driven section in Type-I(b) configuration (see Figure \ref{typeI}b). Face seals are sufficient to seal the driven section, and the low mass of the piston can help obtain faster retraction speeds. This configuration has been demonstrated in miniature shock tubes for driver pressures of up to 102 bar\cite{Tranter_2013,Lynch_2016}. The closure element in Type-I(c), as shown in Figure \ref{typeI}c, is a sliding sleeve on which the opposing forces for the sleeve movement are minimum due to the smaller exposed area. Downey et al.\cite{Downey_2011} employed this configuration to operate at high operating pressures of about 200 bar. Figure \ref{typeI}d shows a configuration with a rubber membrane. Yang et al.\cite{Yang_1994}, and Hariharan et al.\cite{Hariharan_2010} used this configuration and obtained short opening times due to the quick retraction of the membrane. The cycle life and the limited pressure range of operation were some drawbacks of using the rubber membrane. Figures \ref{typeI}e and \ref{typeI}f show configurations of a piston-based valve for vertical shock tubes. In Type-I(e) configuration, the piston retraction is against gravity. The shock wave travels vertically downwards while the direction of piston movement and shock wave propagation is vice-versa in Type-I(f) configuration. The retraction force has to overcome the piston's weight in Type-I(e) configuration\cite{Teshima_1995,Kosing_1999, Alvarez_2015}. This configuration is ideal for vertical systems in studying shock wave interaction with liquids. Type-I(f) and I(g) configurations are used for the generation of annular shock waves using a piston\cite{Watanabe_1995} and membrane \cite{Hosseini_2000}, respectively.

\subsubsection{\label{sec3c2:type2}Type-II driver-driven configurations}
The driver section is in line with the driven section in Type-II driver-driven configurations. Unlike the conventional diaphragm-type shock tube, the cross-section of the driver section is larger than the driven section. There is no major flow redirection compared to the type-I configurations. Therefore, there are fewer losses due to flow turning at the valve location. Generally, a streamlined path is provided in the valve to minimize resistance to flow. The variants of Type-II configuration are shown in Figure \ref{typeII}. Type-II(a) is a cap-type configuration, as shown in Figure \ref{typeII}a, which is adopted by many commercial valve manufacturers\cite{Svete_2020_1,Sembian_2020}. This design allows the valve to be directly mounted in shock tubes that had previously utilized a diaphragm by simply replacing the diaphragm section. The front face of the cap has a conical shape for streamlining the flow. The manufacturing tolerances and material selection in this design are critical to the operation of the valve. Type II(b) configuration is similar to Type II(a), but the driven section has a variable cross-section (see Figure \ref{typeII}b). The sleeve-type design (shown in Figure \ref{typeII}c) was reported by Heufer et al.\cite{Heufer_2012}. Compared to the cap-type and piston-type designs, one of the main disadvantages of this design is that the overall stroke length required to open the valve is much larger. Therefore, the valve is much longer, and a larger volume of gas has to be released from the actuating chamber during the operation. Figure \ref{typeII}d shows a piston-type configuration in which the piston retracts into an actuating chamber behind it. Numerous researchers have adopted this design because of its simplicity\cite{Muirhead_1964,Ikui_1977, Maeno_1980, Onodera_1992,Shiozaki_2005,Hariharan_2010,Zhang_2020}. The designs used by Tranter et al.\cite{Tranter_2008} and Fuller et al.\cite{Fuller_2019} are similar to those given in Figure \ref{typeII}d expect that the closure element does not retract into the actuating chamber but remains in the driver chamber. The schematics given in the figure mainly portray the driver-driven configurations. Figure \ref{Tranter2008} shows the actual depiction of Tranter et al.'s design\cite{Tranter_2008}.

\subsubsection{\label{sec3c3:type3}Type-III driver-driven configurations}
In Type-III configuration, a part of the driver section is annular to the driven section, and the rest of the driver section is in line with the driven section. Therefore, the flow of the driver gas is a combination of both scenarios seen in Type-I and Type-II configurations. Figure \ref{typeIII} shows the schematic diagrams of Type-III configurations. The piston-type (see Figure \ref{typeIII}a) is more popularly used as compared to the membrane-type configuration (see figure \ref{typeIII}b). The flow field in both cases is similar, except that the faster retraction of the membrane can lead to expansion waves in the region where the membrane is initially inflated. Supporting grids were used to limit the membrane's movement and eliminate these expansion waves due to the rapid movement of the membrane. A number of double-sliding piston design concepts use Type-II(a) configuration\cite{Yamauchi_1987,Hurst_1993,Matsui_1994,Udagawa_2012,Udagawa_2015} while the membrane-based design is reported by Garen et al.\cite{Garen_1974} and Udagawa et al.\cite{Udagawa_2007}.

\subsubsection{\label{sec3c4:type4}Type-IV driver-driven configurations}
In this configuration, the driver gas undergoes a 90$^\circ$ flow turn as the driver and driven sections are perpendicular to each other. Figure \ref{typeIV} shows the schematic diagrams of Type-IV configurations. A long driver section is typically necessary to delay the rarefaction wave's interaction with the reflected shock wave. The 90$^\circ$ bend at the valve location may cause the rarefaction wave to interact with the contact surface too early and yield a decelerating shock. Hence, the alignment of the driver tube at a right angle is a disadvantage, as opposed to the straight-through flow path in configurations with in-line mounting. In Type-IV(a) and IV(b), the piston movement is horizontal, while the movement is vertical in Type-IV(c) and IV(d). The driven section is vertical in type IV-(a) and IV-(d), while the driver section is vertical in type IV-(b) and IV-(c). The diaphragmless shock tubes proposed by Bredin et al.\cite{Bredin_2007} and Miyachi et al.\cite{Miyachi_2012} utilized Type IV(a) configuration. McGivern et al.'s bellow-actuated shock tube\cite{McGivern_2019} has a vertical driver section attached to a 31.8 mm diameter driven section. Abe et al.\cite{Abe_2015} and Ojima et al.\cite{Ojima_2001} employed  Type IV-(c) and IV-(d) configurations, respectively.

\subsubsection{\label{sec3c5:type5}Type-V driver-driven configurations}
The best way to minimize resistance to the gas flow in the shock tube is by having the driver and the driven section of the same cross-section and in line with each other. Such a configuration ensures minimal losses in shock formation and propagation. In type V configuration (see Figure \ref{typeV}), the driver-driven chambers are in line, and the driver and driven sections' cross-sections are the same. Type V-(a) configuration, as shown in Figure \ref{typeV}a,  has a single closure element that moves in the transverse direction. In type V-(a) configuration, the valve opening is asymmetrical about the shock tube axis. This design aspect of type-V(a) configuration makes it distinct from all the other configurations. The closure element is composed of multiple individual components in type V-(b) (see Figure \ref{typeV}b). Abe et al.\cite{Abe_1997} used two opposing pistons in their design. The design proposed by Samimi et al.\cite{Samimi_2020} used a shutter configuration composed of either three or six blades. A substantial advantage of type V-(b) configuration is that the fast-acting valve opens from the center of the tube similar to the diaphragm-type shock tube operation. The transverse motion of the closure elements increases the loads on the seals due to the differential pressure. Ensuring proper sealing between multiple closure elements in type-V(b) is also challenging.

\begin{figure*}
\includegraphics[scale=0.18]{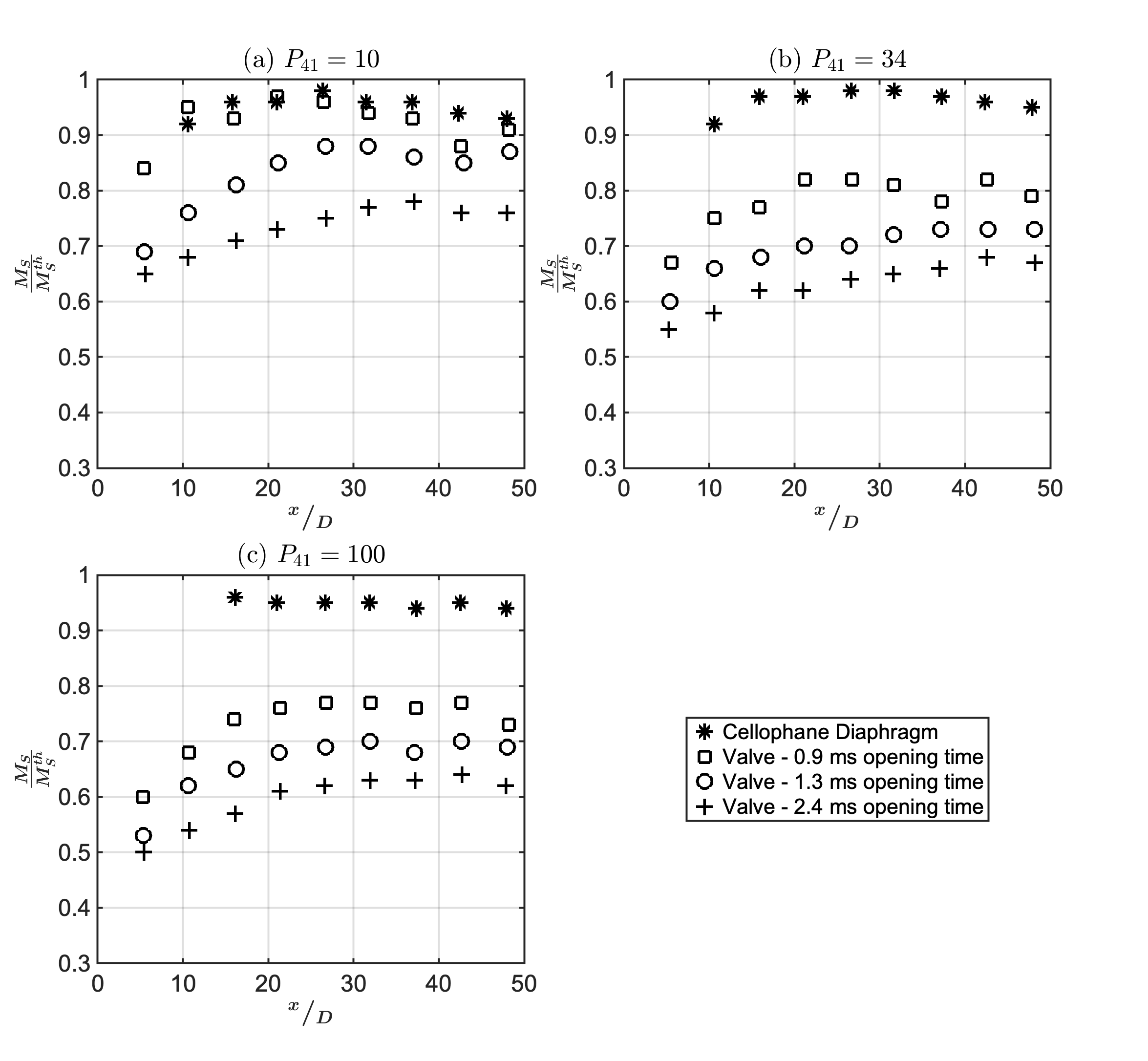}
\caption{\label{Top}Variation of the normalized shock Mach number along the diaphragm-type and diaphragmless shock tube ($M_S$ represents the experimental shock Mach number and $M_S^{th}$ represents the theoretical shock Mach number obtained from 1D shock tube relations). (a) $P_{41} = 10$ (b) $P_{41} = 34$ (c) $P_{41} = 100$. {\color{black}(Reprinted with permission from The Japan Society of Mechanical Engineers: Bulletin of JSME, Ikui et al.\cite{Ikui_1979}, copyright 1979)}}
\end{figure*}

\section{\label{sec4:performance}Performance of fast-acting valves}
The performance of diaphragmless shock tubes in terms of the opening time, shock formation distance, operating range, and shock wave conditions are discussed in the following sections. 

\subsection{\label{sec4a:opening}Opening time of valve}
The opening time of a fast-acting valve is an important parameter that determines the performance of the diaphragmless shock tube. It is the time taken by the closure element to move a distance that creates an opening with an area equal to the cross-sectional area of the driven section of the shock tube. Ideally, the opening time should be short so that a supersonic flow is initiated in the shock tube. The inability of commercial valves to produce such a condition through quick action makes them unsuitable for use in shock tubes. Therefore, customized fast-acting valves have to be designed for diaphragmless shock tubes. Determining the opening time of a fast-acting valve requires a dedicated optical measurement system. Since the closure element is enclosed in the driver or driven section, obtaining the opening time during the system's operation becomes challenging. Instead, in many cases, the movement of the closure element is monitored without mounting the driven section. Although this method does not give the valve's actual opening time, a rough idea can be obtained. In one of the early reports, the opening time was estimated using two laser beams and phototransistors as light detectors to track the retraction of the membrane\cite{Garen_1974}. The opening time was estimated to be 460 $\mu$s. Hariharan et al.\cite{Hariharan_2010} measured retraction speeds of about 8 m/s using high-speed photography for their membrane-based fast-acting valve. For a miniature 1 mm and 3 mm shock tube, the opening time of the rubber membrane valve was 46.6 $\mu$s and 117.2 $\mu$s, respectively\cite{Udagawa_2008}. 

Piston-based diaphragmless valve concepts have slower speeds compared to rubber-membrane systems. Ikui et al.\cite{Ikui_1977} performed a detailed investigation of the opening time as a function of pressures used in different chambers. They reported opening times of less than ten milliseconds for a 38 mm square shock tube. Ikui and co-workers\cite{Ikui_1979} also compared the variation of the shock Mach number along a diaphragm-type and diaphragmless shock tube at different $P_{41}$ and valve opening times (see Figure \ref{Top}). It is clear from the plots that the longer the opening time, the slower the shock propagation velocity. Also, at higher values of $P_{41}$, the deviation of the valve performance compared to the diaphragm-mode of operation is significant. Short opening times ($<$ 2 milliseconds) are relatively simple to achieve in smaller diameter tubes ($<$ 10 mm) as the moving parts are lighter. The solenoid-based HRRST facility\cite{Tranter_2013} has opening times of less than 1 millisecond for a 6.35 mm shock tube. The direction of movement of the closure element also has a vital role in the valve's opening time. Axial opening of the valve is preferred to gate-type opening because for the same distance traveled by the closure element, the flow area is more in the case of axial opening than transverse opening (gate-type opening). An example is the sleeve-based systems where the closure element has to travel a longer distance to completely open the valve (Sleeve moves about 100 mm stroke in the design reported by Heufer et al.\cite{Heufer_2012}). Abe et al.\cite{Abe_1997} reported an opening time of 1.79 milliseconds for a diaphragmless shock tube system that used an opposing piston. The short opening time results from the piston traversing only half the diameter of the shock tube and the high pressures to control the valve (greater than 150 bar).

\subsection{\label{sec4b:shockformation}Shock formation distance}
\begin{figure}
\includegraphics[scale=0.38]{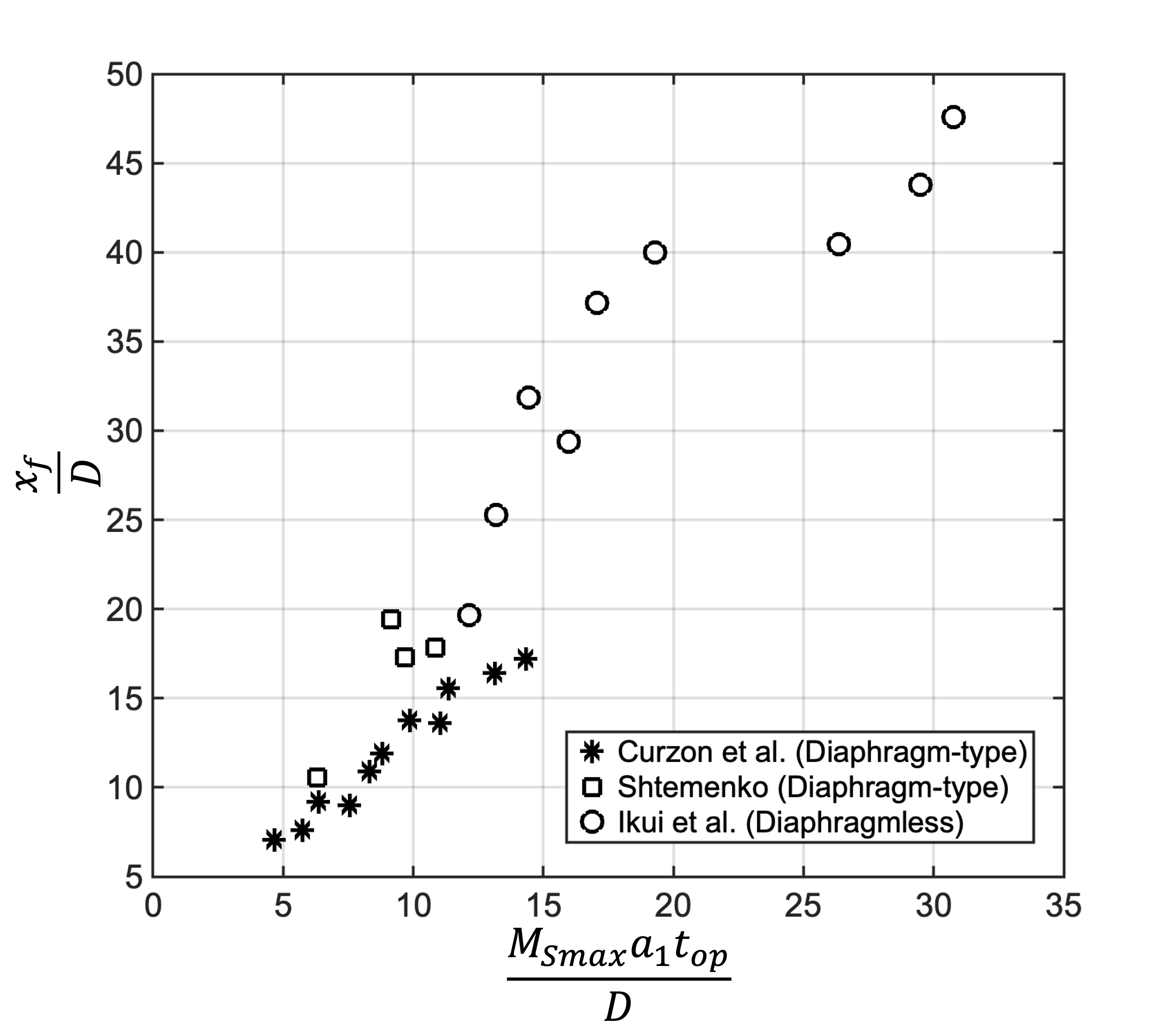}
\caption{\label{TopvsXf}Dependence of shock formation distance ($x_f$) on the opening time ($t_{op}$) for diaphragm-type and diaphragmless shock tubes. {\color{black}(Reprinted with permission from The Japan Society of Mechanical Engineers: Bulletin of JSME, Ikui et al.\cite{Ikui_1979}, copyright 1979)}}
\end{figure}

Since measuring the opening time of the valve is challenging, an alternative method to evaluate the valve's performance is by measuring the shock formation distance. The shock formation distance is directly dependent on the opening time of the diaphragm in a conventional diaphragm-type shock tube\cite{Rothkopf_1974,Pakdaman_2016,Janardhanraj_2021}. Although fast-acting valves do not open from the center of the tube (unlike the diaphragm rupture), a linear dependence between the opening time of the valve and the shock formation distance can be expected in a diaphragmless shock tube as well. This assumption holds good because the piston analogy for shock wave formation holds good for diaphragmless shock tubes. Ikui et al.\cite{Ikui_1979} showed that the dependence of opening time and shock formation distance for diaphragm-type and diaphragmless shock tubes is linear (see Figure \ref{TopvsXf}). The term $M_{Smax}$ refers to the maximum shock Mach number obtained during the measurements. The shock formation distance ($x_f$) also depends on several other parameters indicated by the empirical and functional relationships reported in various studies.

\begin{equation}
    x_f = t_{op}.a_1.f(P_{41}) \cite{Ikui_1969}
\end{equation}
\begin{equation}
    x_f = K_1.s.t_{op} \cite{Simpson_1967}
\end{equation}
where $t_{op}$ is the opening time, $a_1$ is the speed of sound of the gas in the driven section, $s$ is the shock speed, $P_{41}$ is the ratio of initial pressure in driver and driven sections, and $K_1$ is a constant of proportionality. Verifying a fully developed shock wave is relatively simple compared to measuring the valve's opening time. It is generally done by mounting several pressure transducers along the driven section of the shock tube. The amplitude of the pressure trace and the calculation of the shock speed by the time-of-flight method can confirm the shock formation in the shock tube. The non-dimensional shock formation length is represented by the ratio of the shock formation length and the hydraulic diameter of the shock tube. The hydraulic diameter ($d_h$) of a shock tube is defined as,
\begin{equation}
    d_h = \frac{4A}{PR}
\end{equation}
where $A$ is the area and $PR$ is the perimeter of the cross-section. Typically, the shock formation distance in a diaphragm-type shock tube varies between 15-20 hydraulic diameters\cite{Rothkopf_1976}. Longer shock formation distances are expected in diaphragmless shock tubes because of the slower opening times. Onodera reported a non-dimensional shock formation length of 65 for a composite piston-based diaphragmless shock tube\cite{Onodera_1992}. The annular driver arrangement for a diaphragmless shock tube described by Alvarez and co-workers had a consistent shock formation at 41 diameters from the driver in the 165 mm diameter driven tube\cite{Alvarez_2015}. The diaphragmless shock tube based on a hydraulic brake pad mechanism had a non-dimensional shock formation length between 20 and 40\cite{Kosing_1999} for different piston materials. A comparative study between a piston-based and membrane-based diaphragmless shock tube showed that the latter had a shorter shock formation distance\cite{Hariharan_2010}.

\begin{figure*}
\includegraphics[scale=0.3]{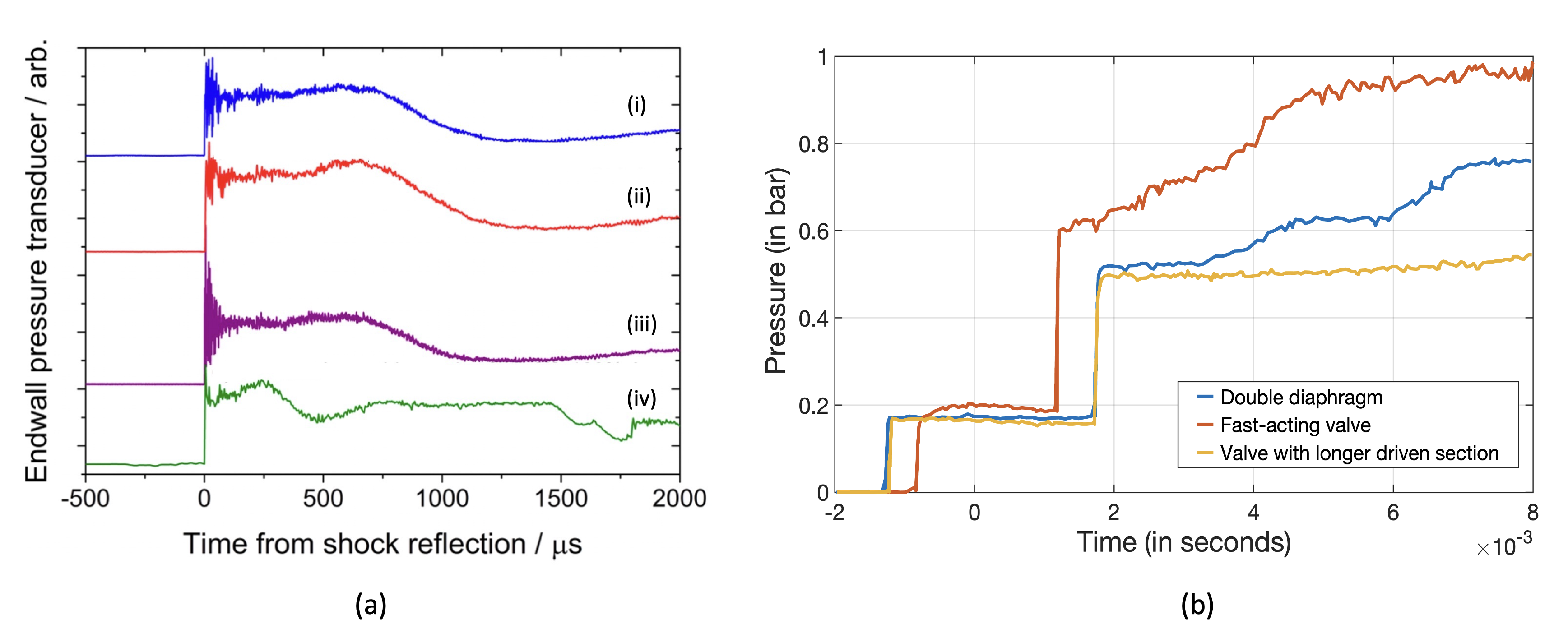}
\caption{\label{AmerLynch} Comparison of shock profiles obtained using the diaphragmless and diaphragm-type shock tubes. (a) Experimental plots reported by Lynch et al.\cite{Lynch_2016}. (i) 12.7 mm bore, modified solenoid driver\cite{Lynch_2016}, 100\% Ar (ii) 12.7 mm bore, modified solenoid driver\cite{Lynch_2016}, 0.003 C$_2$H$_3$F$_3$/Ar (iii) 12.7 mm bore, diaphragm driver, 100 \% Ar (iv) 6.35 mm bore, solenoid driver\cite{Tranter_2013}, 100\% Ar, driven tube length shorter than that in the 12.7 mm shock tube. {\color{black}(Reprinted with permission from AIP Publishing: Review of Scientific Instruments, Lynch P. T.\cite{Lynch_2016}, copyright 2016)} (b) Experimental plots reported by Amer et al.\cite{Amer_2021}. {\color{black}(Adapted with permission from Amer et al.\cite{Amer_2021})}}
\end{figure*}

Lynch et al.\cite{Lynch_2016} compared the shock pressure profiles obtained in a miniature shock tube using diaphragms and a fast-acting valve (shown in Figure \ref{AmerLynch}(a)). The pressure profile obtained in non-reactive conditions using the diaphragmless driver (shown in blue in Figure \ref{AmerLynch}(a)) is similar to the one obtained in the reactive conditions (shown in red in Figure \ref{AmerLynch}(a)). However, a small hump is noticed after about 350 $\mu$s compared to the shock profile obtained in the diaphragm-mode of operation (shown in magenta in Figure \ref{AmerLynch}(a)). The valve opening characteristics are a possible reason for this observation. Non-ideal effects are more pronounced in the 6.35 mm diaphragmless shock tube that has a shorter driven section length (shock profile shown in green in Figure \ref{AmerLynch}(a)). {\color{black} An essential aspect concerning the shock profile and duration is the intended application of the shock tube. If the primary information is extracted before distortions appear in the pressure profile, then the shape of the rest of the signal is of little consequence.} Amer et al.\cite{Amer_2021} showed an improved shock profile with increased driven tube length. The difference in the shock profiles between the double-diaphragm and diaphragmless-mode operations in a 5 m long shock tube (driver length of 2 m and driven length of 3 m) is quite prominent. When the overall length of the shock tube is changed to 10 m (driver length of 3 m and driven length of 7 m), the shock profile resembles the one obtained in the diaphragm-mode of operation. The reflected shock pressure is relatively flat compared to that obtained in the double-diaphragm experiments.

\begin{figure*}
\includegraphics[scale=0.15]{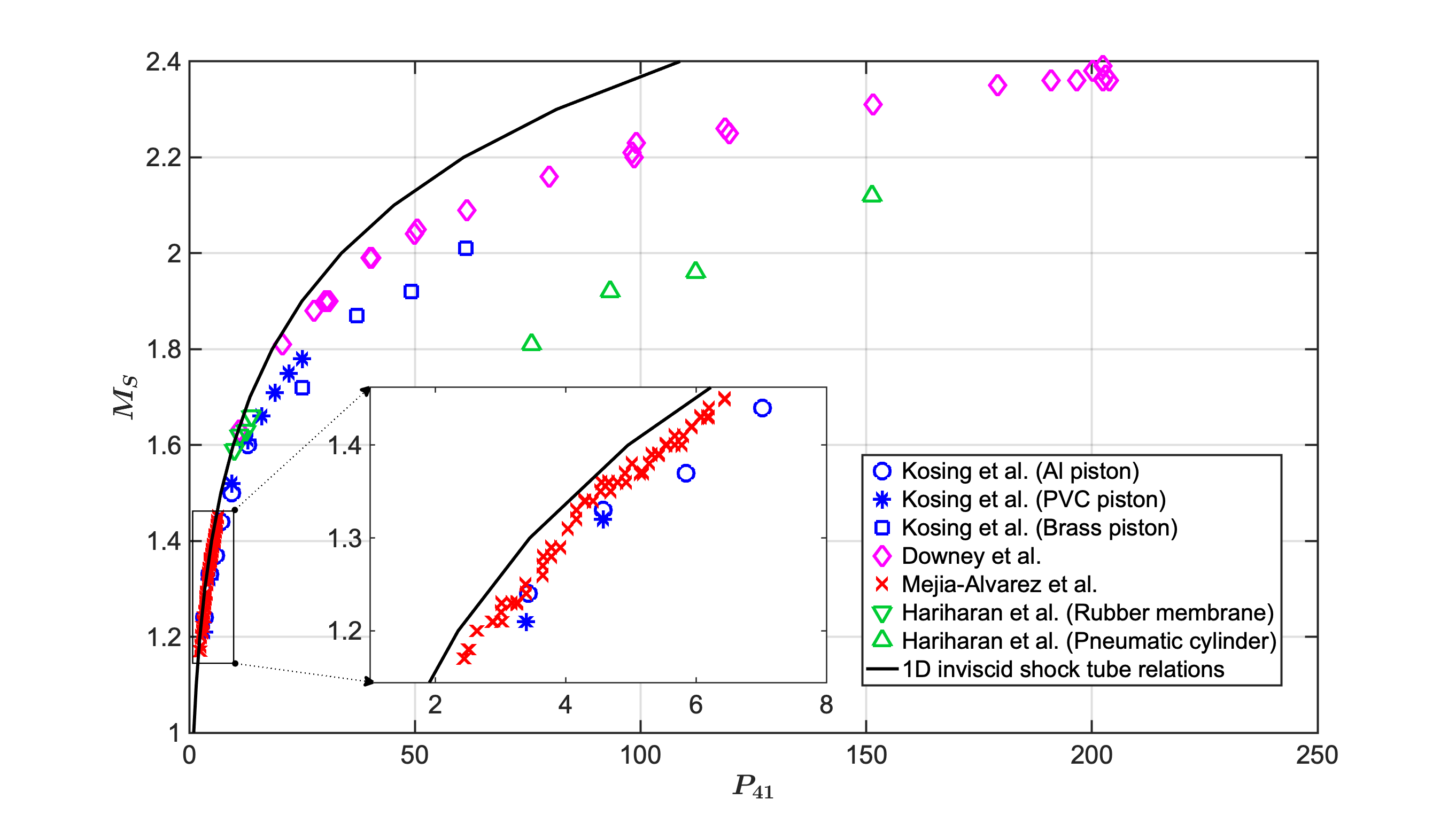}
\caption{\label{P41vsMs}Driver efficiency for diaphragmless designs reported in literature for air as working gas. {\color{black}(Adapted with permissions from Springer Nature: Shock Waves Mejia-Alvarez et al.\cite{Alvarez_2015} copyright 2015, from Springer Nature: Shock Waves Kosing et al.\cite{Kosing_1999} copyright 1999, from Springer Nature: Shock Waves Downey et al.\cite{Downey_2011} copyright 2011, and from Springer Nature: Shock Waves Hariharan et al.\cite{Hariharan_2010} copyright 2010)}}
\end{figure*}

\begin{figure}
\includegraphics[scale=0.46]{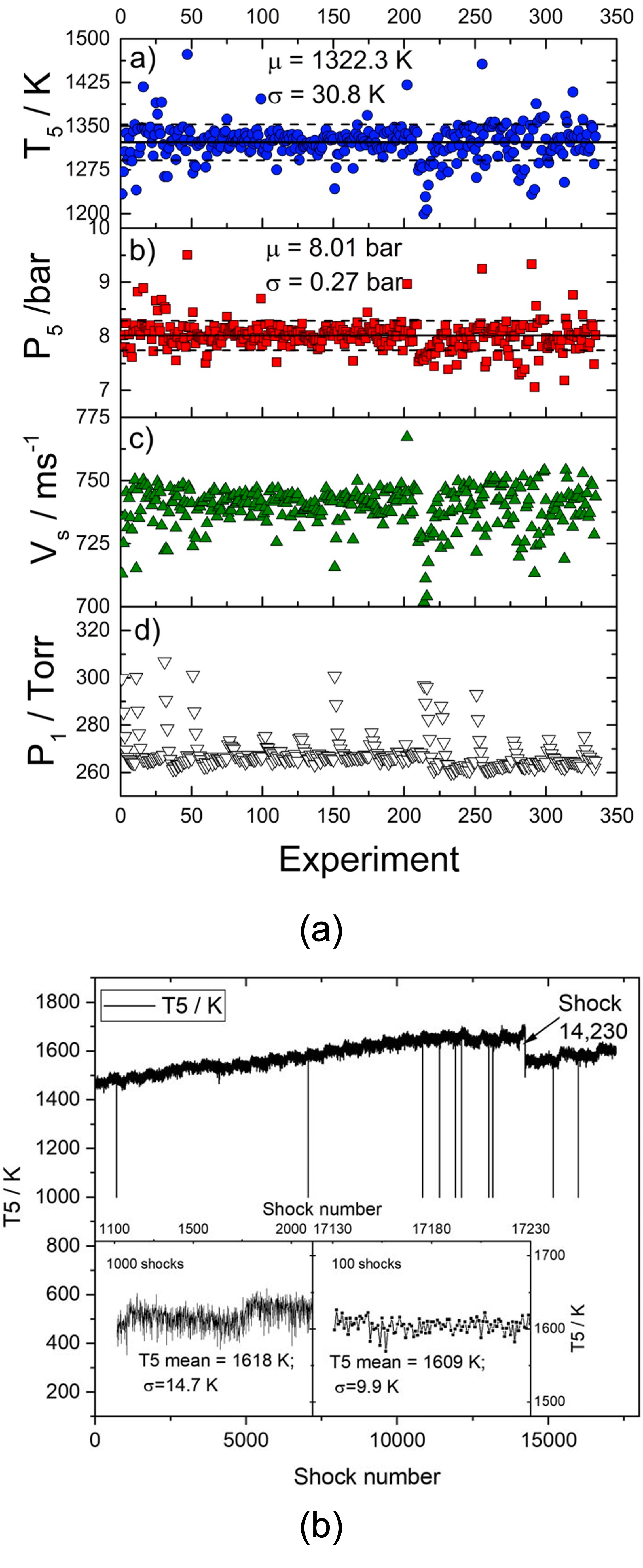}
\caption{\label{highrepetition}(a) Repeatability of signals obtained by Lynch et al.\cite{Lynch_2016} in a miniature diaphragmless shock tube operated at 0.25 Hz. Solid lines represent $\mu$ (mean value) while dashed lines represent $\pm \sigma$ (standard deviation) from $\mu$. {\color{black}(Reprinted with permission from AIP Publishing: Review of Scientific Instruments, Lynch P. T.\cite{Lynch_2016}, copyright 2016)} (b) Reflected shock conditions from 17000 experiments as reported by Tranter and co-workers\cite{Tranter_2020}. The insets show the mean and standard deviation of the reflected shock temperature for 1000 and 100 runs acquired at 1 Hz repetition rate. {\color{black}(Reprinted with permission from AIP Publishing: Review of Scientific Instruments, Tranter et al.\cite{Tranter_2020}, copyright 2020)}}
\end{figure}

\subsection{\label{sec4c:range}Operating range, repetition rate and reliability}
The operating range of diaphragmless shock tubes depends on the design features (driver-driven configuration, closure, actuation, and control elements) used in the fast-acting valve. Table \ref{tab:table1} shows the maximum driver pressure used in the various diaphragmless shock tubes. It is essential to mention here that the value of maximum driver pressure listed in the table is simply the highest value of pressure used for the experiments. This value does not necessarily indicate the upper operating limit of the valve. Table \ref{tab:table1} shows that diaphragmless shock tubes can operate at pressures as low as 1 bar to pressures of up to 200 bar. A helpful way to estimate the efficiency of the diaphragmless shock tubes is by determining the shock Mach number as a function of the initial conditions in the shock tube. It is generally seen that the efficiency of diaphragmless drivers is lower than diaphragm-type shock tubes\cite{McGivern_2019}. Figure \ref{P41vsMs} compares the efficiency for different diaphragmless shock tubes. The experimental data are plotted against the one-dimensional inviscid relation shown in equation \ref{eq1} (represented as a solid line in the figure). Experimental data from only a few reports have been used in the plot as the working gas differs for different studies. Figure \ref{P41vsMs} shows that the performance of the diaphragmless shock tubes significantly deviates from the predictions beyond $P_{41}=50$. Therefore, the performance of diaphragmless shock tubes at higher initial pressures needs to be improved.

The turnaround time between experiments is significantly reduced by using diaphragmless drivers. Incorporating a mechanism to bring the closure element back to its original position improves the turnaround time further. Using a secondary diaphragm to run a diaphragmless shock tube is a significant disadvantage in terms of the repetition rate\cite{Oguchi_1976,Ikui_1976,Udagawa_2007}. Amer and co-workers compared the turnaround times in shock tubes operated in single-diaphragm mode, double-diaphragm mode, and diaphragmless mode\cite{Amer_2021}. They found that the operating time for their diaphragmless shock tube was about 6 min compared to about 32 min and 48 min for the single-diaphragm and double-diaphragm modes, respectively. Miniature diaphragmless shock tubes having a high-repetition-rate of 4-5 Hz have also been demonstrated\cite{Tranter_2013,Shiozaki_2005}. Figure \ref{highrepetition} shows the repeatability obtained in a high-repetition-rate diaphragmless shock tube operated at 0.25 Hz. The cost and lifetime of the fast-acting valves are a couple of key points to consider while designing diaphragmless shock tubes. Since most fast-acting valves are customized designs, developed in laboratories and used solely for research purposes, their cost is not reported in the literature. The cost of commercial valves can be about a few thousand US dollars depending on the diameter of the valve. The high-repetition rate solenoid actuated driver valve has been tested for several thousand experiments\cite{Tranter_2020}. The commercial fast-acting valve manufacturer claims the valve's lifetime to be at least 5 million shots. The scalability of the diaphragmless shock tube to large diameters is essential in some applications. Commercial valves are currently available up to a maximum diameter of 80 mm.

\subsection{\label{sec4d:kineticperf}Performance in reflected shock mode}
While shock tubes have several different applications, one of the primary uses is the study of chemical kinetics by the combustion community, which is often done in the reflected shock mode (utilizing the $T_5$ and $P_5$ conditions). In such experiments, key performance characteristics are the attenuation rate of the incident shock wave, the temporal dependence of the reflected shock pressure (d$P_5/$d$t$), and the duration of test-time. The attenuation of the incident shockwave is particularly important in determining the reflected shock temperature that is vital for chemical kinetics studies\cite{Campbell_2017}. In general, keeping the incident shock attenuation rate as low as possible is desirable to avoid a large axial temperature gradient. Similarly, d$P_5/$d$t$ should be minimal to perform kinetic studies at conditions fairly close to the reflected shock ($T_5$, $P_5$). Other performance aspects of diaphragmless shock tubes pertaining to chemical kinetics studies include the range of Mach numbers that can be obtained and the potential for tailoring to extend the reflected-shock test times. The range of Mach numbers obtained in the shock tube is directly related to the operating pressure range, as discussed in the previous section. A recent study investigated these reflected-shock characteristics by employing a newly developed diaphragmless shock tube to study combustion chemistry\cite{Janardhanraj_2022}. The average d$P_5/$d$t$ in the reflected shock region was about 2.5\%/ms. The velocity error was about ±0.224\%, and the attenuation rate was about 0.41\%/m. These values are comparable to those obtained using diaphragm-type shock tubes\cite{Campbell_2014}. In this work, the driver gas was tailored using nitrogen gas (18\% by volume), as described in detail by Campbell et al.\cite{Campbell_2015}, to obtain longer test times for some of the fuel mixtures. 

\subsection{\label{sec4e:limitations}Improving valve performance}
While diaphragmless shock tubes present several advantages over conventional shock tubes, there is still significant scope for performance improvement. The limitations discussed below are specific to certain design concepts. A valve design that addresses all such limitations without compromising performance and efficiency would be the most desirable.  
\begin{itemize}
    \item \textbf{Shortening opening times} - One of the main limitations of a diaphragmless shock tube, compared to a conventional shock tube, is that the time scales of the diaphragm rupture process cannot be matched. The opening time of fast-acting valves has to be improved to reduce the shock formation distance in the shock tube.
    \item \textbf{Avoid flow turning} - The configurations of the driver and the driven sections in a diaphragmless shock tube are not necessarily similar to the conventional shock tube (except for very few design concepts employing Type-V driver-driven configuration). The flow turning at the location of the fast-acting valve can lead to losses that will affect the strength of the shock wave produced and hence lowers the efficiency of the diaphragmless driver.
    \item \textbf{Minimal flow obstruction} - The components of the fast-acting valve (closure, actuation, and control elements) obstruct the gas flow in the shock tube. Complex flow interactions lead to undesirable effects in the observation window of the diaphragmless shock tube.
    \item \textbf{Higher operational pressures} - The operation range of diaphragmless shock tubes is limited because larger forces must be overcome at higher pressures to move the closure element. The actuators controlling the movement of the closure element become more bulky and expensive. 
    \item \textbf{Reduce shock wave attenuation} - The presence of obstacles at the driver-driven interface or the perpendicular orientation of driver-driven sections can result in the reflected expansion waves catching up with the moving shock front earlier than expected. This scenario can lead to faster attenuation of the shock wave.
    \item \textbf{Avoid damage to seals} - The amount of wear and tear of a seal is dependent on the valve design. In most designs, the seals on the closure element are breached during the operation of the diaphragmless shock tube. Repeated operation of the diaphragmless shock tube can wear and tear these seals and other moving parts. Therefore, there is a need for constant maintenance of the parts while using diaphragmless shock tubes. 
    \item \textbf{Reduce noise during operation} - Pneumatically-driven fast-acting valves operate by the sudden exhaust of pressurized gases due to the actuating mechanism. This process can be extremely noisy, and sufficient measures must be taken to exhaust these gases.
\end{itemize}

\section{\label{sec5:model}Mathematical model for fast-acting valves}
Mathematical models for several diaphragmless driver configurations have been developed and verified previously. Rego et al. \cite{Rego_2008} developed a numerical model based on the motion equation to describe the piston sliding time against pressure ratio for a double-piston arrangement similar to Oguchi et al.'s design. Alvarez et al.\cite{Alvarez_2015} developed a model to describe the variants of the Oguchi et al. designs. Portaro et al.\cite{Portaro_2015} analyzed the performance characteristics of the sleeve-based diaphragmless shock tube driver proposed by Downey et al.\cite{Downey_2011} using computational fluid dynamics (CFD) simulations. A CFD study was also performed to evaluate the performance of Heufer et al.'s design\cite{Heufer_2012} to give insights into the complex interaction of the sliding piston with the transient flow. Udagawa and co-workers developed a numerical model using the Runge–Kutta–Gill method to understand the motion of the rubber membrane-based valve\cite{Udagawa_2008}. In another work, Udagawa et al. suggested a simplified model for a double-sliding piston diaphragmless driver. This section presents a generalized model to describe fast-acting valves and applies to two popular driver configurations as test cases.

A typical fast-acting valve consists of a moving piston to block the driver and the driven section, a mechanism to pull the piston as quickly as possible, and a provision to engage the piston to barricade the sections between runs. In every fast-acting valve, the mechanism by which the piston is released is different. However, the pneumatically or electropneumatically operated piston is preferred in most cases. Following are the key requirements for a fast-acting valve used in a shock tube:
\begin{enumerate}
    \item Short opening time (on the order of a few milliseconds)
    \item Lightweight and high-strength piston
    \item Simple and cost-effective auxiliary systems
    \item Ability of components to withstand extreme environments like high operating pressure, temperature, and corrosive gases. 
\end{enumerate}
Mathematically, one can design a fast-acting valve based on a simple approach to derive the vital geometrical parameters per the operational requirements, considering the abovementioned criteria. 

\begin{itemize}
    \item Firstly, the minimum distance ($x$) moved by the piston has to be estimated (see Figure \ref{fig:valve_opening}). The effective diameter ($d$) of the driven section determines the value of $x$. Consider a piston with a circular cross-section that seals the driven section passage. The retraction of the piston to open the valve thus traces a cylindrical pathway whose circumference offers the entry of fluid flow into the driven section. The circumference of the cylindrical pathway and the effective flow diameter required can be thus equated to extract the piston displacement as,
    \begin{align} \label{eq:for_x}
        \begin{split}
        x\pi d &= \frac{\pi}{4}d^2,\\
        x &= d/4.
        \end{split}
    \end{align}
    The distance moved by the piston is determined by matching the flow area open to the driven section using this approach. 
    
    \item Secondly, the acceleration ($a_p$) required to move the piston of a certain mass ($m_p$) is expressed in terms of the distance moved ($x$) and the total time taken for this motion(this is essentially the valve opening time $t_{op}$). From Newtonian mechanics, the motion equation can be derived for $a_p$. 
    
    \item Thirdly, a free-body diagram of the piston helps determine the forces acting on the different faces. The force contributions due to friction, self-weight, seat loading and other design-specific loads (as explained in section \ref{sec3c:act}) must be included. The force required to move the piston can be related to the acceleration as,
    \begin{equation} \label{eq:for_Fa}
        F \geq m_pa_p.
    \end{equation}
    The piston's mass ($m_p$) can be reduced by having a high strength-to-weight ratio material or material with low density ($\rho_p$). There are various choices, including Teflon, polyurethanes, hard-plastics, aluminum, and carbon-fiber-reinforced thin-metal shells for the lightweight piston material. 
    
    \item Finally, the mechanism to generate $F$ is decided and engineered. Popular choices are electromagnetic force, spring assistance, and pneumatic drives. Here, a pneumatically operated piston is considered.
\end{itemize}

\subsection{\label{sec5a:case1}Case I: Annular driver configuration}
\begin{figure*}
\includegraphics[width=0.8\textwidth]{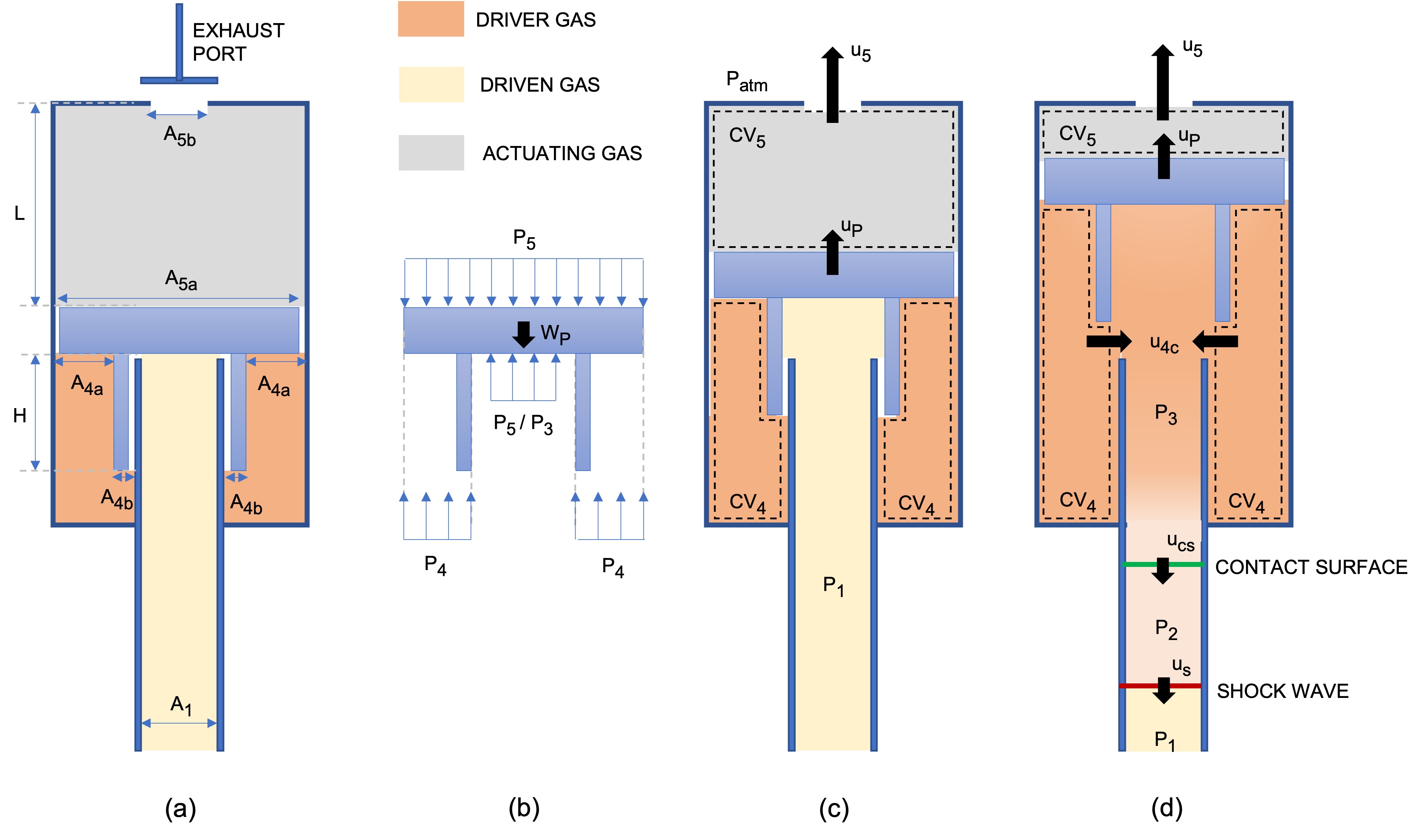}
\caption{\label{fig:alvarez model} Schematic diagrams of the model developed by Mejia-Alvarez et al.\cite{Alvarez_2015} (a) Relevant geometric parameters. (b) free body diagram of the piston; $P_1$ contributes before valve opening; $P_3$ replaces $P_1$ after valve opening. (c) Conditions before seals are breached. (d) Conditions after seals are breached. {\color{black}(Adapted by permission from Springer Nature: Shock Waves, Mejia-Alvarez et al.\cite{Alvarez_2015}, copyright 2015)}}
\end{figure*}

The schematics of the model developed by Alvarez et al. are shown in Figure \ref{fig:alvarez model}. A piston with a lip configuration is shown in the figure, although the model was extended to cases where a piston with a plug and sleeve was used as a closure element. Relevant geometric parameters are shown in Figure \ref{fig:alvarez model}a. The position, speed and acceleration of the piston are easily related as,
\begin{equation}
    a_p = \frac{\mathrm{d}u_p}{\mathrm{d}t} = \frac{\mathrm{d}^2x}{\mathrm{d}t^2}
\end{equation}

The free-body diagram of the pressures acting on the different faces of the piston before and after breaching the seals is shown in Figure \ref{fig:alvarez model}b. Figures \ref{fig:alvarez model}c and \ref{fig:alvarez model}d show the conditions before and after the seals are breached. The pressure evolution in the control volumes $CV_4$ and $CV_5$ are determined by two ordinary differential equations as the gas is released through the back pressure valve. The acceleration of the closure element for the case of piston with a lip is obtained as,
\begin{equation}
    \frac{\mathrm{d}u_p}{\mathrm{d}t} = \frac{1}{m_p}[(A_{4a}+A_{4b}).P_4+(A_t+A_w).P_3-A_{5a}.P_5-w_p]
\end{equation}
For a piston with a plug, the acceleration is,
\begin{equation}
    \frac{\mathrm{d}u_p}{\mathrm{d}t} = \frac{1}{m_p}[(A_{4a}+A_{4b}+A_w).P_4+A_t.P_3-A_{5a}.P_5-w_p]
\end{equation}
For a sleeve, the acceleration is,
\begin{equation}
    \frac{\mathrm{d}u_p}{\mathrm{d}t} = \frac{1}{m_p}[A_{4b}.P_4-A_{5a}.P_5-w_p]
\end{equation}
where $w_p$ is the weight of the piston. The equations for acceleration and pressure evolution are solved numerically to obtain the velocity and opening time of the diaphragmless valve.

\subsection{\label{sec5b:case2}Case II: Inline driver configuration}

\begin{figure*}
\includegraphics[width=0.7\textwidth]{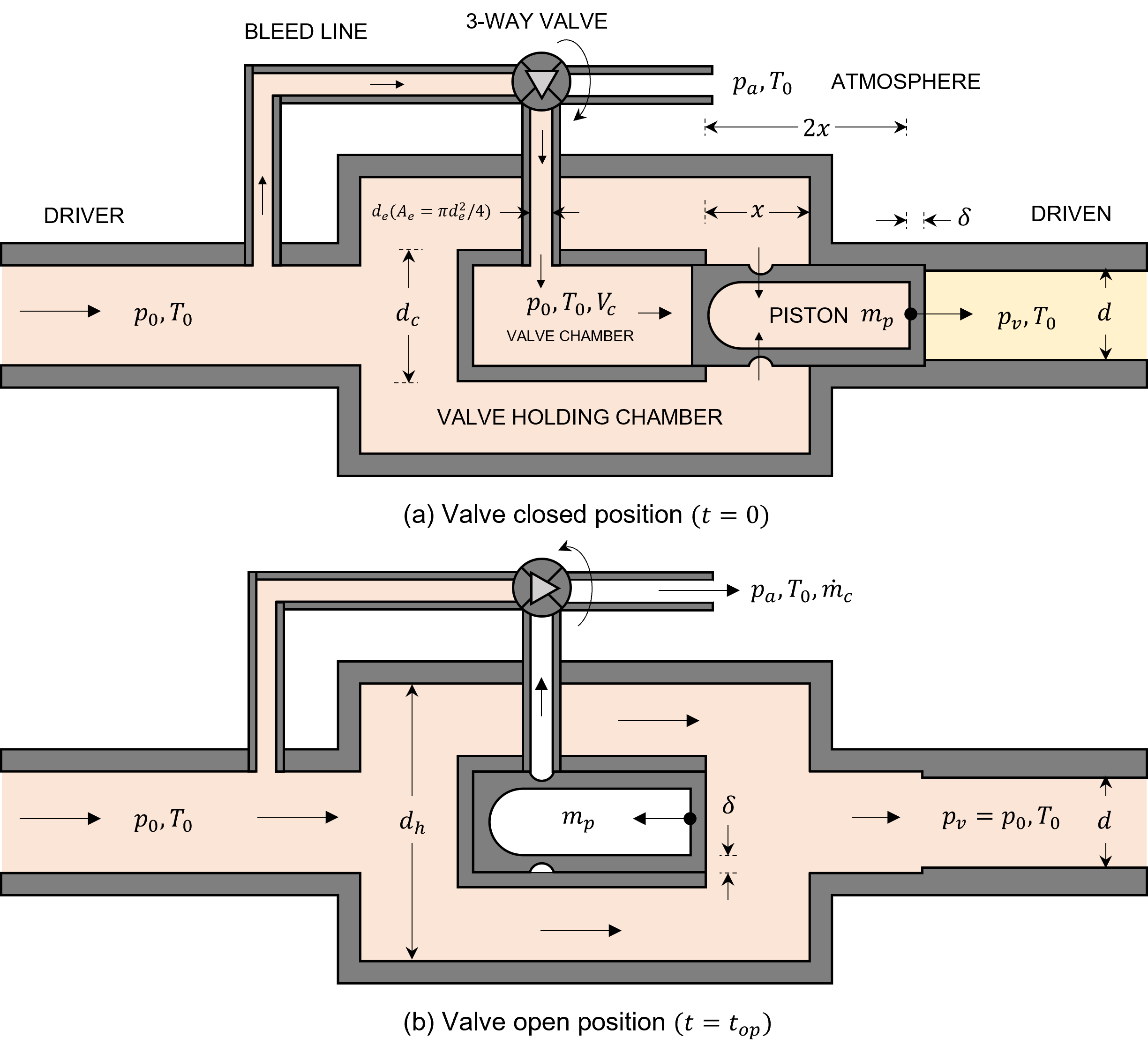}
\caption{\label{fig:valve_opening} Schematics showing the vital geometrical and physical parameters involved in the (a) closing and (b) opening operation of a typical fast-acting valve (Type-IIa, however, the parameters mostly remain the same).}
\end{figure*}

An internally held pneumatically operated annular fast-acting valve during the closing and opening is shown in Figure \ref{fig:valve_opening}. The piston is initially at rest, so the initial velocity is $u=0$. The piston should not throttle the flow to enter the driven section as it reduces $t_t$ significantly. Hence, a part of the piston is extended so that it runs $x$ times deep into the driven section instead of exactly closing in front of the driven section. In this manner, the tip of the piston, upon retracting from the driven section entrance, already possesses sufficient velocity and quickly allows the driver gas to enter the driven section. The desired $a_p$ is then calculated as,
\begin{align} \label{eq:for_ap}
\begin{split}
    2x &= ut_v + \frac{1}{2}a_pt_v^2,\\
    a_p &= \frac{4x}{t_v^2}.
    \end{split}
\end{align}

For the present configuration, after considering the hemispherical dome-end and perforation, $m_p$ is calculated as,
\begin{align} \label{eq:for_mp}
    \begin{split}
    m_p &\lesssim \frac{\pi \rho_p}{4} \left[d^2(2x+2\delta ) - ...\color{white}\frac{1}{1}\right] \\
    & \color{white}\left[\color{black}...(d-2\delta )^2\left(2x-\frac{d-2\delta }{2}\right)-\frac{(d-2\delta )^3}{3}\right]\color{black}.
    \end{split}
\end{align}
The mass per unit density of the piston is calculated by subtracting the volume of the outer cylinder from the sum of the inner cylinder and the inner hemispherical dome-end. Additionally, the piston mass variation due to the peripheral perforation is considered negligible. Considering the calculated $a_p$ and $m_p$ in the previous steps, the net force is given by
\begin{equation} \label{eq:for_Fb}
    F \gtrsim \frac{4xm_p}{t_v^2}. 
\end{equation}

The piston enclosed in the valve chamber retracts quickly upon exhausting the gas pressure in the chamber, as the surrounding driver gas and the suction pressure produce the necessary force to move the piston. The time scale of gas exhaustion should be smaller than the valve opening time scale. The valve's performance can be further improved by exhausting the gas into a pre-evacuated chamber (typically $\leq$ 0.1  Torr). Smaller the valve chamber volume ($V$), the quicker the exhaustion and, thus, the piston's movement. The total time ($t_0$) required to empty $V$ is calculated from the mass conservation law or continuity equation in fluid dynamics as,
\begin{align} \label{eq:for_t0}
    t_0 &= \frac{2}{k(\gamma-1)}\int_{r^*}^r(r^2+1)^{(2-\gamma)/(\gamma-1)}dr + t_c, 
\end{align}
where,
\begin{align*}
    r^* &= \sqrt{\frac{\gamma-1}{2}},\\
    r &= \sqrt{\left(\frac{p_0}{p_a}\right)^\frac{\gamma-1}{\gamma}-1},
    \end{align*}
    \begin{align*}
    k &= \frac{-\gamma A_e}{V_c}\left(\frac{2\gamma R T_0}{\gamma-1}\right)^\frac{1}{2}\left(\frac{p_0}{p_a}\right)^\frac{\gamma-1}{2\gamma},\\
    t_c &= \frac{V_c}{A_e\sqrt{\gamma R T_0}}\left(\frac{2}{\gamma-1}\right)\left(\frac{\gamma+1}{2}\right)^{\frac{\gamma+1}{2(\gamma-1)}}\left[\left(\frac{p_0}{p_c}\right)^{\frac{\gamma-1}{2\gamma}}-1\right].
\end{align*}
Here, $A_e$ represents the exhaust piping cross-sectional area ($\pi d_e^2/4$), $\gamma$ is the specific heat ratio, $R$ (J/kg/K) is the specific gas constant, and $p_c$ is the critical total pressure required to choke the flow in the exhaust duct. Value of $V_c$ in the considered case is calculated as,
\begin{equation} \label{eq:for_Vc}
    V_c = \frac{\pi}{4}d^2(2x+2\delta).
\end{equation}

For the current case, other geometrical parameters like the valve-holding chamber dimension are calculated based on $d$ and $d_c$, where $d_c$ is the outer diameter of the valve chamber. It has to be noted that $d_c$ is larger than $d$. The effective diameter and the annular diameter of the valve-holding chamber can be equated to avoid local choking. Assuming $d_c \sim d$, $d_h$ is given as,
\begin{align} \label{eq:for_dh}
\begin{split}
    \frac{\pi}{4}(d_h^2 - d_c^2) &= \frac{\pi}{4}d^2,\\
    d_h^2 &= d^2 + d_c^2 \sim 2d^2,\\
    d_h &\gtrsim \sqrt{2}d.
    \end{split}
\end{align}

The piston is held in the valve chamber as a concentric slider with a 0.05-0.1 mm tolerance. The sliding piston offers negligible frictional force and allows light gas leakage for the considered short-time operation. During the valve-closed operation, the gas fills the valve holding chamber and valve chamber with the same driver pressure ($p_0$). The body of the piston is perforated or made of small holes such that the driver gas fills it. Upon releasing the 3-way valve, a fast-acting pneumatic valve with an opening time of 3-30 ms (e.g., Festo\textsuperscript{\textregistered} fast-switching valves-MH series), the valve chamber gas exhausts into the atmosphere rapidly. As mentioned earlier, the valve performance can be further improved by exhausting the gas into a pre-evacuated chamber. The net force acting on the piston ends is the same due to how they are shaped. The end that faces the valve chamber is hemispherical, thus the net force $F_v$ is 
\begin{equation} \label{eq:for_Fv}
    F_v = (p_0-p_a)\left[2\pi\left(\frac{d-2\delta}{2}\right)^2\right].
\end{equation}
Similarly, the end that faces the driven section is flat, and thus the net force $F_d$ is
\begin{equation} \label{eq:for_Fd}
    F_d = (p_0-p_v)\left[\frac{\pi}{4}(d-2\delta)^2\right].
\end{equation}
For $p_0>p_a$ and $p_v\sim 0$, the net resulting force acting on the sides of the piston is the same ($F_v = F_d$). Given that $F_v >F$, the piston moves back into the valve chamber upon the actuation of the 3-way valve resulting in the opening of the fast-acting valve. In summary, as long as the time scales of valve chamber depressurization and of opening the 3-way valve are less than the desired valve-opening time, the fast-acting valve operates fine. 

Let us consider a simple case to understand the design and sizing of parameters for a typical fast-acting valve for an inline driver configuration type which is the most commonly used arrangement in aero/gas-dynamic testing facilitates. Figure \ref{fig:valve_opening} shows the vital design parameters. The desired input parameters are given in Table \ref{tab:table3}.
\begin{table}
\caption{\label{tab:table3} Input parameters for the fast-acting valve design.}
\begin{ruledtabular}
\begin{tabular}{ll}
\textbf{Parameters} & \textbf{Values} \\
\midrule
Effective tube diameter ($d$) & 25 mm\\
Working gas ($\gamma,R$) & air (1.4, 287 J/kg/K)\\
Driver pressure ($p_0$) \& temperature ($T_0$) & 2 bar, 300 K\\
Driven pressure ($p_v$) \& temperature ($T_v$) & $\sim$0 bar, 300 K\\
Ambient pressure ($p_a$) \& temperature ($T_a$) & 1 bar, 300 K\\
Desired uniform piston wall-thickness ($\delta$) & 3 mm\\
Piston material density - Teflon\textsuperscript{\textregistered} ($\rho_p$) & 2200 kg/m$^2$\\
Desired valve opening time ($t_v$) & 3 ms\\
\end{tabular}
\end{ruledtabular}
\end{table}
From equations \ref{eq:for_x}-\ref{eq:for_Fa}, and \ref{eq:for_ap}-\ref{eq:for_Fd}, the key design parameters are computed and tabulated in Table \ref{tab:table4}.
\begin{table}
\caption{\label{tab:table4} Output parameters for the fast-acting valve design.}
\begin{ruledtabular}
\begin{tabular}{ll}
\textbf{Parameters} & \textbf{Values} \\
\midrule
Equivalent displacement ($x$) & 6.35 mm\\
Piston acceleration required ($a_p$) & 2822.22 m/s$^2$\\
Mass of the piston ($m_p$) & 14.7 g\\
Force needed to move the piston ($F$) & 41.46 N\\
Holding chamber diameter ($d$) & 35.9 mm\\
Holding chamber volume ($V_c$) & 9.48 $\times$ 10$^{-6}$ m$^3$\\
Pressure and vacuum side force ($F_d=F_v$) & 59.9 N\\
\end{tabular}
\end{ruledtabular}
\end{table}
The valve design will be effective if a suitable material for the casings and the piston is chosen. The number of fill holes in the piston rod can be as many as the surface of the piston-rod supports. A notable advantage is that the gas fills the piston-rod interior rapidly, making the piston lightweight, e.g., $m_p$ = 14.7 g in the present case. 

Overall, the mathematical model described will come in handy while developing customized fast-acting valves based on the desired application and is a useful guide to follow prior to the fabrication of the valve. An important point to note is that the opening times calculated by the above procedure provide a first-order approximation of how quickly the valve might open. Several additional factors that influence the valve opening time are difficult to estimate accurately. Some useful recommendations based on the mathematical model are to have a lightweight closure element, minimize the surfaces on the closure element that lead to unwanted forces opposing the net retraction force, reduce the friction due to seals and incorporate a small clearance distance so that the closure element accelerates before breaching the seal. Other recommendations to improve the valve's aerodynamics include streamlining the flow and driver-driven configurations to ensure minimal flow turning. Material selection is also important in valve design. For example, high-enthalpy aerodynamic testing facilities generally operate at temperatures above ambient. Commercial fast-acting valves have limitations while continuously operating above 400-500 K. Suitable thermal resistant materials can be employed in these cases, and the fast-acting valve design can be tailored appropriately to suit the application needs.

\section{\label{sec6:applications}Shock wave applications using diaphragmless shock tubes: Current trends}
The advantages of diaphragmless shock tube drivers made them a desirable candidate to replace the diaphragm-type mode of operation. Additionally, they have opened a range of new applications which were not previously possible with conventional shock tubes. The new possibilities that have emerged due to the high reproducibility, fast repetition, cleaner flow characteristics, and precise control are described in the subsequent sections.

\subsection{\label{sec6a:kinetics}Reliable studies in shock wave chemistry and physics}
\begin{figure}
\includegraphics[scale=0.5]{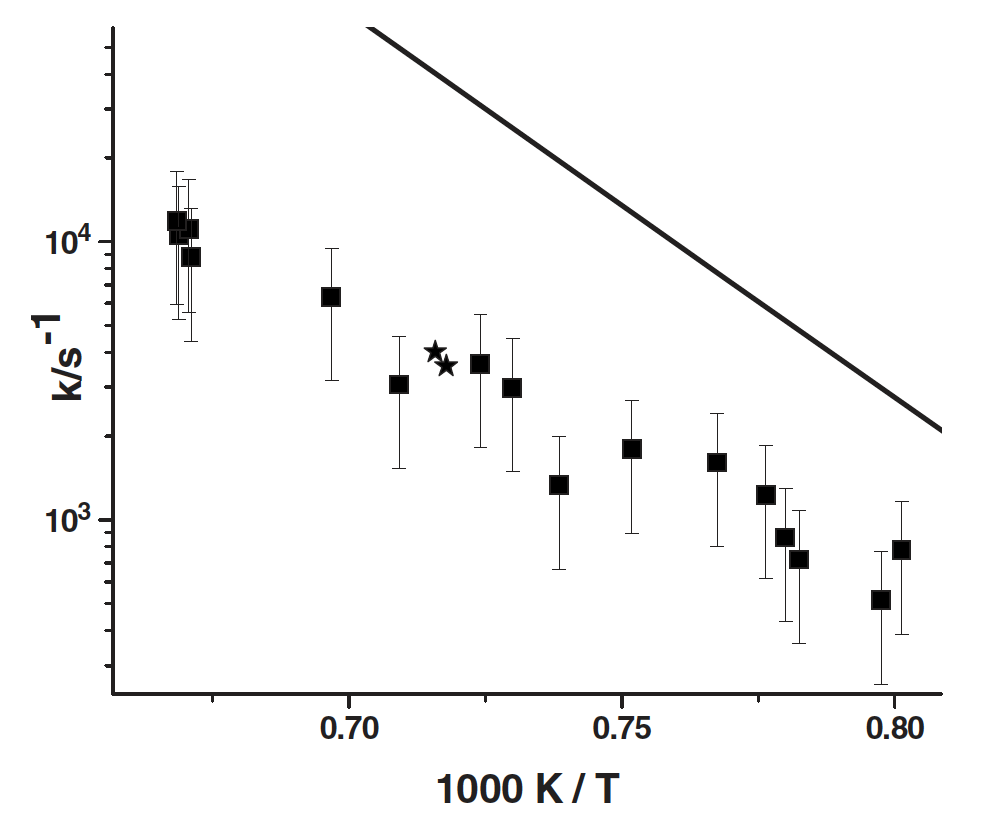}
\caption{\label{TranterRatecoeff} Rate coefficients for the dissociation of fluoroethane reported by Tranter et al.\cite{Tranter_2008}. Solid squares are data points obtained from diaphragm-mode of operation\cite{Giri_2008}, solid stars represent data obtained from diaphragmless driver section and solid line represents $k_\infty$\cite{Giri_2008}. {\color{black}(Reprinted with permission from AIP Publishing: Review of Scientific Instruments, Tranter et al.\cite{Tranter_2008}, copyright 2008)}}
\end{figure}

Shock tubes can produce conditions that cover a wide temperature and pressure range and provide time scales suitable for studying various chemical reactions. Some shock tube investigations include temperature dependence of the reaction rate constants, emission/absorption spectra of various reaction intermediates, ignition delay times of pure substances and distillate fuels, and mechanisms of many elementary reactions. In these studies, the shot-to-shot fluctuation of the thermodynamic properties behind the reflected shock in a conventional diaphragm-type shock tube can affect the quality of the acquired kinetic data. Since several gas properties are monitored at a fixed condition, comparing data at the same condition becomes a problem if the experiments cannot be precisely replicated. Therefore, using diaphragmless shock tubes in chemical kinetics studies is advantageous, and the unique diaphragmless shock tube design concepts described in this review facilitate these investigations. Yamauchi et al.'s diaphragmless shock tube was used for emission spectra studies in reproducible shock heated conditions\cite{Yamauchi_1987}. This facility was also used to improve the kinetic data at higher temperatures up to 3197 K\cite{Koshi_1990,Hsiao_2002}. Matsui and co-workers\cite{Matsui_1994} used a combination of a diaphragmless piston-actuated shock tube, excimer laser photolysis, and Atomic Resonance Absorption Spectrometry (ARAS) to study individual elementary reactions. The bellow-based diaphragmless shock tube developed by Tranter and co-workers\cite{Tranter_2008} demonstrated the ability of a diaphragmless shock tube to obtain kinetic data in the low-pressure limit. A comparison between the rate coefficients obtained using the diaphragmless and diaphragm-modes is shown in Figure \ref{TranterRatecoeff}. Janardhanraj et al.\cite{Janardhanraj_2022} also demonstrated using a diaphragmless shock tube to measure ignition delay times in methane and n-hexane oxidation. The level of control on the operating conditions allowed the post-shock pressure to be constrained to a much narrower range for a wide range of temperatures than a conventional diaphragm-type shock tube. The diaphragmless shock tube was also coupled to a TOF-MS (Time-of-flight Mass Spectrometer), which allowed signal averaging over multiple experiments\cite{Giri_2008}. This application was, again, not possible with a conventional shock tube. The work by Shaik et al.\cite{Shaik_2021} studied fluid dynamic effects in miniature tubes using synchrotron-based X-ray absorption, which is not possible with a conventional shock tube. Some other important studies that demonstrate the use of diaphragmless shock tubes in chemical kinetics include investigations of the decomposition of styrene\cite{Sikes_2021}, reactions of propyl radicals\cite{Banyon_2021} and butyl radical isomers\cite{Randazzo_2020}.

The HRRST (high repetition rate shock tube) facilitated high temperature, high pressure, gas-phase experiments at facilities such as synchrotron light sources where space is limited, and many experiments need to be averaged to obtain adequate signal levels\cite{Tranter_2013}. The diaphragmless shock tube facility was designed to generate reaction conditions of $T > 600$ K, $P < 100$ bars at a cycle rate of up to 4 Hz. The benefits of shock tube/TOF-MS research were extended to include synchrotron-sourced PI-TOF-MS (Photo Ionization Time-of-flight Mass Spectrometer) as well\cite{Lynch_2015}. The HRRST facility has also been widely used for other studies, which include chemical thermometry using 1,1,1-trifluoroethane dissociation\cite{Lynch_2016}, ignition delay measurements\cite{Tao_2018}, kinetic modeling of ignition\cite{Tao_2019}, and pyrolysis studies using double imaging photoelectron/photoion coincidence spectroscopy\cite{Nagaraju_2021}. High-repetition shock waves produced by an automated diaphragmless shock tube have also facilitated molecular beam scattering experiments\cite{Shiozaki_2005}. Diaphragmless shock tubes have been successfully used in gas dynamic laser (GDL) experiments\cite{Oguchi_1978_1,Maeno_1980,Rego_2007_1,Rego_2007_2} as, in addition to the high reproducibility of signals, they also limit the impurities inside the shock tube. The diaphragmless shock tube compresses and pre-heats a mixture instantaneously to elevated pressure and temperature in these experiments. The mixture is delivered to a nozzle where the rapid cooling of the hot gas creates a population inversion and produces a laser medium. The precise control over the shock wave generation process in the diaphragmless shock tubes also allows for the synchronized operation of two shock tubes, as demonstrated by Maeno and Oguchi\cite{Maeno_1980}. In conventional shock tube operation, the water vapor from ambient air enters the shock tube during diaphragm replacement and freezes on the shock tube walls. The use of diaphragmless shock tubes can avoid water vapor contamination and is ideal for studies of low-temperature gases\cite{Maeno_1985}.

\subsection{\label{se6b:materials}Exploring shock waves in miniature scales}
{\color{black}The interest in exploring shock wave phenomena in millimeter regimes springs from numerous interdisciplinary applications of shock waves in biology and medicine. Unlike large-scale shock tubes, the flow in miniature shock tubes is dominated by wall effects, and the attenuation of shock waves at these length scales has been investigated in recent years.}  Most of the studies utilize channels mounted at the end of a large diameter shock tube to study the propagation of shock waves in small-scales\cite{Brouillette_2003,Zhang_2017_1,Zhang_2017_2}. It would be more pertinent to produce the shock waves in a miniature shock tube as it would not be practical to use a large-scale shock tube for miniature shock wave applications. {\color{black}As highlighted earlier (in section \ref{sec1:intro}), the smaller exposed area in miniature shock tubes poses a problem for the choice of diaphragm material and thickness. Therefore, a fast-acting valve is the best method to produce shock waves in miniature shock tubes. The studies performed in the 6.35 mm shock tube \cite{Tranter_2013,Lynch_2015}, 12.7 mm shock tube \cite{Lynch_2016, Tao_2018, Lynch_2017, Tao_2019}, and 8 mm shock tube\cite{Nagaraju_2021} are great examples of the use of fast-acting valves in miniature shock tubes.} Udagawa and co-workers\cite{Udagawa_2007} used a rubber membrane-based diaphragmless shock tube with an internal diameter of 1 mm to study the formation and propagation of shock waves in small-scales tubes. They observed weak dispersed shock waves using laser differential interferometry to detect the shock wave densities. They extended their work to observe shock waves and contact surfaces in a 3 mm internal diameter small-scale shock tube over a range of pressure conditions using the interferometric signal measurement with the collimated and polarized laser beams\cite{Udagawa_2009}. The precise control of the operating pressure conditions in the diaphragmless shock tube allowed measurements over a range of pressures. This study confirmed that the shock waves do not propagate at constant velocity in this mini-shock tube. The attenuation of the shock waves in a 1 mm shock tube was also studied by a completely transparent driven section, generating the shock wave using a diaphragmless driver\cite{Garen_2009}. The propagation characteristics of the shock wave in small diameter shock tubes were studied using a double-piston diaphragmless driver based on Oguchi's design\cite{Udagawa_2012}. 

\begin{figure*}
\includegraphics[width=0.9\textwidth]{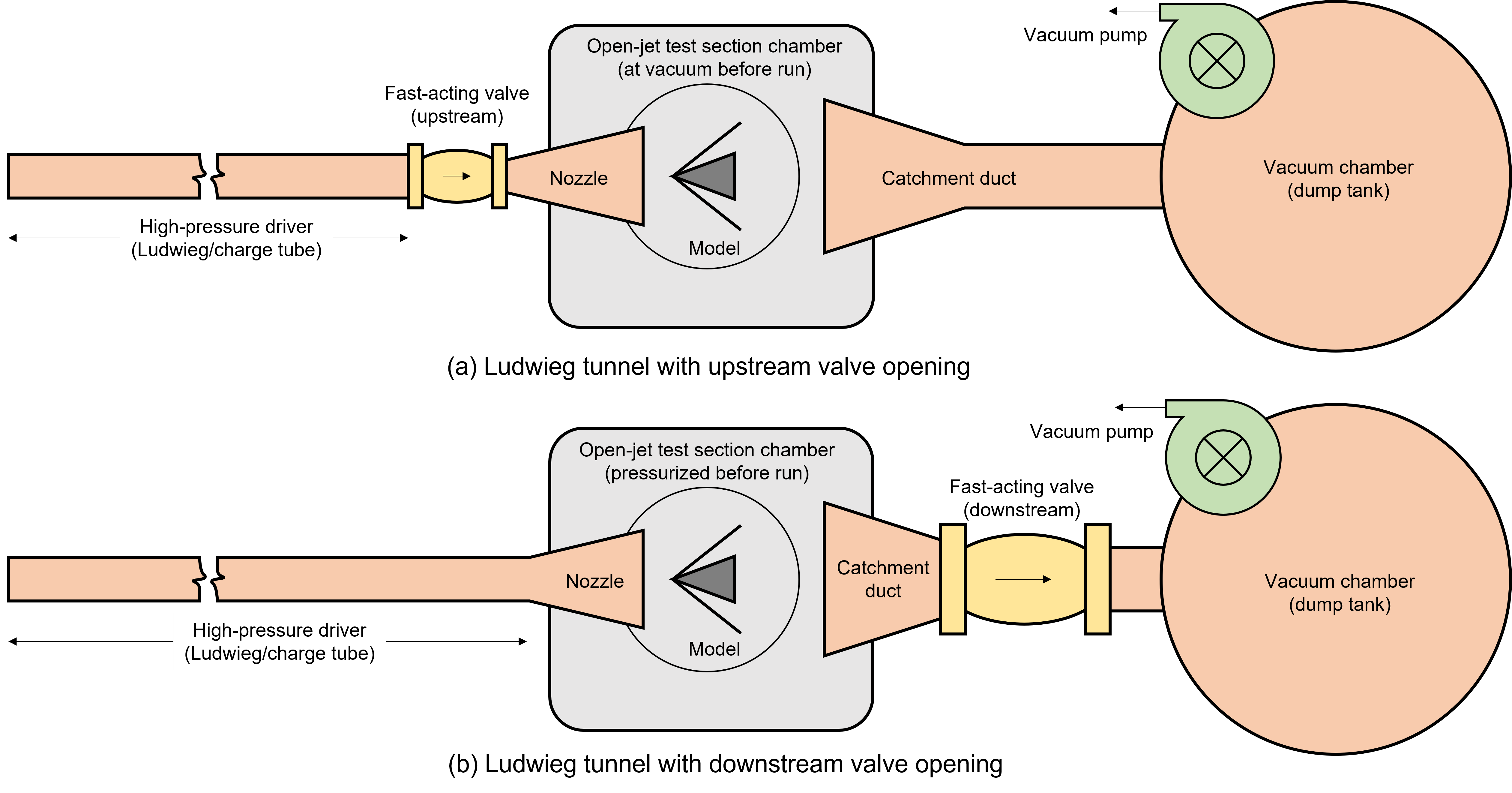}
\caption{\label{fig:ludwieg_schema} Typical schematics of Ludwieg tunnel to realize hypersonic flow in the test-section in two different configurations with the fast-acting valve placed (a) upstream and (b) downstream. Some of the vital components that constitute to the construction of the hypersonic Ludwieg tunnel are also shown in the schematics.}
\end{figure*}

\begin{figure*}
\includegraphics[width=\textwidth]{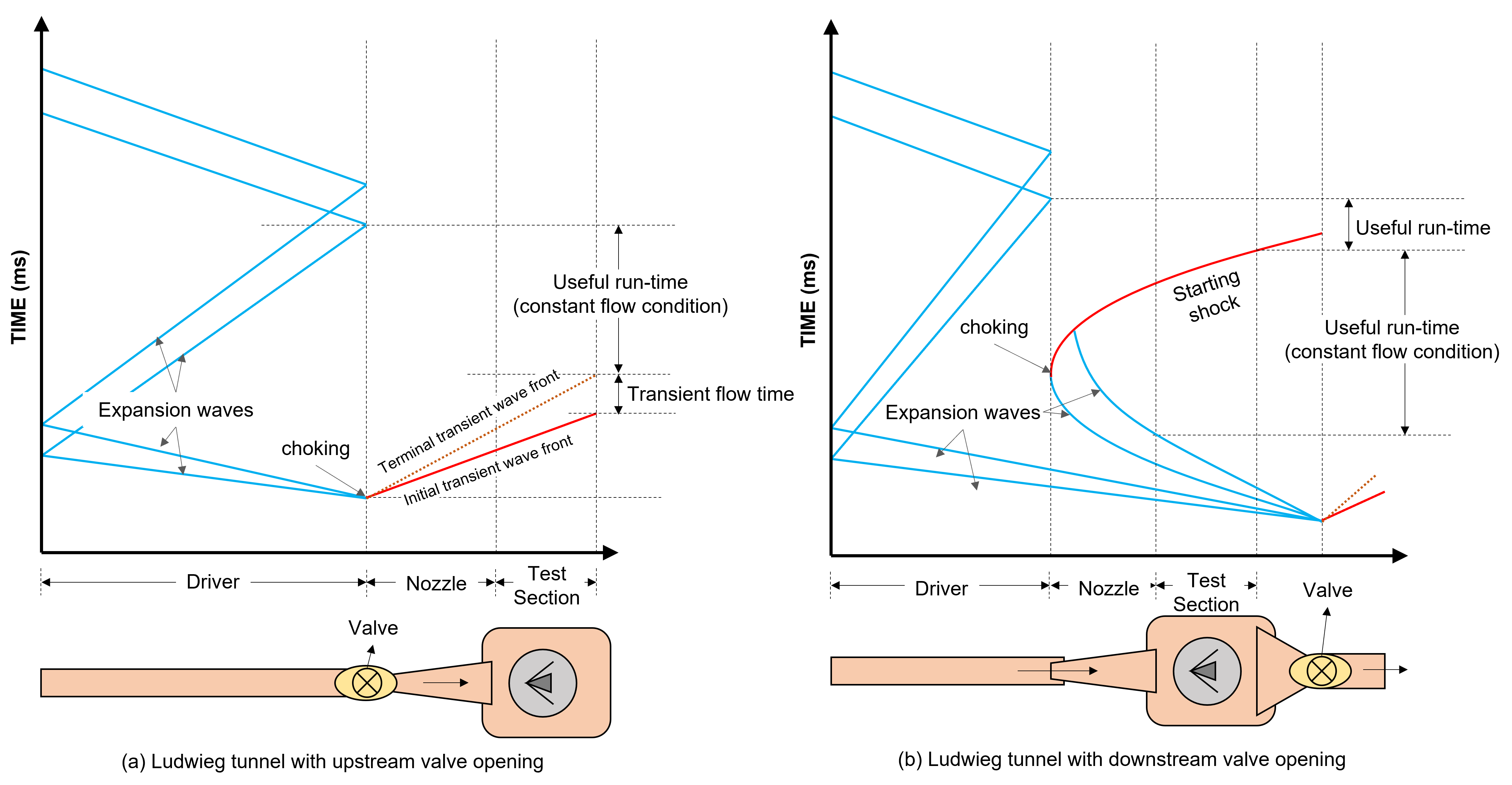}
\caption{\label{fig:xt_ludwieg} Typical trajectory of shock wave, interface, and expansion fans across the different segments of a hypersonic Ludwieg tunnel at two different operational mode: (a) upstream and (b) downstream placement of the fast-acting valve. The $x-t$ diagram indicates the attainment of useful run-time obtained during a test, typically on the order of milliseconds (ms).}
\end{figure*}

\begin{table*}
\caption{\label{tab:table2} Some of the Ludwieg tube/tunnel type aerodynamic testing facilities that use fast-acting/diaphragmless valves and some vital design parameters\footnote{Values are gathered from the published literature, and official websites of the testing facilities. Some values are not available directly in the literature and they are left unfilled. Nomenclature: $d_m$-mounting tube diameter, $t_{op}$-valve opening time, $t_{ft}$-flow test time, $T_0$-stagnation temperature, $p_0$-stagnation pressure, $M_\infty$-freestream Mach number}.}
\begin{ruledtabular}
\begin{tabular}{lccccccc}
Location \footnote{TUB-Technical University of Braunschweig, UB-University of Bremen, ZARM-Center of Applied Space Technology and Microgravity (in English), DLR-German Aerospace Center (in English), UTA-University of Texas at Arlington, AFRL-Air Force Research Laboratory, UM-University of Maryland (under construction), USAFA-United States Air Force Academy, DUT-Delft University of Technology, KAU-King Abdulaziz University, KU-Kyushu University, IISc-Indian Institute of Science, IIT-Israel Institute of Technology.} & $d_m$ (mm) & Valve Configuration \footnote{U-upstream and D-downstream of the test-section, RO-rapid opening.} & $t_{op}$ (ms) & $t_{ft}$ (ms) & $T_0$ (K) & $p_0$ (bar) & $M$\\
\midrule
TUB, Germany\cite{Estorf2004,Wolf2005,Wolf2007,Stephan2012,Stephan2013,Wagner2018,Igra_2016} & 200 & Type-II(a):U & 23 & 80 & 500 & 3-30 & 3-6\\
UB, Germany\cite{Renken1998} & - & Type-II(a):U & - & 80-130 & 900 & 100 & 11\\
ZARM-UB, Germany\cite{Rickmers2008} & 336 & Type-II(a):U & - & 120 & 1000 & 50 & $<$1 \\
DLR, Germany\cite{Rosemann1995,Costantini2015,Costantini2016} & 800 & Type-II(c):D & - & 100-1000 & 105 & 12.5 & 0.3-0.95 \\
UTA, USA\cite{Balcazar2011} & 353 & Type-II(c):D & 65 & 120 & 300 & 45 & 0.5-1.2 \\
AFRL, USA\cite{Kimmel2017} & 247.65 & Type-II(a):U & 20 & 80 & 500 & 7-27 & 6\\
UM, USA\cite{Chung2018} & - & Type-II(a):U & $\sim$5 & 100 & 1600-1800 & 60 & 6\\
USAFA, USA\cite{Cummings2012,Decker2015} & 247.65 & Type-II(a):U & $\leq$15 & 100 & 300-673 & 10-40 & 6\\
USQ, Australia\cite{Currao2020,Currao2021,Birch2018,Buttsworth2010,Buttsworth2010LudwiegTF} & 247.65 & Type-II(a):U & $\leq$15 & 100 & 300-673 & 10-40 & 6\\
DUT, The Netherlands\cite{Mathijssen2015} & 40 & Type-II(a):U & 2.1-9 & - & 573 & 1-8.5 & - \\
DUT, The Netherlands\cite{Schrijer2010} & 49.25-59 & Type-II(a):U & 10 & 100 & 579 & 2.2-88 & 6.4 10.5 \\
KAU, Saudi Arabia\cite{Juhany2006,Juhany2007} & 394.4 & Type-II(c):D & 10-17 & 95 & 300 & 2-20 & 1.9 \\
KU, Japan\cite{Matsuo1978} & 60 & RO gate-valve:D & 10-17 & 30 & 300 & 3 & 2-4 \\
IISc, India\cite{MVRao2019,Sugarno_2022} & 50 & Type-II(a):U & 10 & 35 & 300 & 10-40 & 8 \\
Technion-IIT, Israel\cite{karthick_2022,Karthick_2022b} & 25 & Type-II(a):U & $\leq$1 & 12.5 & 300-500 & 2-10 & 6 \\
\end{tabular}
\end{ruledtabular}
\end{table*}

\subsection{\label{sec6c:aerodynamics}Automating aerodynamic ground test facilities}

Many ground testing impulse facilities incorporate a diaphragm burst to initiate the flow for aerodynamic studies. Shock tunnels, free-piston shock tunnels, and Ludwieg tunnels have started replacing the diaphragm with fast-acting valves to automate the facility and avoid damage to test models by the impact of diaphragm fragments. A Ludwieg tube/tunnel is a short-duration aerodynamic test facility that produces high-speed flows between 100-1000 m/s for a 10-1000 milliseconds test duration. The total flow temperature can be maintained between 105 K and 1800 K. An automated Ludwieg tunnel (see Figure \ref{fig:ludwieg_schema}) consists of five essential components: 1. Driver or Ludwieg tube where the test gas is stored at high pressure, 2. Fast-acting valve whose response time is on the order of a few milliseconds (3-10 ms), 3. A converging-diverging (CD) nozzle accelerates the test gas to hypersonic velocity, 4. A test-section chamber to house the model under investigation, and 5. A vacuum or dump tank is kept at low pressure (on the order of millibars). Sometimes the test section and the vacuum tank are constructed as a single piece. The high pressure in the driver tube and the low pressure in the vacuum chamber facilitate an under-expanded hypersonic jet that emerges from the CD nozzle for a short duration once the fast-acting valve opens (see Figure \ref{fig:xt_ludwieg}). The expansion wave propagates in the driver tube to establish a constant total pressure and temperature ($p_0$, $T_0$) in the tube. The reflected expansion waves reach the fast-acting valve to complete the test duration given by,
\begin{equation}
    t_r = \frac{2L}{a},
\end{equation}
where $t_r$ is the run-time, $L$ is the driver tube length, and $a$ is the sound velocity at that temperature.

A fast-acting valve at high-enthalpy conditions is preferred in modern hypersonic testing facilities like the Ludwieg or shock tunnels. However, the high stagnation temperature encountered at the end of the tube poses limitations on the material wear and tear of the fast-acting valve. Recently, researchers at the University of Maryland\cite{Chung2015,Chung2018} opted to develop a high-enthalpy hypersonic short-duration tunnel by coupling the free-piston and Ludwieg tube with a fast-acting valve. The facility can reach Mach numbers between 5 to 7 with a temperature of 1600-1800 K. Many of the existing free-piston or Ludwieg facilities that study scramjet engines at high-enthalpy\cite{Jachimowski1992,COPPER1964,Aso2005,Paull1995,Stalker2005,Srinivasan2018,Wendt1996,Boyce2000,Hongru1999,BenYakar2002,Ridings,Nagamatsu1959,Stalker1996,OSGERBY1970} using rupture-diaphragms can be upgraded to fast-acting valves for better automation, high repeatability, and clean flow conditions.

Several fast-acting valve concepts have been reported for Ludwieg tunnels consolidated in Table \ref{tab:table2}. The fast-acting valve's location in the facility plays a significant role in altering the run time (see Figures \ref{fig:ludwieg_schema} and \ref{fig:xt_ludwieg}). Mounting the valve downstream of the test section ensures a clean and quiet flow upstream. Such a flow field is necessary for studying hypersonic transition-related research. However, the test section is pressurized, and mounting sensors pose difficulties. Ludwieg tunnel with a fast-acting valve is helpful for a variety of reasons. Following are some of the significant advantages: 1. simpler construction, 2. easy to automate operations and instrumentation, 4. ensuring remote operations and enhanced safety protocol, 5. lower downtime between successive runs, 6. ability to produce moderate to high enthalpy, 7. economically cheaper compared to other aerodynamic facilities, 8. repeatability of flow conditions for statistical consistency, 9. reduced turbulence intensity levels, and 10. quiet upstream conditions.

\begin{figure}
\includegraphics[width=\columnwidth]{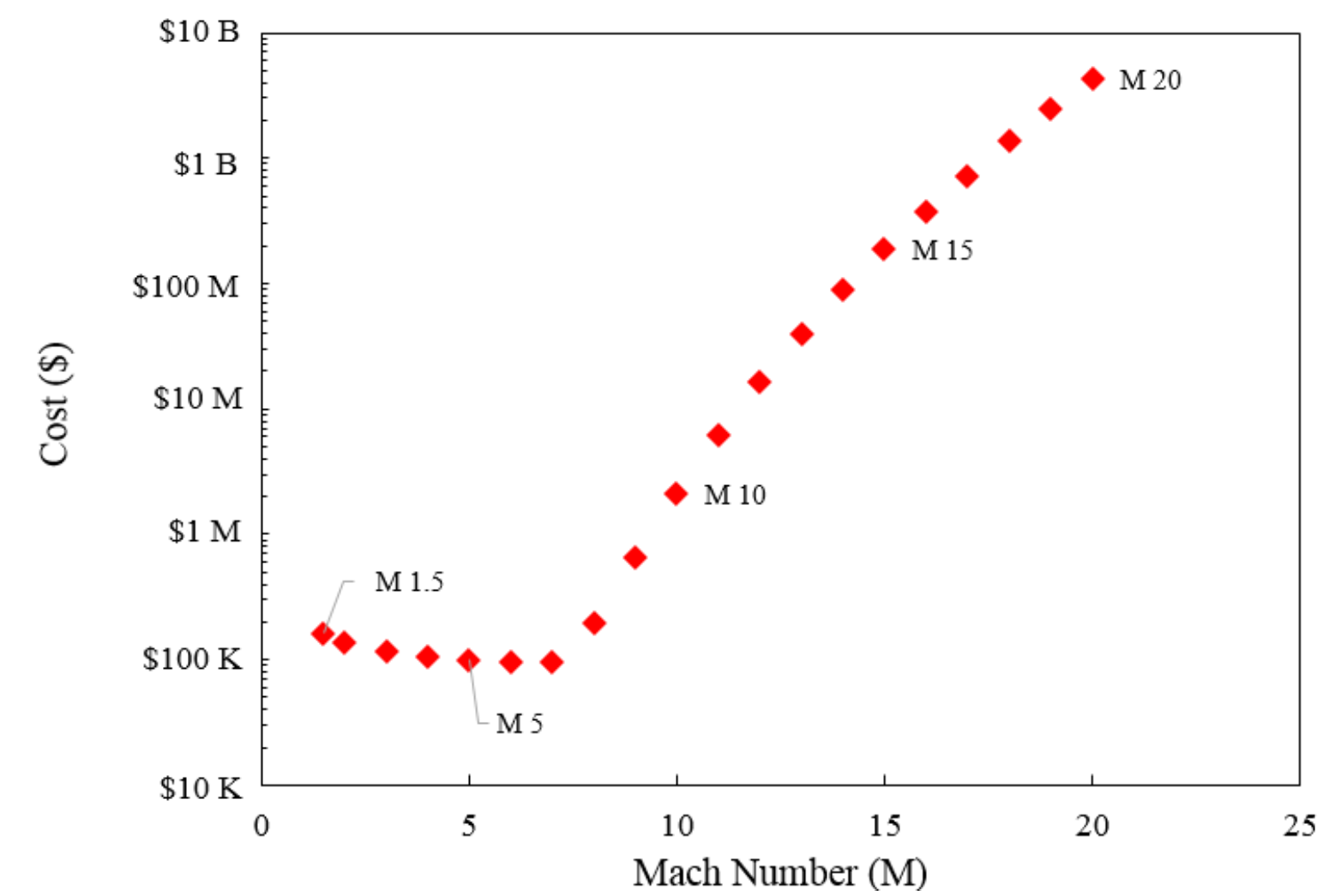}
\caption{\label{fig:cost_analysis} Cost analysis proposed by Eugene and Combs\cite{Eugene_2019} for their newly commissioned Ludwieg tube of a particular test-section configuration and exhaust conditions {\color{black}(Reprinted with permission from Eugene and Combs\cite{Eugene_2019})}.}
\end{figure}

The report of Eugene and Combs\cite{Eugene_2019} easily highlights the effect of cost requirements in Ludwieg type of facilities, either with diaphragm or fast-acting valves. A typical cost analysis graph for the present-day Ludwieg facility from their work is shown in Figure \ref{fig:cost_analysis}. A typical Ludwieg tunnel construction for $M_\infty=6$ costs only about 97,000 to 122,000 USD (excluding vacuum pump, tank and instrumentation) which runs for about 75 ms. On the other hand, from the archives of NASA history \cite{nasa_archives}, a typical construction of a moderate hypersonic blow-down wind tunnel of a similar Mach number with considerable run time was estimated to be around 30 million USD in 1949 itself. After excluding instrumentation costs, while using the commercially available Russian fast-acting valve ISTA\textsuperscript{\textregistered} KB-20-10, the complete construction of a typical hypersonic Ludwieg tunnel (75 mm diameter open jet test-section, $M_\infty=6$, and $t_r=12.5$ ms) is estimated to cost about 30,000 USD.  

\begin{figure}
\includegraphics[width=0.95\columnwidth]{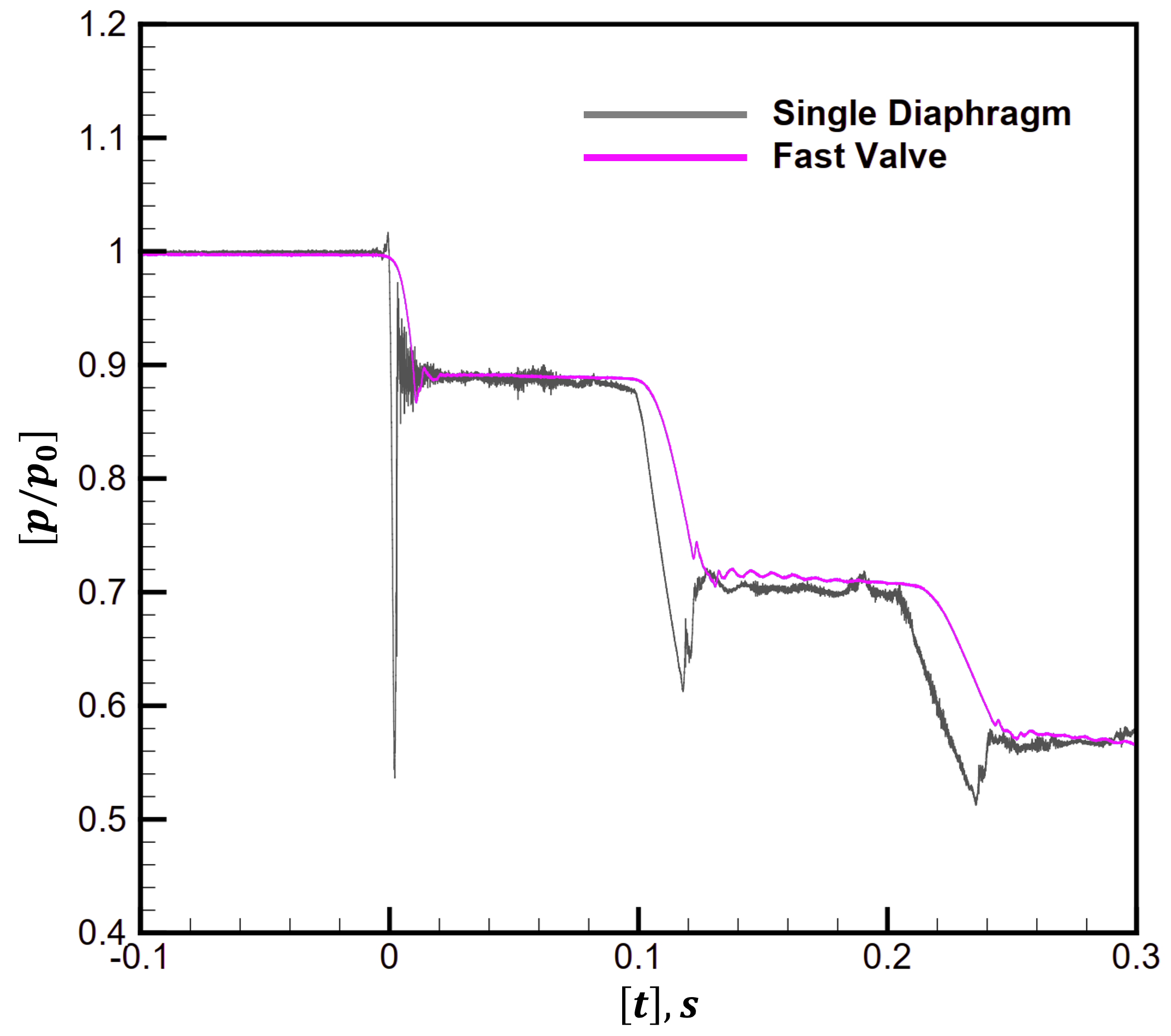}
\caption{\label{fig:low_noise} A typical run time captured signal captured by monitoring the driver tube pressure while using the diaphragm and fast-acting valve as reported in the work of Kimmel \etal \cite{Kimmel2017} {\color{black}(Reprinted with permission from AIAA: Kimmel et al.\cite{Kimmel2017}, copyright 2017)}. The $x$-axis represents the time in seconds, and the $y$-axis represents the dimensionless pressure, where $p_0$ is the initial driver pressure.}
\end{figure}

Regarding automation of the Ludwieg tunnel, the ease of assembly and operation is vividly shown in Figure 11 of Eugene \etal \cite{Eugene_2020b}. The low volume of high-pressure tubes and the relatively low run time reduce the operational risk in Ludwieg tunnels. Besides, advanced optical diagnostics use laser-based systems in addition to the existing high acquisition rate pressure sensors, which require proper timing to repeatedly sample the data during the desired part of the run time. Given the fast-acting valve replacing the diaphragm, almost continuous repeatable bursts are ensured from a remote operating location after proper automation.

{\color{black}The initial performance studies of the AFRL Ludwieg tunnel by Kimmel \etal \cite{Kimmel2017} highlight the ability of the tunnel to produce fairly consistent flow conditions as shown in Figure \ref{fig:low_noise}}. Diaphragms often produce transient flows as they are shredded into pieces during the rupture. The short times are sometimes 10-20\% of the whole run time itself and compensate for efficient data acquisition. On the other hand, fast-acting valves reduce the transient times by order and offer consistency during repeated runs. In short-duration facilities like the Ludwieg tunnels, repeatability is important as the tunnels run times are small enough for advanced optical diagnostics to collect turbulent flow statistics.

\begin{figure}
\includegraphics[width=\columnwidth]{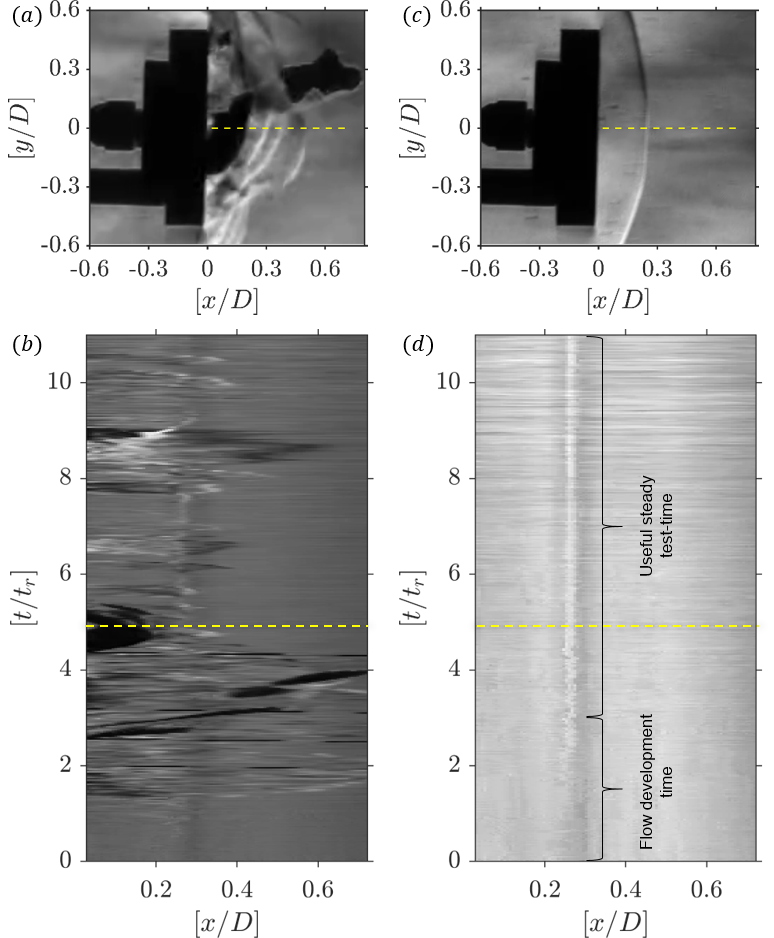}
\caption{\label{fig:paper_valve} Instantaneous schlieren imaging showing (a) the influence of the paper diaphragm in the flow field and (c) a clean flow with a steady detached shock (flow is from right to left at a freestream Mach number of $M_\infty$=6). The respective $x-t$ diagrams constructed by piling up the intensity variation along the dotted yellow line at each time instants along the center for (a) and (c) is shown in (b) and (d). The $y$-axis represents the dimensionless time, where $t_r$ is a reference time of 1 ms, and the $x$-axis represents the dimensionless spatial units, where $D$ is the diameter of the disc. The dotted yellow line particularly in (c) and (d) mark the respective time instant in which the instantaneous images (a) and (c) are taken.}
\end{figure}

The usefulness of the flow quality during the test-time can be appreciated while comparing the usage of a paper diaphragm and a fast-acting valve in a Ludwieg tunnel. For simple pitot measurements on a flat-disc, the standing shock should be stationary for an unsteady flow. If the flow has free disturbances from the non-uniform breaking of the diaphragm or its chunks, then the standing shock is no longer stationary but oscillates. In the Schlieren image shown in Figure \ref{fig:paper_valve}a, the paper diaphragm enters the test-section during the test-time and contaminates the flow to a greater extent. The shock is completely unsteady, and the oscillations can be seen in the $x-t$ diagram shown in Figure \ref{fig:paper_valve}b. On the other hand, while using the fast-acting valve, the flow is steady (Figure \ref{fig:paper_valve}c). The $x-t$ diagram shows the achieved useful steady test-time, marked by the presence of the detached shock as a solid vertical line. 

\subsection{\label{sec6e:sensor}Precise calibration of sensors}
Dynamic calibration of pressure sensors using shock tubes is a notable method due to their inherent capability to generate pressure pulses of desired amplitude and fast rise time\cite{Downes_2014,Qiang_2015}. The amplitude of step jump in pressure can be analytically calculated using the well-known shock tube relations. The pressure jump, $P_{51}$, behind the reflected shock wave in a shock tube is given by,
\begin{equation}
    P_{51} = \left[\frac{2\gamma _1M_S^2-(\gamma _1-1)}{\gamma _1+1}\right] \left[\frac{(3\gamma_1-1)M_S^2-2(\gamma _1-1)}{(\gamma _1-1)M_S^2+2}\right]
\end{equation}
where $\gamma$ is the specific heat ratio, $M_S$ is the incident shock Mach Number. Svete et al.'s method characterizes the dynamic performance of pressure transducers with a relative expanded uncertainty of less than 0.025\cite{Svete_2020_1,Svete_2020_2}. Their diaphragmless shock tube has the potential to be a primary time-varying pressure calibration standard to generate pressure steps with the desired magnitude. Experiments in the diaphragmless shock tube with a commercial fast-acting valve were performed for pressure steps with magnitudes ranging from approximately 0.83 MPa to 1.32 MPa. Sembian et al.'s novel technique using converging shock waves pushed the upper limit of the calibration to medium-high pressure range as there is an increasing need for a traceable dynamic calibration standard across wider pressure ranges\cite{Sembian_2020}. Using the converging shock waves technique, pressure pulses with peak amplitudes in the range of 30–40 MPa, with $<3.4\%$ uncertainty based on numerical reference profile, were realized.

\subsection{\label{sec6f:industrial}Novel industrial applications}
Diaphragmless shock tubes have been used for a variety of novel industrial applications. Miyachi et al.'s diaphragmless driver was an attempt to address the environmental destruction of the marine ecosystem caused by micro-organisms in ship ballast water\cite{Miyachi_2012}. The basic idea was to kill marine bacteria in a large amount of ballast water by treating it with strong pressure pulses and free radicals created from the collapse of microbubbles. The work on ballast water treatment was extended to the rapid-opening valve developed by Abe et al.\cite{Abe_2015}. The pneumatic-cylinder-based diaphragmless shock tube proposed by Hariharan et al. was designed for a variety of industrial applications\cite{Hariharan_2010}. The most notable application was the sandal oil extraction enhancement using the diaphragmless shock tube\cite{Jagadeesh_2008}. The rate and quantity of oil extracted from the sandalwood specimen exposed to shock waves were higher than conventional methods\cite{Arun_2005}. The amount of oil extracted in different sandalwood samples after shock wave loading showed an increase of $4.56-58.51\%$ compared to the oil extracted from samples not exposed to shock waves. The rate of oil extraction in the samples exposed to shock waves was substantially enhanced during the initial 30 min of the extraction process compared to nondestructive oil extraction techniques. 

\subsection{\label{sec6d:misc}Miscellaneous applications}
Diaphragmless shock tubes have been used to study shock-induced deformation of metals like aluminum, copper, and brass\cite{Nagaraja_2012,ObedSamuelraj_2018,Kubsad_2012}. Precise control over the driver pressure and repeatable shock conditions are specific advantages of using diaphragmless shock tubes in these studies. Fast-acting valves have been used for other combustion applications apart from shock tubes. As an example, a fast-acting pneumatic valve was developed for injecting hydrogen into the combustion chamber of a scramjet (supersonic combustion ramjet) engine model\cite{Morgan_1983}. This valve had an adjustable opening time with a minimum value of 10 ms. Although a commercial solenoid valve could achieve these opening times easily, customized fast-acting valve was designed to meet the space restrictions. This valve has been used for injection applications in a number of studies in transient shock tunnels\cite{Byrne_2000,Gardner_2002,Wendt1996}. Shock waves have been extended to biological and biomedical engineering research in recent years\cite{Loske_2017}. Shock tubes have been used as blast simulators to study blast-induced neurotrauma in laboratory scales. Precise control over the repeatability of pressure pulses is essential while working with biological samples. Diaphragm fragments propagating with the flow can contaminate and sometimes even be detrimental to biological samples placed at the end of the shock tube. These drawbacks are overcome by using diaphragmless shock tubes to produce low-intensity shock waves. The table-top diaphragmless shock tube developed by Swietek et al. for traumatic brain injury (TBI) studies is an economical and safe alternative to using conventional shock tubes\cite{Swietek_2019}. Teshima's vertical diaphragmless shock tube was used to study biomechanical and biological effects of high-pressure pulses on microorganisms and DNA\cite{Teshima_1995}. The facility could produce shock waves with an amplitude of 10 MPa, and a pulse width of about 25 microseconds at high repetition. Divya Prakash et al., for the first time, performed an \textit{in vivo} study of shock wave treatment of biofilms in a murine model using 50 mm diaphragmless shock tube\cite{Gnanadhas_2015}. In combination with antibiotic therapy, the studies showed that shock waves could treat lung and skin infections caused by bacteria.

\section{\label{sec7:conclusions}Concluding remarks and future scope}
Using fast-acting valves in shock tubes presents many advantages over the conventional mode of operation. The various diaphragmless driver designs developed over the last half-century have been compiled and discussed in detail in this review. Critical design features like the type of closure element, actuators, control parameters, driver-driven configurations, and valve opening time are explained. The following conclusions can be drawn from this review:
\begin{itemize}
    \item The opening time of the fast-acting valve is a vital parameter that determines the shock wave parameters obtained at the end of the shock tube. The opening time in fast-acting valves reported to date cannot match the time-scale of a diaphragm rupture. At higher pressure ratios ($P_{41}$), the deviation of the shock Mach number is significant as compared to the diaphragm-type mode of operation. 
    \item The shock formation distance is directly proportional to the opening time of the fast-acting valve. Diaphragmless shock tubes have longer shock formation distances than diaphragm-type shock tubes; therefore, longer driven sections are required. 
    \item The efficiency of a diaphragmless driver is comparable to a diaphragm-type mode of operation at lower pressure ratios ($P_{41}\leq50$). 
    \item Lighter closure elements are preferable in fast-acting valves. A cap made of high-strength composites is a good choice for faster retraction and a wide operating range.
    \item Pneumatic or electro-pneumatic actuators have fast action over long stroke movement of closure elements compared to hydraulic and purely electrical actuators. 
    \item Inline mounting of the driver and driven sections in diaphragmless shock tubes is a desirable configuration because of its closeness to replicating the wave system in a conventional shock tube and the ease of converting existing conventional shock tubes to diaphragmless shock tubes.
\end{itemize}
A mathematical model to describe the motion of the closure element in the fast-acting valve has been developed and presented. The versatility of diaphragmless shock tubes as a reliable high-temperature wave reactor, an automated aerodynamic test facility, a precise pressure calibration standard, a tool for biological research, and disruptive technology for innovative industrial applications has been illustrated with several examples. 

The design improvement of paramount importance in diaphragmless shock tubes is to shorten the opening time of fast-acting valves without inducing undesirable flow features in the shock tube. Advances in actuation technology promise more robust and compact actuators to realize quicker opening times in the future. Nevertheless, diaphragmless shock tubes in the present development stage provide limitless opportunities for applications in untapped interdisciplinary fields of science and technology. Novel applications of shock tubes are being demonstrated in material synthesis and modification, food preservation and quality enhancement, cell transformation, gene therapy, wound healing, etc. With these areas of shock wave applications being continuously explored, diaphragmless shock tubes have a very important role for many years.

\section*{Acknowledgments}
The work of authors from King Abdullah University of Science and Technology (KAUST) was funded by the baseline research funds at KAUST.

\nocite{*}
\section*{References}
\bibliography{References}
\onecolumngrid
\PRLsep
\end{document}